\documentclass[fleqn,usenatbib]{mnras}
\usepackage{lineno}

\usepackage{newtxtext,newtxmath}

\usepackage[T1]{fontenc}
\usepackage{orcidlink}

\DeclareRobustCommand{\VAN}[3]{#2}
\let\VANthebibliography\thebibliography
\def\thebibliography{\DeclareRobustCommand{\VAN}[3]{##3}\VANthebibliography}

\usepackage{graphicx}	
\usepackage{amsmath}	
\usepackage{bm}
\usepackage[mathscr]{euscript}
\DeclareMathOperator*{\argmax}{arg\,max}

\title[FlowSN]{FlowSN: Neural Simulation-Based Inference under Realistic Selection Effects applied to Supernova Cosmology}

\author[B. M. Boyd et al.]{Benjamin M. Boyd$^{1}$\thanks{E-mail: \href{mailto:bmb41@cam.ac.uk}{bmb41@cam.ac.uk}}\orcidlink{0000-0002-0622-1117
}, Kaisey S. Mandel$^{1,2}$\orcidlink{0000-0001-9846-4417},
Matthew Grayling$^{1}$\orcidlink{0000-0002-6741-983X}, Ayan Mitra$^{3,4}$\orcidlink{0000-0002-9436-8871},\newauthor Richard Kessler$^{5,6}$\orcidlink{0000-0003-3221-0419}, Maximilian Autenrieth$^{1,2}$\orcidlink{0009-0006-2068-5950}, Aaron Do$^{1}$\orcidlink{0000-0003-3429-7845}, Madeleine Ginolin$^{1}$\orcidlink{0009-0004-5311-9301}, Lisa Kelsey$^{1}$\orcidlink{0000-0003-0313-0487},\newauthor Gautham Narayan$^{3,4,7,8}$\orcidlink{0000-0001-6022-0484}, 
Matthew O’Callaghan$^{1}$\orcidlink{0009-0007-4567-7751},
Nikhil Sarin$^{1}$\orcidlink{0000-0003-2700-1030}, and Stephen Thorp$^{1}$\orcidlink{0009-0005-6323-0457} \\ 
$^{1}$Institute of Astronomy and Kavli Institute for Cosmology, University of Cambridge, Madingley Road, Cambridge, CB3 0HA, UK\\
$^{2}$Statistical Laboratory, DPMMS, University of Cambridge, Wilberforce Road, Cambridge, CB3 0WB, UK\\
$^{3}$Center for Astrophysical Surveys, National Center for Supercomputing Applications, Urbana, IL 61801, USA
\\
$^{4}$Department of Astronomy, University of Illinois Urbana-Champaign, 1002 West Green Street, Urbana, IL 61801, USA\\
$^{5}$Kavli Institute for Cosmological Physics, University of Chicago, Chicago, IL 60637, USA\\
$^{6}$Department of Astronomy and Astrophysics, University of
Chicago, 5640 South Ellis Avenue, Chicago, IL 60637, USA\\
$^{7}$Illinois Center for Advanced Studies of the Universe, University of Illinois Urbana-Champaign, Urbana, IL 61801, USA\\
$^{8}$NSF-Simons SkAI Institute, 875 North Michigan Avenue, Chicago, IL 60611, USA}

\date{Accepted XXX. Received YYY; in original form ZZZ}

\pubyear{\the\year{}}

\begin{document}
\label{firstpage}
\pagerange{\pageref{firstpage}--\pageref{lastpage}}
\maketitle

\begin{abstract}
We present \textit{FlowSN}, a statistical framework using simulation-based inference (SBI) with normalising flows to account for selection effects in observational astronomy. Failure to account for selection effects can lead to biased inference on global parameters. An example is Malmquist bias, where detection limits result in a sample skewed towards brighter objects. In Type Ia supernova (SN Ia) cosmology, these selection effects can systematically shift the inferred posterior distributions of cosmological parameters, necessitating the development of robust statistical frameworks to account for the biases. SBI enables us to implicitly learn probability distributions that are analytically intractable to calculate. In this work, we introduce a novel approach that employs a normalising flow to learn the non-analytic selected SN likelihood for a given survey from forward simulations, independent of the assumed cosmological model. The resulting likelihood approximation is incorporated into a hierarchical Bayesian framework and posterior sampling is performed using Hamiltonian Monte Carlo to obtain constraints on cosmological parameters conditioned on the observed data. The modular learnt likelihood approximation can be reused without retraining to evaluate different cosmological models, providing a key advantage over other SBI approaches. We demonstrate the performance of this methodology by training and testing the SBI technique using realistic LSST-like SNANA simulations for the first time. Our \textit{FlowSN} approach yields accurate posterior estimates on cosmological parameters, including the dark energy equation of state $w_\text{0}$, that are an order of magnitude less biased than those obtained with conventional techniques and also exhibit improved frequentist calibration.
\end{abstract}




\section{Introduction}
Type Ia supernovae are thermonuclear explosions that occur in binary systems involving at least one white dwarf. Owing to the relative uniformity of their peak luminosities, they can be used as ``standardisable candles'' in observational cosmology. Distances are inferred by comparing the modelled absolute magnitude of a supernova with its observed apparent magnitude. When combined with measurements of supernova redshifts arising from the expansion of the Universe, these distance estimates can be used to construct a Hubble diagram and constrain cosmological models \citep{riess1998,perlmutter1999,brout2022,des5yr}.

It has long been established that the absolute magnitudes of Type Ia supernovae correlate with several observable properties. In particular, supernovae that appear bluer in colour are typically brighter, as are those whose light curves evolve more slowly around maximum light \citep{phillips}. These correlations are conventionally described using a linear standardisation relation \citep{tripp1998}, which links the absolute magnitude of a supernova to its peak apparent brightness, colour at maximum light and the characteristic timescale of its light curve evolution. The relevant light curve parameters are commonly estimated using light curve fitters such as SALT \citep{guy2007,guy2010,taylor2021,kenworthy2021,kenworthy2025} or BayeSN \citep{mandel2009, mandel2011, thorp2021,mandel2022,ward2023,grayling2024}. The dependence of absolute magnitude on apparent colour and light curve shape is governed by global population-level parameters that can be inferred jointly with the absolute brightness normalisation and cosmological parameters, by analysing the full supernova sample together with their measured redshifts. Observed quantities are subject to measurement uncertainty and differ from their underlying latent values.

Selection effects are important factors to consider in any population-level analysis in observational astronomy and failure to do so can lead to biased inference of global parameters of interest. Examples of analyses where selection comes into play include quasar luminosity functions \citep{schmidt1968}, galaxy star-formation rates \citep{kauffmann2003}, gravitational wave population properties \citep{abbott2021} and exoplanet occurrences \citep{petigura2013,hu2023}. One of the most significant selection effects is Malmquist bias \citep{malmquist1922}, whereby flux-limited surveys preferentially detect apparently brighter objects. In supernova surveys, this leads to an enhanced detection probability for events that are bluer in colour or exhibit more slowly evolving light curves, since these properties are correlated with higher luminosity. Additional selection effects can arise from survey design and observing strategy, including the choice of photometric filters used for detection. Moreover, supernovae with longer-lived light curves are observable over extended time intervals and are therefore more likely to be detected, further biasing the observed sample relative to the underlying population. Selection cuts are also often applied requiring a spectroscopic redshift and classification. Robustly accounting for selection effects in supernova samples is essential in modern cosmology, as they directly impact inferences of the Universe’s expansion history and the nature of dark energy. Current cosmological debates centre on whether the resulting tensions signal new fundamental physics or instead arise from residual selection-driven systematics in the data and analysis \citep{desi2025,efstathiou2024,vincenzi2025,dhawan2025,wiseman2026}. This motivates a critical review of existing analysis techniques and the exploration of newer, interpretable methodologies.

The most popular method currently used in analyses to mitigate these biases is BEAMS with Bias Corrections \citetext{BBC, \citealt{kessler2017}}, which relies on detailed survey simulations produced with the \textit{SNANA} framework \citep{kessler2009}. SNANA provides a comprehensive simulation environment for modelling Type Ia supernova surveys, including the generation of supernova populations under different cosmological models while incorporating relevant physical effects such as host-galaxy association, dust extinction and light curve contamination from host-galaxy light. The simulations capture the full diversity of the supernova population by modelling variations in light curve shapes, colours and intrinsic scatter. The SNANA suite also simulates realistic survey pipelines that reflect observing strategies, cadence, filter sensitivity, photometric noise and the procedures used for light curve fitting. Observational selection effects arising from detection thresholds, survey downtime and gaps in cadence are naturally captured, while instrumental and calibration systematics can be included in a flexible manner. 

In one-dimensional BBC, simulated supernovae are divided into redshift bins and each event is analysed using the same fitting procedure applied to real data. Within each bin the average difference between the fitted and true distances is computed to derive a redshift-dependent (1D) bias-correction. This correction is then applied to the observed supernova sample before constructing the Hubble diagram and performing cosmological inference. BBC can also be extended to multiple dimensions, for example by additionally binning supernovae according to colour and light curve timescale for a 3D bias-correction. The method is flexible enough to account for more complex effects, such as contamination from non-Ia transients, effectively weighting the likelihood of each supernova according to its probability of being a Type Ia event \citep{beams,mitra2025}. A key limitation of BBC is that the computed bias-corrections depend on the cosmological model assumed in the simulations. The analysis is performed in multiple stages, with the inferred distances bias-corrected prior to fitting the cosmological model. Furthermore, there is ongoing debate as to whether binning on the Hubble diagram introduces systematic biases into parameter inference \citep{brout2021bin,kessler2023,karchev2025}. 

\cite{camilleri2024} investigated the impact of using different fiducial cosmological parameters to the true parameters when calculating BBC bias corrections. They found that this mismatch does introduce biases in the direction of the $\Omega_{m0}-w_0$ degeneracy, though these biases are inferior in comparison to statistical uncertainties in present-day sample sizes. The work, however, concedes that with future survey sample sizes, the bias induced by BBC may become the dominant systematic. To remedy this, \cite{camilleri2024} proposes an iterative approach to calculating bias corrections with different fiducial cosmologies determined by a previous fit.

An alternative approach to statistical modelling of Type Ia supernovae is using a hierarchical Bayesian framework \citep[e.g.][]{mandel2009,mandel2011,mandel2014,scalzo2014,mandel2017,mandel2022,thorp2021,grayling2024,sarin2026}. These models jointly describe the population distribution of supernova properties and the measurement process for individual supernovae, and can be extended to perform inference on cosmological parameters \citep{march2011,shariff2016}. Subsequent developments incorporated Malmquist bias by explicitly modelling the probability that a supernova is detected as a function of its observed properties \citep{rubin2015,march2018,rubin2023,rubin2026,hoyt2026}. The detection probability is assumed to take a simple analytic form, such as a step function or more often a smooth sigmoid function, allowing the likelihood to be analytically tractable. This yields a unified probabilistic framework in which population-level parameters and cosmological parameters are inferred consistently whilst accounting for selection effects. These approaches rely on the assumption that the survey selection function can be adequately described by a simple analytic form, which may not hold for realistic surveys \citep{rix2021}.

Simulation-based inference \citetext{SBI, \citealt{cranmer2020}} provides a complementary approach by enabling direct use of detailed survey simulations that capture complex and non-linear selection effects. In this regime, simulations are used to learn often intractable probability distributions that can be evaluated with real observed data. This can be achieved using neural density estimation \citep{rezende2015,papamakarios2021} or neural ratio estimation \citep{hermans2020}. Previous applications of simulation-based inference to Type Ia supernova cosmology have focused on learning the posterior distribution or likelihood-to-evidence ratio conditioned on the entire observed supernova sample \citep{alsing2019,karchev2023b,karchev2023,karchev2024,karchev2025,karchev2025cigars}. Given the size of current supernova datasets, these methods typically rely on data compression and require training across a wide range of cosmological parameter values for a fixed cosmological model, leading to significant computational cost. When applying simulation-based inference to population-level analyses, it is important to account for the hierarchical structure of the problem, as well as the variability in sample sizes and data lengths. This has motivated the development of hierarchical simulation-based inference architectures that are explicitly designed to respect these constraints \citep{heinrich2023,ocallaghan2025,karchev2025,ocallaghan2026,darvishi2026}.

In our previous work \citep{boyd2024}, we proposed a methodology that combines simulation-based inference with hierarchical Bayesian modelling for Type Ia supernova cosmological analyses. The approach uses simulation-based inference with normalising flows \citep{tabak2010density,tabak2013family} to train a non-analytic auxiliary density estimator that learns the observed SNe distribution, incorporating selection effects directly from simulations, while remaining independent of any assumed cosmological model. The learnt density is then used in hierarchical Bayesian models to infer cosmological parameters and other population parameters. The hierarchical structure allows the same likelihood to be reused across different cosmological models and accommodates varying sample sizes without requiring retraining. The method was validated using a simplified forward model where the exact analytical likelihood was derived for comparison.

In this work, we generalise the methodology proposed in \cite{boyd2024} and apply it to realistic survey simulations. First, we demonstrate close agreement with the tractable analytical likelihood solution of a simplified forward model, performing bias assessments and frequentist calibration checks over one hundred random survey realisations. We then train the model and perform inference using realistic SNANA survey simulations under different cosmologies whilst comparing against BBC, showing that our proposed method reduces bias and exhibits superior frequentist calibration.

In Section~\ref{sec:method}, we define the general \textit{FlowSN} statistical model, detailing the training of our normalising flow likelihood approximation and the assumptions used for posterior sampling of global parameters. We validate our methodology against a simplified forward model in Section~\ref{sec:toy}, comparing our results to posteriors obtained from a tractable analytical likelihood. In Section~\ref{sec:SNANA}, we apply the \textit{FlowSN} methodology to realistic LSST-like SNANA simulations \citep{kessler2009} and compare our results with the BBC framework \citep{kessler2017}. Finally, in Section~\ref{sec:conc}, we summarise our conclusions and discuss future avenues for the work. 

\section{Method}
\label{sec:method}

In this section we define the \textit{FlowSN} methodology. In Section~\ref{sec:forward_model}, we define the general realistic forward model starting with cosmology and supernova population parameters and resulting in our observed data post-selection. In Section~\ref{sec:hbm}, we outline the assumptions made to construct the selection-aware likelihood and sample the posterior on global parameters. Section~\ref{sec:flowsn} describes how the \textit{FlowSN} methodology uses a normalising flow to approximate the likelihood that accounts for selection effects, while Section~\ref{sec:implementation} discusses the practical implementation of this.

\subsection{Realistic General Forward Model}
\label{sec:forward_model}
We first define the assumed probabilistic generative model that includes the uncertain and stochastic effects that result in our observed light curve summary statistics after selection. We indicate parameters termed latent variables associated with an individual supernova with the subscript $s$. This distinguishes them from population-level parameters shared across all SNe, referred to as global parameters.

\subsubsection{Redshift Generation}
The true CMB-frame redshift $z_s$ for the SN with index $s$ is drawn from a distribution that is composed of a volumetric rate $R(z)$ and a differential co-moving volumetric component $\text{d}V_c/\text{d}z$, such that
\begin{equation}
\label{eq:samp_z}
z_s \sim P(z|\ \bm{C}) \propto \frac{R(z)}{1+z}\ \bigg|\frac{\text{d}V_c(z;\,\bm{C})}{\text{d}z}\bigg|. 
\end{equation}
In our work, we assume the volumetric rate to be a power-law $R(z) \propto (z+1)^{\xi_{\text{rate}}}$ defined from $z_{\text{low}}< z_s < z_{\text{max}}$ where $\xi_{\text{rate}}$ is a fixed rate parameter which describes its steepness. The $(1+z_s)$ factor in the denominator accounts for time dilation \citep{dilday2008}, while $V_c$ is the co-moving volume which itself is a function of redshift and cosmological parameter values $\bm{C}$ \citep{hogg1999}.

The  observer-frame or heliocentric redshift $z_s^{\text{hel}}$ arises from the perturbation of the CMB-redshift by peculiar velocities $v^{s}_\text{pec}$ and the observer's velocity relative to the CMB rest-frame projected along the line of sight to each SN $v^{s}_{\odot}$ \citep{davis2011,carr2022}.
\begin{equation}\label{eq:zhel}
z_s^{\text{hel}} = \bigg(z_s+1\bigg)\bigg(\frac{v^s_{\odot}}{c}+1\bigg)\bigg(\frac{v^s_{\text{pec}}}{c}+1\bigg)-1.
\end{equation}
A SN's peculiar velocity is often not known and usually assumed to be normally-distributed according to $v^s_{\text{pec}}~\sim~N(0,\sigma_{\text{pec}}^2)$ where $\sigma_{\text{pec}}\approx 300~\text{km s}^{-1}$. We can calculate $v^{s}_{\odot}$ using the SN's coordinates and our Sun's speed relative to the CMB ($\approx 369~\text{km s}^{-1}$). The speed of light in $\text{km s}^{-1}$  is denoted as $c$.

We then measure that heliocentric redshift according to
\begin{equation}\label{eq:zhel_hat}
\hat{z}_s^{\text{hel}}\sim P(\hat{z}^{\text{hel}}|\ z_s^{\text{hel}},\sigma_{z,s} ), 
\end{equation}
where $\sigma_{z,s}$ is the associated measurement uncertainty. Spectroscopic redshift uncertainties are very small $\sigma_{z,s} \approx 10^{-5}$. When the redshift is determined using photometric galaxy data only, the measurement uncertainties are larger and the likelihood may be non-Gaussian or even multimodal \citep{autenrieth2024improved}. We can convert the measured heliocentric redshift into the measured CMB-frame redshift $\hat{z}_s$ using 
\begin{equation}
     \hat{z}_s = \frac{(1+\hat{z}^{\text{hel}}_s)}{\Big(1+\frac{v^s_{\odot}}{c}\Big)}-1.
\end{equation}

\subsubsection{Luminosity Distance}

The luminosity distance is calculated using \citep{davis2011} 
\begin{equation}\label{eq:dl}
d^s_\text{L} = \bigg(z_s+1\bigg)\bigg(\frac{v^s_{\odot}}{c}+1\bigg)\bigg(\frac{v^s_{\text{pec}}}{c}+1\bigg)^2 d_c(z_s;\ \bm{C}),
\end{equation}
where the transverse co-moving distance $d_c$, in units of parsecs, is a function of the true CMB-frame redshift $z_s$ and cosmological parameter values $\bm{C}$. Both velocities enter the distance equation to account for Doppler shifting. The peculiar velocity is accounted for a second time due to relativistic beaming \citep{davis2011}.

The distance modulus $\mu$ is a deterministic function of the luminosity distance 
\begin{multline}
\label{eq:mu_func}
\mu(z, v_{\text{pec}}, v_{\odot};\, \bm{C}) = 5 \log_{10}\bigg(\frac{d_{\text{L}}}{10 \ \text{pc}}\bigg) \\= 5 \log_{10}\Bigg(\frac{1}{10\text{ pc}}
\bigg(z+1\bigg)\bigg(\frac{v_{\odot}}{c}+1\bigg)\bigg(\frac{v_{\text{pec}}}{c}+1\bigg)^2 d_c(z;\ \bm{C})\Bigg).
\end{multline}

\subsubsection{Latent Supernova Variables}
The true light curve shape $x_s$, sometimes referred to as stretch $x_1$, or $\theta$, depending on the model, is drawn from a population distribution \begin{equation}\label{eq:xpop}
x_s \sim P(x|\ z_s,\bm{\theta}_x).
\end{equation}
The population distribution is parametrised by  $\bm{\theta}_x$, which might include the centre and width of the distribution. There have been various studies into the shape of these population distributions \citep{guy2007,betoule2014,dimitriadis2025,ginolin2025a}. We also explicitly condition on $z_s$ as studies have shown that population distributions evolve as a function of cosmic time \citep{nicolas2021}. 

Similarly, we draw the SN's true $B-V$ peak apparent colour $c_s$ from a population distribution
\begin{equation}\label{eq:cpop}
c_s \sim P(c|\ x_s, z_s,\bm{\theta}_c).
\end{equation}
We parametrise the distribution with $\bm{\theta}_c$, which can influence the shape of the colour distribution \citep{ginolin2025b}. We condition on $x_s$ to allow for correlations with stretch and also on $z_s$ to consider the population distribution evolving with time \citep{popovic2025}. The red (positive) tail of the population distribution is also influenced by dust reddening, which may also be modelled explicitly \citep[e.g.][]{jha2007,mandel2011,mandel2017,mandel2022,burns2014,popovic2021,brout2021,thorp2022,thorp2024,karchev2024,grayling2024,ginolin2026}.

We then construct the latent apparent magnitude (before measurement uncertainty) $m_s$ using a simple linear regression model known as the \cite{tripp1998} formula
\begin{equation}\label{eq:tripp}
 m_s = \mu_s + M_0 + \alpha x_s + \beta c_s + \epsilon_s,
\end{equation}
where $\mu_s=\mu(z = z_s,v_{\text{pec}}=v^s_{\text{pec}}, v_{\odot}=v^s_{\odot}; \, \bm{C})$ evaluating Eq.~\eqref{eq:mu_func}. The residual scatter around the model that is unaccounted by measurement errors is $\epsilon_s\sim N(0,\sigma_{\text{res}}^2)$ with variance $\sigma_{\text{res}}^2$. The offset $M_0$ is the expected absolute magnitude for a SN
with light curve shape $x_s=0$ and colour $c_s=0$, which is degenerate with the value of the Hubble constant $H_0$. If the value of $H_0$ is not known, we can instead infer the quantity $\mathcal{M}_0 = M_0 - 5 \log_{10}h$ where $h = H_0/100  \text{ km s}^{-1} \text{ Mpc}^{-1}$. If $H_0$ is known, we can fix this in our analysis to infer $M_0$ directly, as we do in this work. The $\alpha$ coefficient captures the "broader-brighter" width-luminosity relation \citep{phillips} and the $\beta$ coefficient captures the "redder-dimmer" colour-luminosity relation. The $\alpha$ coefficient is often modelled as a single constant, although it has been observed to have a non-linear stretch dependence \citep{larison2024,ginolin2025a}. The $\beta$ coefficient parametrises the apparent colour-magnitude effect, which may be broken into the physically-distinct contributions of intrinsic variations and dust extinction \citep{mandel2017}. Furthermore, the model is often extended to include an achromatic "mass-step" \citep{kelly2010,sullivan2010,lampeitl2010,childress2013,ginolin2026}, where SNe Ia in more massive host galaxies appear brighter. Our method allows for arbitrary extensions to this linear model, however, for simplicity in this work we use its general form as shown in Eq.~\eqref{eq:tripp}.

\subsubsection{Light Curve Data and Summary Statistic Fits}
The simulated latent variables $\bm{d}_s=(m_s,c_s,x_s)^T$ (with the addition of dust parameters if explicitly modelled) can be used to determine the supernova's spectral energy distribution (SED) \citep[e.g.][]{mandel2022,kenworthy2021} . The SED $F(T, \lambda_o ;\ \bm{d}, z^{\text{hel}})$ is defined as a continuous function of observer-frame time $T$ and observer-frame wavelength $\lambda_o$,  as well as a function of the latent variables and observed heliocentric redshift. To determine the latent flux $\mathrm{f}_{s,i}$ for observation $i$ of SN $s$, taken at observer-frame time $T_{s,i}$, the SED is integrated over passband $X$'s transmission function $\mathcal{T}_X(\lambda_o)$ such that
\begin{equation}
\label{eq:flux_int}
    \mathrm{f}_{s,i} = 10^{0.4\times Z_{s,i}} \times \Bigg (\frac{\int^{\infty}_0 \lambda_o \cdot F(T_{s,i}, \lambda_o;\bm{d}_s, z^{\text{hel}}_s) \cdot \mathcal{T}_X(\lambda_o) \cdot \text{d} \lambda_o}{\int^{\infty}_0 \lambda_o\cdot F_{\text{std}}(\lambda_o)\cdot \mathcal{T}_X(\lambda_o)\cdot  \text{d} \lambda_o}\Bigg),
\end{equation}
where $F_{\text{std}}(\lambda_o)$ is the spectral flux density of a spectrophotometric standard expressed in $\text{ergs cm}^{-2}$ $\text{s}^{-1}$ $\text{\AA}^{-1}$ \citep[e.g.][]{damodel, narayan2019}. The zeropoint $Z_{s,i}$ is determined by a reference magnitude in the passband $Z_{s,i}= 27.5 + m^{\text{std}}_{s,i}$. The SED models typically convert time and wavelength into rest-frame \citep{mandel2009,kenworthy2021}. The rest-frame time (phase) $t_{s,i}$ since peak $B$ band brightness corresponding to observation $i$ of SN $s$  is calculated using $t_{s,i} \equiv (T_{s,i} - T^{\text{max}}_s)/(1+z_s^{\text{hel}})$, where $T^{\text{max}}_{s}$ is the estimated observer-frame time of peak. Similarly, the rest-frame wavelength $\lambda_r$ may be calculated using $\lambda_r \equiv \lambda_o/(1+z^{\text{hel}}_s)$.

There are various stochastic instrumental effects that influence the observed light curve data point $\hat{\mathrm{f}}_{s,i}$. For convenience, these random effects are assumed to result in a Gaussian measurement uncertainty such that 
\begin{equation}\label{eq:lc_eq}
\hat{\mathrm{f}}_{s,i} \sim N(\mathrm{f}_{s,i}, \sigma_{\mathrm{f},s,i}^2),
\end{equation}
where $\sigma_{\mathrm{f},s,i}$ is the associated measurement uncertainty in the flux. The observed light curve data for SN $s$, consisting of $N_s^{\text{obs}}$ photometric points, is represented as $\bm{\hat{\mathrm{f}}}_s = \{\hat{\mathrm{f}}_{s,i}\}_{i=1}^{N_s^{\text{obs}}}$ and observer-frame times $\bm{T}_s = \{T_{s,i}\}_{i=1}^{N_s^{\text{obs}}} $ .

Most cosmological analyses rely on light curve summary statistics determined by a specialist SN Ia fitter \citep{guy2007,guy2010,burns2011,kenworthy2021,mandel2022,grayling2024}.
 The fitters use the light curve data and heliocentric redshift $\hat{z}^{\text{hel}}_s$ to determine the observed summary statistics $\bm{\hat{d}}_s$. This is often done with maximum likelihood estimation such that
\begin{equation}\label{eq:salt_fit}
\argmax_{\bm{d}_s} P(\bm{\hat{\mathrm{f}}}_s |\bm{T}_s, \bm{d}_s, \hat{z}^{\text{hel}}_s) \rightarrow \bm{\hat{d}}_s, \bm{\Sigma}_s,
\end{equation}
where $\bm{\Sigma}_s$ is the $3\times3$ measurement covariance between the summary statistics associated with each light curve fit, which can be determined by the inverse Hessian of the negative log-likelihood with respect to $\bm{d}_s$ evaluated at $\hat{\bm{d}}_s$. Alternative methods for light curve fitting include MCMC \citep[e.g.][]{mandel2009} and variational inference \citep{uzsoy2024}.

\subsubsection{Selection}
\begin{figure}
\centering
\includegraphics[width=1.0\columnwidth]{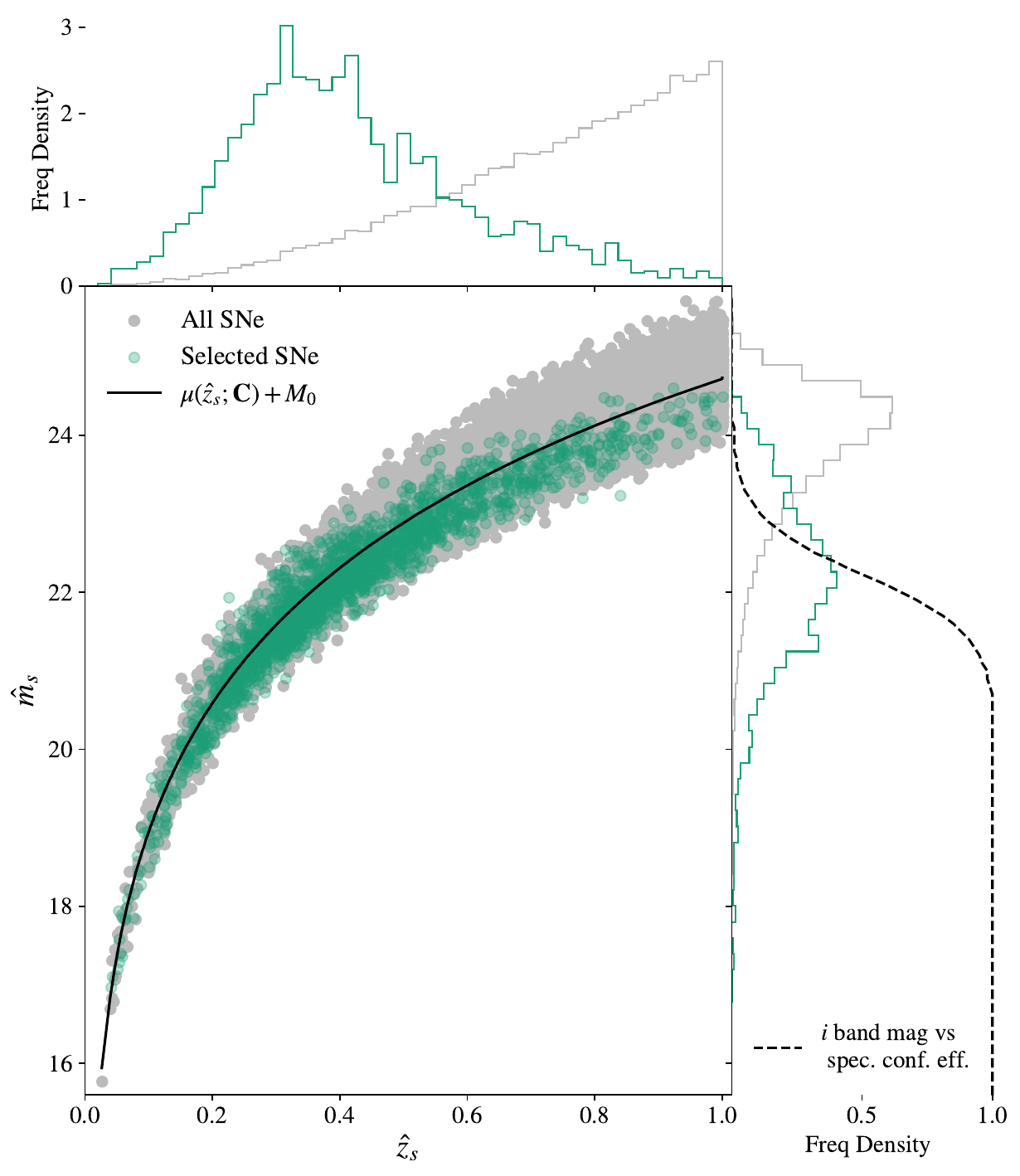}
    \vspace{0cm}
    \caption{Plot illustrating the impact of selection effects in realistic SNANA survey simulations, modelling a spectroscopic LSST Deep Drilling Fields supernova sample. At high redshift, Malmquist bias causes the selected supernovae to systematically lie below the true underlying cosmology relation curve. The histograms are normalised to unit area. The dashed curve represents the spectroscopic confirmation selection efficiency, corresponding to the mock 4MOST selection function used in \citet{plasticc}, which is one of the multiple layers of selection in our simulations.} 
    \label{fig:SNANA_data}
\end{figure}

Each survey has a different selection function that might be dependent on the light curve data, summary statistics or heliocentric redshift. We define
\begin{equation}\label{eq:detection}
I_s = \begin{cases}
0 & \text{not selected}\\
      1& \text{selected}\end{cases},
\end{equation}
where $I_s$ is an indicator variable denoting whether supernova $s$ is selected. An illustration of the effects the selection function can have on our selected data distributions is shown in Figure~\ref{fig:SNANA_data}. A SN may not be selected either because it is not observed in a given survey, or because it is observed but fails to pass the selection cuts applied prior to analysis, such as those listed in Appendix~\ref{app:SNANA_cuts}. Hereafter, for simplicity, symbols with ``$\text{obs}$'' subscript or superscript denote SNe that are both observed and pass selection cuts.

\subsection{Posterior Construction}
\label{sec:hbm}

In this subsection, we define the modelling assumptions underlying posterior inference of the global parameters.

The posterior of the global parameters of interest $\bm{\Theta}=(\bm{C},\alpha,\beta,M_0,...)$ conditioned on all the data, framed as a joint inference problem, is 
\begin{multline}
\label{eq:joint_post}
P( \bm{\Theta} |\ \bm{\hat{D}}_\text{obs}, \bm{\hat{Z}}_\text{obs}, \bm{I}_\text{obs} ) \propto \\P( \bm{\hat{D}}_\text{obs} |\  \bm{\hat{Z}}_\text{obs}, \bm{I}_\text{obs}, \bm{\Theta} ) P(\bm{\hat{Z}}_\text{obs}|\ \bm{I}_\text{obs}, \bm{\Theta}) P(\bm{\Theta} |\ \bm{I}_\text{obs}),
\end{multline}
where $\bm{\hat{D}}_\text{obs}~=~\{\bm{\hat{d}}_s\}_{s=1}^{N_{\text{SN}}^{\text{obs}}}$ denotes the observed summary statistics for the selected sample and $\bm{\hat{Z}}_\text{obs}~=~\{\hat{z}_s, \hat{z}_s^{\text{hel}}\}_{s=1}^{N_{\text{SN}}^{\text{obs}}}$ denotes the redshift measurements associated with the selected SNe. The selected sample is represented by the indicator vector of ones $\bm{I}_\text{obs}~=~\{I_s = 1\}_{s=1}^{N_{\text{SN}}^{\text{obs}}}$. The term $P(\bm{\Theta} |\ \bm{I}_\text{obs})$ is the prior on our global parameters, including $\bm{C}$, conditional on only the selected SNe $\bm{I}_\text{obs}$. The summary statistic likelihood for all the observed  SNe combined is $P( \bm{\hat{D}}_\text{obs} |\  \bm{\hat{Z}}_\text{obs}, \bm{I}_\text{obs}, \bm{\Theta} )$. The term $P(\bm{\hat{Z}}_\text{obs}|\ \bm{I}_\text{obs}, \bm{\Theta}) \equiv \mathcal{L}_{\bm{\hat{Z}}_{\text{obs}}}(\bm{\Theta})$ is an observed redshift likelihood. Traditional analyses estimate cosmological parameters by performing a regression on the Hubble diagram of distance as a function of redshift. The regression subtly assumes that the observed redshift likelihood $\mathcal{L}_{\bm{\hat{Z}}_{\text{obs}}}(\bm{\Theta})$ has a weak and negligible dependence on global parameters, including $\bm{C}$, in comparison to the constraining power of the SN distances. The assumption then implies
\begin{equation}
\label{eq:regression_post}
P( \bm{\Theta} |\ \bm{\hat{D}}_\text{obs},  \bm{\hat{Z}}_\text{obs}, \bm{I}_\text{obs} ) \propto P( \bm{\hat{D}}_\text{obs} |\ \bm{\hat{Z}}_\text{obs}, \bm{I}_\text{obs},\bm{\Theta} ) P(\bm{\Theta}| \  \bm{I}_\text{obs}),
\end{equation}
where $\bm{\hat{Z}}_\text{obs}$ becomes a covariate. For this work, we keep this assumption and model the problem as a regression so we can easily compare with BBC \citep{kessler2017}, which frames the problem in the same way. In future versions of this work, as we move to photometric redshifts, we will include a redshift likelihood. Our modelling choices also assume $P(\bm{\Theta} \mid \bm{I}_{\text{obs}}) = P(\bm{\Theta})$, implying that the number of selected SNe is non-informative for the global parameters, including cosmological parameters. This neglects the weak constraining power one can gain from selected SN counts which has been incorporated in other models \citep{karchev2025}. 

The combined summary statistic likelihood for the $N_{\text{SN}}^{\text{obs}}$ selected SNe can be written as the product of the individual observed summary statistic likelihoods, such that
\begin{equation}
\label{eq:big_lik}
P( \bm{\hat{D}}_\text{obs} |\ \bm{\hat{Z}}_\text{obs}, \bm{I}_{\text{obs}}, \bm{\Theta} ) =  \prod_{s=1}^{N_{\text{SN}}^{\text{obs}}} P\big(\bm{\hat{d}}_s\big|\,I_s=1,\hat{z}_s,\hat{z}_s^{\text{hel}},\bm{\Theta}\big).
\end{equation}
Each SN’s observed summary statistics are assumed to be conditionally independent of those of other SNe. We also assume that the selection probability for a given SN is conditionally independent of the selection of the others. There may be some covariance systematics between SNe, however, these relationships can be incorporated as conditional likelihood information to maintain the hierarchical structure of the model.

The observed summary statistic likelihood for an individual SN is defined as the marginal likelihood that is a direct function of observed redshifts and global parameters, such that
\begin{multline}
\label{eq:marj_lik}
P(\bm{\hat{d}}_s| \  I_s =1,\hat{z}_s,\hat{z}_s^{\text{hel}},\bm{\Theta}) 
\\=\int^{\infty}_{-\infty} P(\bm{\hat{d}}_s|\ I_s =1, \bm{d}_s,\hat{z}_s^{\text{hel}}) P(\bm{d}_s|\ I_s=1,
\hat{z}_s,  \bm{\Theta}) \ \text{d} \bm{d}_s,
\end{multline}
where $P(\bm{d}_s |\, I_s=1,\hat{z}_s, \bm{\Theta} )$ is the selected population distribution conditioned on the SN's observed redshift and global parameters, which can be derived using Eq.~\eqref{eq:xpop} to Eq.~\eqref{eq:tripp}. The influence of selection and measurement uncertainty on the observed summary statistic distribution is captured by $P(\bm{\hat{d}}_s|\ I_s =1, \bm{d}_s,\hat{z}_s^{\text{hel}})$.

Treating the cosmological inference as a regression imposes constraints on how we deal with redshift uncertainties. Since the true redshift $z_s$ is uncertain, the true distance $\mu_s$ is also uncertain. We write the population distribution as
\begin{multline}
\label{eq:pop_marj}
   P(\bm{d}_s |\, I_s=1,\hat{z}_s, \bm{\Theta}) \\ =  \int^{\infty}_{-\infty} P(\bm{d}_s|\ I_s=1, \mu_s, \bm{\Theta}_{\text{SN}} )\, P(\mu_s |\ I_s=1,\hat{z}_s,\bm{\Theta}_{\text{SN}}, \bm{C}) \, \text{d}\mu_s,
\end{multline}
where $P(\bm{d}_s |\, I_s=1, \mu_s, \bm{\Theta}_\text{SN} )$ is the selected population distribution conditioned on the SN's distance modulus and SN population parameters $\bm{\Theta}_{\text{SN}}=(\alpha,\beta,M_0,...)$ excluding $\bm{C}$. For a spectroscopic analysis, the uncertainty on $\mu_s$ given $\hat{z}_s$ is captured by $P(\mu_s |\ I_s=1,\hat{z}_s,\bm{\Theta}_{\text{SN}}, \bm{C})\approx P(\mu_s |\ \hat{z}_s, \bm{C})$, which we derive in Appendix~\ref{app:mu_err}. The derivation results in 
\begin{equation}
\label{eq:mu_z}
P(\mu_s|\ \hat{z}_s, \bm{C})=N\big(\mu_s|\,\mu(z=\hat{z}_s,v_\text{pec}=0,v_{\odot}=v^s_{\odot};\,\bm{C}),\sigma_{\mu|\,z,s}^2\big),
\end{equation}
where $v^s_{\odot}$ is the known motion relative to the CMB. The $\sigma_{\mu|\,z,s}^2$ variance is calculated by taking the partial derivatives of the distance modulus function Eq.~\eqref{eq:mu_func} with respect to redshift and peculiar velocity, such that
\begin{multline}
\sigma_{\mu|\,z,s}^2\approx\bigg| \frac{\partial\mu}{\partial z}\bigg|_{\substack{z=\hat{z}_s\\ v_{\text{pec}}=0}}^2 \bigg(\sigma_{z,s}^2 +\frac{\sigma_{\text{pec}}^2}{c^2}\bigg) + \bigg|\frac{\partial\mu}{\partial v_{\text{pec}}}\bigg|_{\substack{z=\hat{z}_s\\ v_{\text{pec}}=0}}^2\sigma_{\text{pec}}^2 \\
 - 2 \frac{\partial\mu}{\partial z}\bigg|_{\substack{z=\hat{z}_s\\ v_{\text{pec}}=0}} \frac{\partial\mu}{\partial v_{\text{pec}}}\bigg|_{\substack{z=\hat{z}_s\\ v_{\text{pec}}=0}} \frac{\sigma_{\text{pec}}^2}{c}.
\end{multline}
In practice, we compute the derivatives using automatic differentiation packages. In this spectroscopic analysis with $\text{z}_{\text{low}}=0.05$, the assumptions made in the propagation of redshift error are sufficient as the measurement error $\sigma_{z,s}$ is very small and the maximum influence of peculiar velocities on the distance moduli at $\text{z}_{\text{low}}$ are less than $0.1\%$. \cite{karchev2025} showed that using this approach creates bias when applied to photometric redshift errors, highlighting the importance to model the problem as a joint inference. In the future versions of this work, we could incorporate a photometric redshift likelihood, allowing the problem to be treated in a joint inference framework and making the methodology applicable to LSST photometric samples.

Given Eq.~\eqref{eq:pop_marj}, we can expand Eq.~\eqref{eq:marj_lik} by writing
\begin{multline}
\label{eq:marj_lik_long}
P(\bm{\hat{d}}_s| \ I_s =1,\hat{z}_s,\hat{z}_s^{\text{hel}},\bm{\Theta}) \\=\int^{\infty}_{-\infty} \int^{\infty}_{-\infty} P(\bm{\hat{d}}_s|\ I_s = 1, \bm{d}_s,\hat{z}_s^{\text{hel}})  P(\bm{d}_s|\, I_s=1, \mu_s, \bm{\Theta}_{\text{SN}} ) \\ \hspace{4cm}\times P(\mu_s |\ \hat{z}_s, \bm{C}) \ \text{d} \bm{d}_s \ \text{d}\mu_s\\ 
= \int^{\infty}_{-\infty}  P(\bm{\hat{d}}_s|\ I_s = 1,\mu_s,\hat{z}_s^{\text{hel}}, \bm{\Theta}_{\text{SN}}) P(\mu_s |\ \hat{z}_s, \bm{C})    \ \text{d}\mu_s.
\end{multline}
The observed summary statistic distribution after selection given the distance modulus, observed heliocentric redshift and population parameters $P\big(\bm{\hat{d}}_s\big|\,I_s=1,\mu_s,\hat{z}_s^{\text{hel}},\bm{\Theta}_{\text{SN}}\big)$ can be defined using an auxiliary density $g(\bm{\hat{d}}\big|\,m_0,\hat{z}^{\text{hel}},\bm{\Sigma},\bm{\Theta}_{\text{SN}})$, such that
\begin{multline}
\label{eq:full_aux}
P\big(\bm{\hat{d}}_s\big|\,I_s=1,\mu_s,\hat{z}_s^{\text{hel}},\bm{\Theta}_{\text{SN}}\big)\\
=\frac{\int^\infty_{-\infty}P(I_s=1\big|\,\bm{\hat{d}}_s,\hat{z}_s^{\text{hel}})P(\bm{\hat{d}}_s|\, \bm{d}_s,\hat{z}_s^{\text{hel}})  P(\bm{d}_s|\ \mu_s, \bm{\Theta}_{\text{SN}} ) \, \text{d}  \bm{d}_s }{V 
}\\=g\big(\bm{\hat{d}}_s\big|\,m^s_0=M_0+\mu_s,\hat{z}_s^{\text{hel}},\bm{\Sigma}_s,\bm{\Theta}_{\text{SN}}\big) ,
\end{multline}
where the normalisation factor $V$ is defined as 
\begin{multline}
V = \int^\infty_{-\infty}\int^\infty_{-\infty}P(I_s=1\big|\,\bm{\hat{d}}_s,\hat{z}_s^{\text{hel}})P(\bm{\hat{d}}_s|\, \bm{d}_s,\hat{z}_s^{\text{hel}}) \\ \times P(\bm{d}_s|\ \mu_s, \bm{\Theta}_{\text{SN}} )\, \text{d} \bm{d}_s \,\text{d}\bm{\hat{d}}_s.
\end{multline}
The selection probability $P(I_s=1\big|\,\bm{\hat{d}}_s,\hat{z}_s^{\text{hel}})$ is assumed to be a function of the observed heliocentric redshift and observed summary statistics, rather than the light curve data directly, which is common practice in the field \citep{rubin2015,kessler2017}.  In our work we approximate the auxiliary density $g$, written in terms of latent apparent magnitude $m_0^s$, using a normalising flow as described in Section~\ref{sec:flowsn}. For the special case of our simplified forward model outlined in Section~\ref{sec:toy}, we analytically derive $g$  in Appendix~\ref{app:toy_proof} to compare  with our normalising flow approximation.

After defining the auxiliary density $g$, we may include it in the likelihood marginalised over the uncertain distance modulus 
\begin{multline}
\label{eq:marj_lik_mu}
P(\bm{\hat{d}}_s| \ I_s=1,\hat{z}_s,\hat{z}_s^{\text{hel}},\bm{\Theta})  \\ = \int^{\infty}_{-\infty} g\big(\bm{\hat{d}}_s\big|\, m^s_0=M_0+\mu_s,\hat{z}_s^{\text{hel}},\bm{\Sigma}_s,\bm{\Theta}_{\text{SN}}\big)P(\mu_s |\ \hat{z}_s, \bm{C}) \ \text{d} \mu_s.
\end{multline}
In most cases the integral is intractable as $g$ is often non-Gaussian. In practice we instead sample the distance moduli jointly with the global parameters using the joint posterior
\begin{multline}
\label{eq:hmc_post}
P\big(\bm{\Theta}, \bm{\mu} | \, \bm{\hat{D}}_\text{obs}, \bm{\hat{Z}}_\text{obs}, \bm{I}_{\text{obs}}\big)\\ \propto
\Bigg[ \prod^{N_{\text{SN}}^{\text{obs}}}_{s=1}  g\big(\bm{\hat{d}}_s\big|\,m^s_0=M_0+\mu_s,\hat{z}_s^{\text{hel}},\bm{\Sigma}_s,\bm{\Theta}_{\text{SN}}\big)P(\mu_s |\ \hat{z}_s, \bm{C})\Bigg] P(\bm{\Theta}),
\end{multline}
where $\bm{\mu} = \{ \mu_s\}_{s=1}^{N_{\text{SN}}^{\text{obs}}}$. Sampling the joint posterior $N_{\text{MC}}$ times using Hamiltonian Monte Carlo \citep{duane1987} yields
\begin{multline}
\hspace{1cm} \{\bm{\Theta}_i,\bm{\mu}_i \}_{i=1}^{N_{\text{MC}}} \sim P\big(\bm{\Theta}, \bm{\mu} | \, \bm{\hat{D}}_\text{obs}, \bm{\hat{Z}}_\text{obs}, \bm{I}_{\text{obs}}\big)\\
\hspace{3cm}\downarrow \\ \hspace{-5cm}\{\bm{\Theta}_i\}_{i=1}^{N_{\text{MC}}} \sim P\big(\bm{\Theta} | \, \bm{\hat{D}}_\text{obs}, \bm{\hat{Z}}_\text{obs}, \bm{I}_{\text{obs}}\big),\hspace{1.95cm}
\end{multline}
where we regard $\{\bm{\Theta}_i\}_{i=1}^{N_{\text{MC}}}$ as samples from the global parameter posterior $P\big(\bm{\Theta} | \, \bm{\hat{D}}_\text{obs}, \bm{\hat{Z}}_\text{obs}, \bm{I}_{\text{obs}}\big)$ after disregarding the $\{\bm{\mu}_i\}_{i=1}^{N_{\text{MC}}}$
samples. The priors $P(\bm{\Theta})$ used in our experiments can be found in Table~\ref{tab:toy_priors} and Table~\ref{tab:snana_priors}.

\begin{figure}
\centering
\includegraphics[width=1.0\columnwidth]{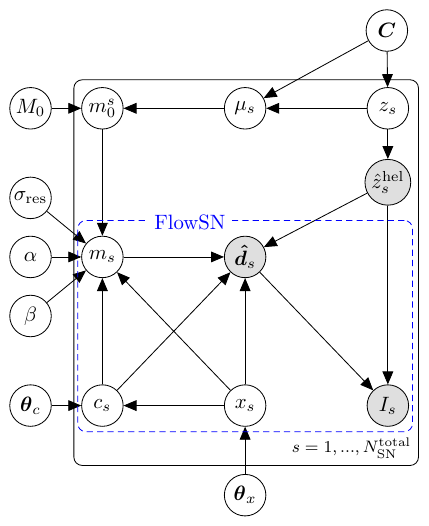}
    \vspace{-0.7cm}
    \caption{Graphical model illustrating the dependencies between
global parameters $\bm{\Theta}=(\bm{C},M_0,\alpha,\beta,\bm{\theta}_c,\bm{\theta}_x)$ to be inferred, supernova-level latent variables $(\mu_s,z_s,m_0^s,m_s,c_s,x_s)$ and observed data shaded
in grey $(\hat{d}_s,\hat{z}^{\text{hel}}_s,I_s)$. The blue dashed area demonstrates the dependencies learnt with the \textit{FlowSN} normalising flow. Each parameter is clearly defined in Section~\ref{sec:method}. Note we only have observed data for selected SNe with $I_s=1$ and train the \textit{FlowSN} normalising flow using only these examples.}
    \label{fig:graphical_model}
\end{figure}

\subsection{Training the Normalising Flow to Approximate the Likelihood}
\label{sec:flowsn}

\begin{figure*}
\centering
	\includegraphics[width=0.95\textwidth]{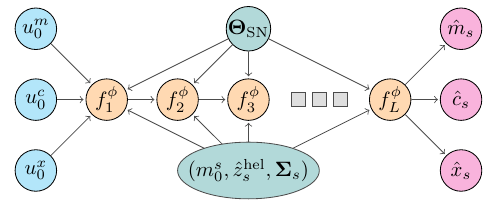}

    \caption{Schematic showing how the parameters interface with the normalising flow that models $q_\phi \big(\bm{\hat{d}}_s\big|\,m^s_0,\hat{z}^{\text{hel}}_s,\bm{\Sigma}_s,\bm{\Theta}_{\text{SN}}\big)$ in order to approximate the auxiliary density $g$ used in the observed data likelihood. The three $u_0$ parameters are each sampled from unit normal distributions and propagated through $L$ transforms $f_l$ with learnable parameters $\bm{\phi}$. These transforms are conditioned on the expected apparent magnitude $m_0^s$ for a given supernova (at zero stretch and colour), observed heliocentric redshift $\hat{z}^{\text{hel}}_s$, the measurement covariances $\bm{\Sigma}_s$, and the supernova population parameters $\bm{\Theta}_{\text{SN}}$. The outputs of $f_L$ are the three observed summary statistics corresponding to the observed apparent magnitude $\hat{m}_s$,  peak $B-V$ colour $\hat{c}_s$ and stretch $\hat{x}_s$.}
    \label{fig:flow_diag}
\end{figure*}

In this subsection we discuss our training process used to train the normalising flow to approximate the three-dimensional SN likelihood used in our $\textit{FlowSN}$ methodology.

The target likelihood is defined using an auxiliary density $g$ such that
\begin{multline}
P\big(\bm{\hat{d}}_s\big|\,I_s=1, \mu_s,\hat{z}_s^{\text{hel}},\bm{\Theta}\big)=\\g\big(\bm{\hat{d}}_s\big|\,m^s_0=M_0+\mu_s,\hat{z}_s^{\text{hel}},\bm{\Sigma}_s,\bm{\Theta}_{\text{SN}}\big),
\end{multline}
 where $\bm{\Theta}_{\text{SN}}=(\alpha, \beta,\sigma_{\text{res}},\bm{\theta}_x,\bm{\theta}_c)$ contains the SN population parameters (excluding cosmological parameters) and $m_0^s$ is the expected apparent magnitude for a SN with zero colour and stretch such that $m_0^s=\mathbb{E}[m_s|\,c_s=0,x_s=0,\mu_s,M_0]$. It is useful to define the auxiliary density in this way to take advantage of the fact that the observed data likelihood depends on $(z_s,M_0,\bm{C})$ only through the combination $m^s_0 = M_0 + \mu(z_s;\bm{C})$.  This same auxiliary density can be reused to define the likelihood for constraining different cosmological models. 
 
To learn our three-dimensional auxiliary density $g\big(\bm{\hat{d}}\big|\, m_0,\hat{z}^{\text{hel}},\bm{\Sigma},\bm{\Theta}_{\text{SN}}\big)$, we use a normalising flow 
\citep{tabak2010density,tabak2013family}. We opt for a normalising flow over a mixture density network \citep{bishop1994} because it provides the flexibility needed to model the non-analytical selection effects inherent in realistic survey simulations. The normalising flow $q_\phi$, used to approximate $g$, is defined as
\begin{equation}
\label{eq:nf}
q_{\bm{\phi}} \big(\bm{\hat{d}}\big|\ \bm{\psi}=(m_0,\hat{z}^{\text{hel}},\bm{\Sigma},\bm{\Theta}_{\text{SN}})\big) = N(\bm{u}_0 |\,\bm{0},\bm{I}) \prod^L_{l=1} \  \biggr\rvert \frac{\partial f_l^{\bm{\phi}}}{\partial \bm{u}_{l-1}}\biggr\rvert^{-1},
\end{equation}
where $\bm{u}_0 \sim N(\bm{0},\bm{I})$, $\bm{u}_l = f_l^{\bm{\phi}}\big(\bm{u}_{l-1}|\, \bm{\psi}=(m_0,\hat{z}^{\text{hel}},\bm{\Sigma},\bm{\Theta}_{\text{SN}})\big)$  and $\bm{\hat{d}} = f^{\bm{\phi}}_L(\bm{u}_{L-1}|\, \bm{\psi})$. We stack $L$ transforms $f$ with learnable parameters $\bm{\phi}$. It is simple to evaluate normalising flows with the chain rule as their architectures are designed to have lower triangular Jacobian matrices, meaning the determinant is the product of diagonal terms. Figure~\ref{fig:flow_diag} illustrates how the parameters are given to the normalising flow. For this work, we use the Masked Autoregressive Flow (MAF) architecture \citep{papamakarios2017masked} that involves stacking Masked Auto-encoder for Distribution Estimation (MADE) transform blocks \citep{germain2015made}. 

The objective is to minimize the negative log-density of the normalising flow with respect to $\bm{\phi}$ at each step over the joint training distribution of the observed summary statistics $\hat{\bm{d}}$ and conditional context $\bm{\psi}$, such that
\begin{equation}
   \min_{\bm{\phi}} \; \mathbb{E}_{(\bm{\hat{d}}_t,\bm{\psi}_t)\sim P_{\text{train}}(\bm{\hat{d}},\bm{\psi})}\Big[ -\log q_{\bm{\phi}}\big(\bm{\hat{d}}_t\big|\ \bm{\psi}_t=(m^t_0,\hat{z}_t^{\text{hel}},\bm{\Sigma}_t,\bm{\Theta}^t_{\text{SN}})\big)\Big],
\end{equation}
where we sample the training data from the joint distribution $P_{\text{train}}(\bm{\hat{d}},\bm{\psi})$. The objective is optimized at each training iteration using stochastic gradient descent in normalising flow parameter-space $\bm{\phi}$, where the expectation is approximated with empirical averages over mini-batches $\{ (\hat{\bm{d}}_t, \bm{\psi}_t) \}_{t=1}^B$ of size $B$

\begin{equation}
    \min_{\bm{\phi}}\; \frac{1}{B} \sum_{t=1}^B -\log q_{\bm{\phi}}\big(\bm{\hat{d}}_t\big|\ \bm{\psi}_t=(m^t_0,\hat{z}_t^{\text{hel}},\bm{\Sigma}_{t},\bm{\Theta}^{t}_{\text{SN}})\big).
\end{equation}
This optimisation is equivalent to minimising a Monte Carlo approximation of the forward Kullback-Leibler (KL) divergence between $g$ and $q_{\bm{\phi}}$ \citep{papamakarios2021}. 

To effectively sample $P_{\text{train}}(\bm{\hat{d}},\bm{\psi})$ and learn the influence the context $\bm{{\psi}_t}$ has on the SN summary statistics $\hat{\bm{d}}_t$, we must vary the context for each event in the training set. This starts by sampling the supernova population parameters for each SN $\bm{\Theta}^t_{\text{SN}}~=~(\alpha^t, \beta^t,\sigma^t_{\text{res}},\bm{\theta}_x^t,\bm{\theta}_c^t)$ such that
$\bm{\Theta}^t_{\text{SN}} \sim P_\text{train}(\bm{\Theta}_{\text{SN}})$. Since we are learning the likelihood, the generative population parameter priors used in the training distribution do not necessarily need to match the inferential priors used in the hierarchical Bayesian model (HBM), $P_\text{train}(\bm{\Theta}_{\text{SN}})\neq P(\bm{\Theta}_{\text{SN}})$, although both distributions need to cover the same range to avoid extrapolation. 

We must also sample observed heliocentric redshifts and latent apparent magnitudes for the training set $(\hat{z}_t^\text{hel},m_0^t)~\sim~P_\text{train}(\hat{z}^\text{hel},m_0)$. Again, this joint distribution does not need to match the real distribution. It was found that oversampling dimmer magnitudes, where Malmquist bias is strongest, improved the normalising flow's ability to learn the auxiliary density. It is also important to sample pairs of $\hat{z}_t^\text{hel}$ and $m_0^t$ that are consistent with all cosmologies that exist in the prior volume. If the simulator allows control over latent variables this can be done by sampling $\hat{z}_t^\text{hel}~\sim~P_\text{train}(\hat{z}^\text{hel})$ and $m_0^t~\sim~P_\text{train}(m_0)$ independently, ensuring no dependence on assumed cosmology. For a simulator such as SNANA \citep{kessler2009}, that only allows one cosmology to be assumed per survey simulation such that $(\hat{z}_t^\text{hel},m_0^t)~\sim~P_\text{train}(\hat{z}^\text{hel},m_0|\ \bm{C})$, we introduce a distance smearing  that overcomes this effect in Appendix~\ref{app:SNANA_sims}. Finally we (implicitly) sample the $3\times3$ measurement covariance $\bm{\Sigma}_t\sim~P_{\text{train}}(\bm{\Sigma})$. In practice, the measurement covariance matrix $\bm{\Sigma}_t$ does not require manual sampling, as it is an output of the light curve fit.

Given the sampled contextual information $\bm{\psi}_t~=~(m^t_0,\hat{z}_t^{\text{hel}},\bm{\Sigma}_{t},\bm{\Theta}^{t}_{\text{SN}})$, the simulator can then be used to sample the true auxiliary density that is learnt by the normalising flow 
\begin{equation}
\bm{\hat{d}}_t\sim g\big(\bm{\hat{d}}\big|\,\bm{\psi_t} =(m^t_0,\hat{z}_t^{\text{hel}},\bm{\Sigma}_t,\bm{\Theta}^t_{\text{SN}})\big).
\end{equation}
This completes the training set $\bm{D}_{\text{train}} = \{ (\hat{\bm{d}}_t, \bm{\psi}_t) \}_{t=1}^{N_{\text{train}}}$ containing $N_{\text{train}}$ examples.
\subsection{Implementation}
\label{sec:implementation}

 It helps in training to normalise the dimensions by first centring the magnitude data $\hat{m}_t \rightarrow \hat{m}_t-m_0^t$. After this, we apply a standardisation to both $\hat{\bm{d}}_t$ and $\bm{\psi}_t$ by transforming each dimension $x_t \rightarrow \frac{x_t - \bar{x}}{s_x}$ using the sample mean $\bar{x}$ and sample standard deviation $s_x$ over the training set. In our examples we stack $L=4$ MADE blocks, permuting the conditional order of the three dimensions in between each block. The architecture of the blocks includes two hidden layers of size 32, amounting to 14,646 total learnable parameters in the simplified forward model example and 13,366 total learnable parameters in the SNANA experiment. The sampling method requires the neural network to be at least second order continuously differentiable so we use GELU activations. We split the training set into batches containing $B=$~8,192 SNe. We use 10\% of the training data as a validation set and stop training when the validation loss gets worse for ten consecutive iterations.  We implement the normalising flow using \textit{JAX} \citep{jax2018github} to allow for compatibility with the \textit{numpyro} \citep{numpyro} HBM. Training takes around 20 minutes on a single NVIDIA A100 GPU.

 For the posterior sampling we use Hamiltonian Monte Carlo \citep{duane1987}, a technique that uses gradient information (something that is also used in neural network backpropagation) for faster convergence on the posterior. We use the No-U-Turn Sampler (NUTS) in \textit{numpyro} \citep{pyro, numpyro}, a probabilistic programming package for automatic differentiation and just-in-time (JIT) compilation built with \textit{JAX} \citep{jax2018github}. We implement the normalising flow using \textit{FlowJAX} \citep{flowjax}. In our redshift-to-distance calculations for the flat $w$CDM model we use \textit{wcosmo}\footnote{\url{https://github.com/ColmTalbot/wcosmo}} (originally made for \citealt{GWPopulation}) and for the non-flat $\Lambda$CDM model we use \textit{jax-cosmo} \citep{jax-cosmo}. Sampling the \textit{FlowSN} HBM posterior takes less than five minutes per chain on a NVIDIA A100 GPU. We sample four chains with 500 warmup and 500 posterior samples, achieving a combined effective sample size of 1,335 on average for the global parameters. When assessing the quality of our chains we diagnose convergence by confirming the Gelman-Rubin $\hat{R}$ statistic is less than 1.02 for each of the inferred parameters \citep{gelman1992,vehtari2021}. We also find that none of the chains result in divergent transitions \citep{betancourt2014,Betancourt2017}. The implementation of \textit{FlowSN} can be found at: 
\url{https://github.com/bayesn/flowsn}
\begin{figure*}
\centering
	\includegraphics[width=0.95\textwidth]{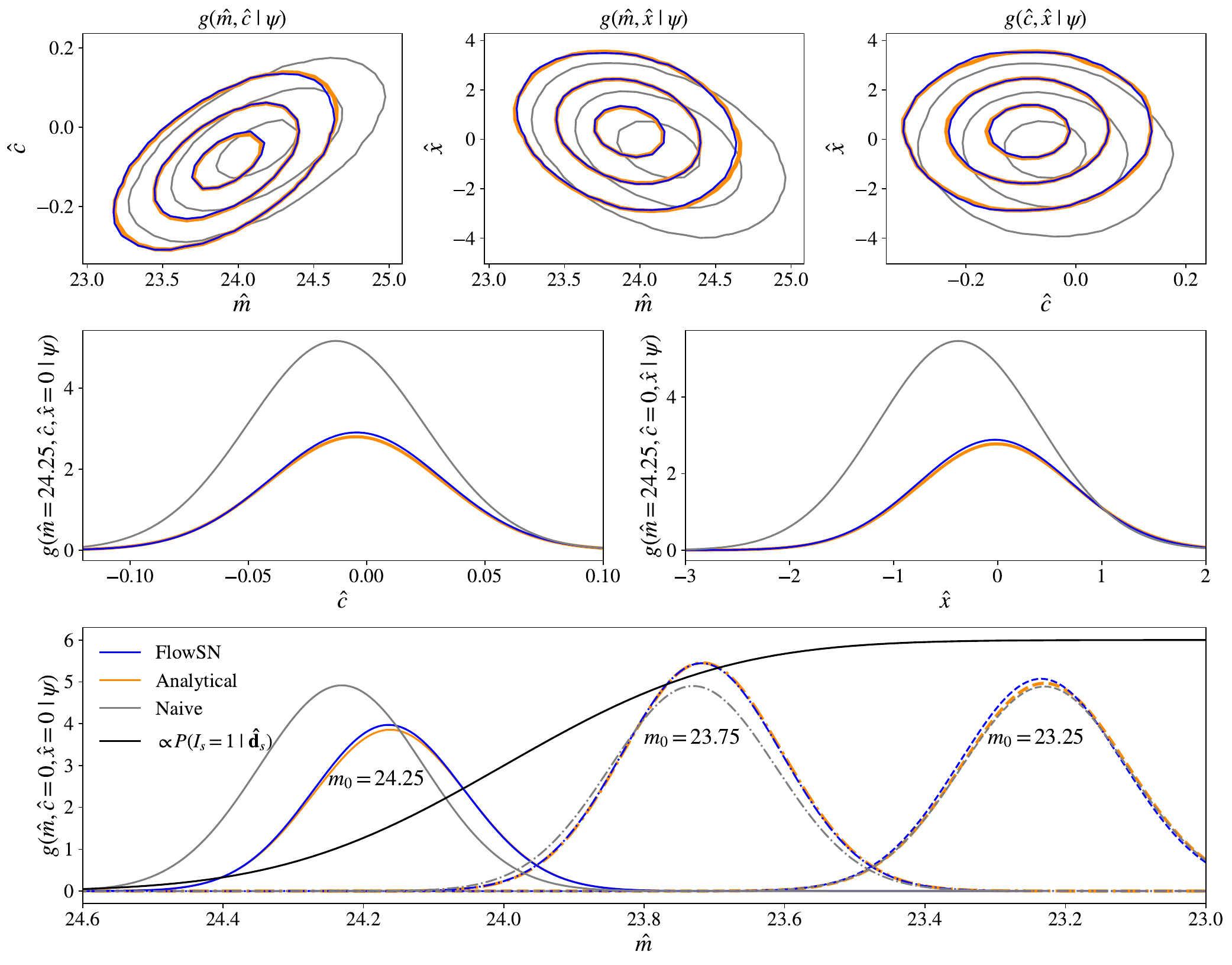}
    \vspace{0cm}
    \caption{Learnt \textit{FlowSN} three-dimensional likelihood approximation in comparison to the analytical solution and naive likelihood solution (that does not account for selection effects) for the simplified forward model. The conditional information is equal to $\bm{\psi}=(m_0,\hat{z}^{\text{hel}},\bm{\Sigma},\bm{\Theta}_{\text{SN}})$. We see as latent apparent magnitude $m_0$ gets dimmer, the analytical and naive likelihood solutions differ more in all three dimensions. Both the analytical and naive solutions are derived in Appendix~\ref{app:toy_proof}.}   
    \label{fig:toy_lik}
\end{figure*}

\begin{figure*}
\centering
	\includegraphics[width=0.95\textwidth]{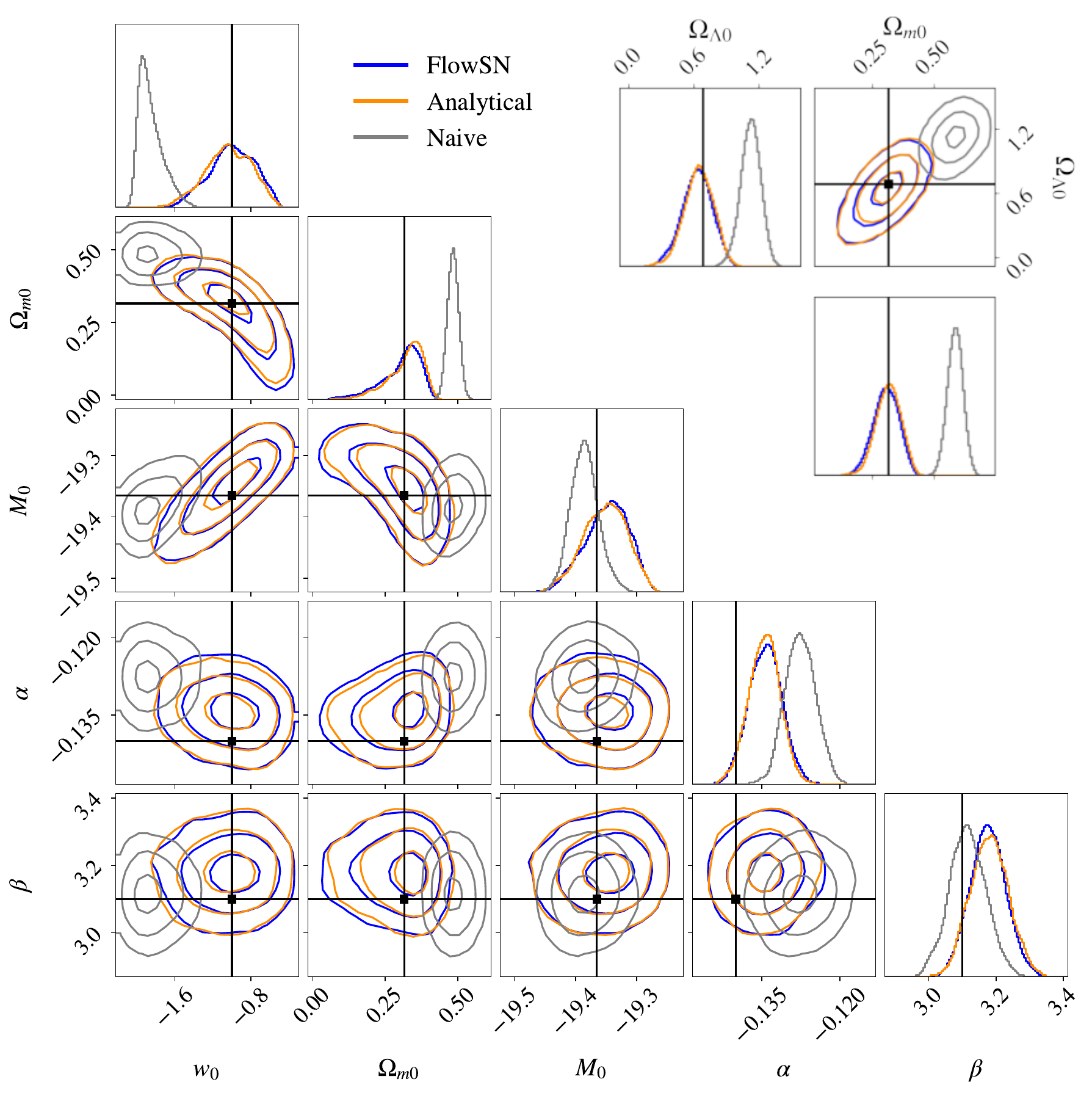}
    \vspace{0cm}    \caption{Posteriors on flat $w$CDM cosmological parameters and SN population parameters from 2000 simulated spectroscopic SNe using the simplified forward model. The top right plot shows constraints on different non-flat $\Lambda$CDM  cosmological parameters inferred without retraining the \textit{FlowSN} normalising flow. We plot the \textit{FlowSN} posterior, the analytical likelihood solution posterior and  the naive solution posterior that does not account for selection effects. Both the analytical and naive solutions are derived in Appendix~\ref{app:toy_proof}. 
    The true global parameter values used in the simulations are illustrated with straight lines. We see that the \textit{FlowSN} and analytical posteriors agree very closely, while the naive posterior is biased, particularly for the cosmological parameters. Full posteriors across eleven global parameters are shown in Figure~\ref{fig:toy_post_full}.}
    \label{fig:toy_post}
\end{figure*}

\section{Analytical Experiments}
\label{sec:toy}

To validate our \textit{FlowSN} methodology, we train and perform inference using simulations from a simplified forward model. The simulations, first described in \cite{boyd2024}, use a Gaussian cumulative distribution function to model the selection probability and assume simple Gaussian stretch and colour population distributions. For this problem we have an analytical likelihood solution to Eq.~\eqref{eq:full_aux}. We derive this analytical likelihood in Appendix~\ref{app:toy_proof} and illustrate it in Figure~\ref{fig:toy_lik}.

\subsection{Simplified Forward Model}
\label{sec:toy_sims}
We simulate our observed redshifts using equations Eq.~\eqref{eq:samp_z} to Eq.~\eqref{eq:zhel_hat}. Our true cosmology is assumed to be a flat $\Lambda$CDM cosmology $\bm{C} = ( \Omega_{m 0} = 0.315, \Omega_{\Lambda 0}  = 0.685, w_0 = -1, w_a = 0) $. We simulate 11,500 SNe between $z_{\text{low}} = 0.05$ and $z_{\text{high}} = 1.1$ with $\xi_{\text{rate}} = 1.5$. We assume a peculiar velocity dispersion of $\sigma_{\text{pec}} = 300$~$\text{km s}^{-1}$ and a negligible spectroscopic redshift uncertainty $\sigma_{z,s} = 0$.

The latent stretch $x_s$ is modelled using
\begin{equation}
\label{eq:toy_xpop}
x_s  = x_0 + \epsilon^s_x,
\end{equation}
where the additional stretch scatter is drawn from  $\epsilon^s_x \sim N(0,\sigma_x^2)$. The latent colour is modelled as
\begin{equation}
\label{eq:toy_cpop}
    c_s = c_0 + \alpha_c x_s + \epsilon^s_c.
\end{equation}
The additional colour scatter is drawn from $\epsilon^s_c \sim N(0,\sigma_{c}^2)$. Correlations between colour and stretch are parametrised by $\alpha_c$. We group stretch and colour population parameters $\bm{\theta}_x = (x_0,\sigma_x)$ and $\bm{\theta}_c = (c_0,\alpha_c,\sigma_c)$. Latent apparent magnitude $m_s$ is calculated using Eq.~\eqref{eq:tripp} with $\sigma_{\text{res}} = 0.1$. 

Instead of simulating the light curve data, we simulate summary statistics $\bm{\hat{d}}_s$ directly from latent variables $\bm{d}_s$ using a multivariate Gaussian distribution such that 
\begin{equation}
\label{eq:salt_like}
\bm{\hat{d}}_s \sim P(\bm{\hat{d}}|\ \bm{d}_s,\hat{z}_s^{\text{hel}}) = N(\hat{\bm{d}} |\, \bm{d}_s,\bm{\Sigma}_s). 
\end{equation}
Making this modelling decision ignores residual effects that the observed heliocentric redshift $\hat{z}_s^{\text{hel}}$ has on the light curve fit, which we find to be important in our experiments on SNANA simulations. We draw the summary statistic fit covariances $\bm{\Sigma}_s$ using the relations in Appendix~\ref{app:cov_samp}, roughly based on those in the Pantheon+ dataset \citep{scolnic2022}. 

Following the simulation of observed SN summary statistics $\bm{\hat{d}}_s$, we pass them through an analytical selection probability to model Malmquist bias \citep{rubin2015,rubin2023,rubin2026} such that
\begin{equation}
\label{eq:cdf}
P(I_s = 1 |\, \bm{\hat{d}}_s, \hat{z}_s^{\text{hel}}) = \Phi \Bigg(\frac{m_{\text{cut}} -(\hat{m}_s + a_{\text{cut}} \, \hat{x}_s + b_{\text{cut}} \, \hat{c}_s)}{\sigma_{\text{cut}}}\Bigg),
\end{equation} 
where $\Phi(\cdot)$ is a unit normal cumulative distribution function. The $m_{\text{cut}}$ parameter represents the limiting magnitude of the survey and $\sigma_{\text{cut}}$ represents the steepness of the drop-off around the limiting magnitude. The coefficient $a_{\text{cut}}$ is included to allow the possibility that longer lasting SNe are more likely to be detected and $b_{\text{cut}}$ allows the colour of the SN to influence the selection probability, if for example, the survey has a preference for detecting bluer events. For the purpose of our simplified forward model simulations, we set $m_{\text{cut}}=24$, $\sigma_{\text{cut}}=0.25$, $a_{\text{cut}}=-0.1$ and $b_{\text{cut}}=-1$. After the selection function is defined, we perform one Bernoulli trial for each SN with their respective selection probability to determine whether they are selected. In each random realisation there are roughly $N_{\text{SN}}^{\text{obs}}\approx$2,000 SNe remaining after selection.

\subsection{Training and Global Parameter Inference}
The global parameters to be inferred are $\bm{\Theta} = (\bm{C},\alpha,\beta,\alpha_c,M_0,\sigma_{\text{res}},c_0,\sigma_c,x_0,\sigma_x)$. The true values of each global parameter, as well as their inferences, are in Table~\ref{tab:toy_results}. When performing inference we sample the posterior Eq.~\eqref{eq:hmc_post} using the global parameter priors in Table~\ref{tab:toy_priors}. For the analytical procedure, we use the exact analytical likelihood solution to Eq.~\eqref{eq:full_aux}, derived in Appendix~\ref{app:toy_proof}. For the \textit{FlowSN} results, we use the likelihood approximation and training procedure defined in Section~\ref{sec:flowsn}. The population parameter training ranges $\bm{\Theta}^t_{\text{SN}} \sim P_\text{train}(\bm{\Theta}_{\text{SN}})$ are in Table~\ref{tab:toy_priors}. We draw summary statistic fit covariances from the relations in Appendix~\ref{app:cov_samp}. To demonstrate that \textit{FlowSN} can be trained independent of a cosmological model, we sample latent $m^t_0$ directly in training using the scheme described in Appendix~\ref{app:toy_sims}.  We oversample dim SNe to ensure there are enough examples of dim objects at the detection limit where the selection effect has the greatest influence. Observed heliocentric redshift does not feature in Eq.~\eqref{eq:salt_like} or Eq.~\eqref{eq:cdf} explicitly, meaning the normalising flow does not need to be conditioned on it for this example. We continuously generate SNe until there are $N^{\text{train}}= 20$~million SNe in our training set after selection.

\begin{table*}
	\centering
	\caption{Average of posterior median estimates over 100 repeats and the standard error on this average for the simplified forward model experiments. Results using the \textit{FlowSN} methodology, as well as those determined with the analytical likelihood solution, are included. Note uncertainties are the standard error on the average so are a factor of 10 smaller than those in individual experiments. The $\sigma$ values are calculated using unrounded averages and standard errors.}

	\label{tab:toy_results}

\begin{tabular}{lcccccc}
\hline 
 
Test &  &  \multicolumn{2}{c}{Flat $w$CDM} &  \multicolumn{2}{c}{$\Lambda$CDM}\\
\hline

Global Parameter & Truth  & \textit{FlowSN}&Analytical & \textit{FlowSN} &Analytical \\
\hline
$w_0$&-1&-1.02 $\pm$ 0.02 (-1.0$\sigma$)&-1.06 $\pm$ 0.02 (-2.7$\sigma$)&Fixed & Fixed \\
$\Omega_{m0}$&0.315&0.304 $\pm$ 0.006 (-1.9$\sigma$)&0.316 $\pm$ 0.005 (0.2$\sigma$)&0.302 $\pm$ 0.005 (-2.5$\sigma$)&0.314 $\pm$ 0.005 (-0.3$\sigma$)\\
$\Omega_{\Lambda 0}$&0.685& Fixed & Fixed &0.657 $\pm$ 0.014 (-2.1$\sigma$)&0.678 $\pm$ 0.013 (-0.5$\sigma$)\\
$M_0$&-19.365&-19.369 $\pm$ 0.003 (-1.5$\sigma$)&-19.372 $\pm$ 0.003 (-2.3$\sigma$)&-19.361 $\pm$ 0.003 (1.3$\sigma$)&-19.364 $\pm$ 0.003 (0.5$\sigma$)\\
$\alpha$&-0.14&-0.1401 $\pm$ 0.0003 (-0.2$\sigma$)&-0.1399 $\pm$ 0.0003 (0.2$\sigma$)&-0.1402 $\pm$ 0.0003 (-0.5$\sigma$)&-0.14 $\pm$ 0.0003 (-0.1$\sigma$)\\
$\beta$&3.1&3.096 $\pm$ 0.005 (-0.8$\sigma$)&3.097 $\pm$ 0.005 (-0.6$\sigma$)&3.099 $\pm$ 0.005 (-0.2$\sigma$)&3.099 $\pm$ 0.005 (-0.1$\sigma$)\\
$\sigma_{\text{res}}$&0.1&0.1001 $\pm$ 0.0003 (0.2$\sigma$)&0.0999 $\pm$ 0.0003 (-0.3$\sigma$)&0.1001 $\pm$ 0.0003 (0.4$\sigma$)&0.1 $\pm$ 0.0003 (-0.2$\sigma$)\\
$c_0$&-0.061&-0.0612 $\pm$ 0.0002 (-1.4$\sigma$)&-0.0604 $\pm$ 0.0002 (3.4$\sigma$)&-0.0612 $\pm$ 0.0002 (-1.3$\sigma$)&-0.0604 $\pm$ 0.0002 (3.7$\sigma$)\\
$\alpha_c$&-0.008&-0.0083 $\pm$ 0.0001 (-1.7$\sigma$)&-0.008 $\pm$ 0.0002 (-0.3$\sigma$)&-0.0083 $\pm$ 0.0001 (-1.7$\sigma$)&-0.0081 $\pm$ 0.0002 (-0.4$\sigma$)\\
$\sigma_c$&0.065&0.0653 $\pm$ 0.0001 (2.0$\sigma$)&0.0651 $\pm$ 0.0001 (0.6$\sigma$)&0.0653 $\pm$ 0.0001 (1.9$\sigma$)&0.0651 $\pm$ 0.0001 (0.5$\sigma$)\\
$x_0$&-0.432&-0.44 $\pm$ 0.003 (-2.8$\sigma$)&-0.442 $\pm$ 0.003 (-3.5$\sigma$)&-0.44 $\pm$ 0.003 (-2.7$\sigma$)&-0.442 $\pm$ 0.003 (-3.4$\sigma$)\\
$\sigma_x$&1.124&1.128 $\pm$ 0.002 (2.2$\sigma$)&1.125 $\pm$ 0.002 (0.6$\sigma$)&1.128 $\pm$ 0.002 (1.8$\sigma$)&1.124 $\pm$ 0.002 (0.2$\sigma$)\\

\hline
\end{tabular}
\end{table*}
\subsection{Results and Discussion}

The first step of the \textit{FlowSN} methodology involves training a normalising flow using the simulations in Section~\ref{sec:toy_sims} following the procedure described in Section~\ref{sec:flowsn}. For these simplified forward model simulations, we have an analytical likelihood solution, derived in Appendix~\ref{app:toy_proof}, meaning we can directly compare the quality of density estimation before we do inference. Figure~\ref{fig:toy_lik} shows the learnt normalising flow likelihood approximation agrees closely with the analytical solution. We see that SNe brighter than $m_{\text{cut}}$, that are uninfluenced by Malmquist bias, have likelihoods that are strongly tied to the naive solution that does not consider selection effects. As we go dimmer, the analytical and naive solutions begin to diverge. The naive likelihood solution assumes a normal density centred at the evaluated latent apparent magnitude, however, the analytical likelihood pushes the density towards brighter magnitudes. We see on the top two rows in  Figure~\ref{fig:toy_lik} that the differences between analytical and naive likelihoods extend to the colour and stretch dimensions, as well as their correlations. The figure shows that the \textit{FlowSN} likelihood approximation is able to recognise all these variations and assigns probability density in almost exact agreement with the analytical solution, providing reassurance before sampling the posterior.

We then use the same single normalising flow likelihood approximation in posterior sampling using Hamiltonian Monte Carlo. Since we only condition the normalising flow on a latent apparent magnitude $m_0^s$, we can infer posteriors on multiple sets of cosmological parameters without the need to retrain. This is a significant advantage over existing SBI approaches to solving this problem, as they would need to be retrained for every cosmological model constrained. To demonstrate this advantage, we sample posteriors for a flat $w$CDM model $\bm{C}_1 = (w_0,\Omega_{\text{m0}})$ and non-flat $\Lambda$CDM model $\bm{C}_2 = (\Omega_{\text{m0}},\Omega_{\Lambda 0})$. For these experiments we perform inference on $N_{\text{SN}}^{\text{obs}} \approx$2,000 selected SNe. Figure~\ref{fig:toy_post} shows the resulting \textit{FlowSN} posteriors for the two cosmological models, plotted along with the posteriors sampled using the analytical and naive likelihoods for this simplified forward model. The figure shows close agreement between the \textit{FlowSN} posteriors and the analytical, recovering the true cosmology, as well as the \cite{tripp1998} formula coefficients in Eq.~\eqref{eq:tripp}, within 2$\sigma$. For the $w$CDM model, \textit{FlowSN} inferred cosmological parameters $w_0 = -1.00\pm 0.23$ and $\Omega_{m0} = 0.30 \pm 0.07$, while the analytical likelihood solution inferred  $w_0 = -1.04\pm 0.24$ and $\Omega_{m0} = 0.31 \pm 0.06$. Meanwhile for the non-flat $\Lambda$CDM model, \textit{FlowSN} inferred  $\Omega_{\Lambda0} = 0.63 \pm 0.14$ and  $\Omega_{m0} = 0.30 \pm 0.05$, with the analytical likelihood solution in close agreement with $\Omega_{\Lambda0} = 0.64 \pm 0.14$ and  $\Omega_{m0} = 0.31 \pm 0.05$. The posteriors using the naive likelihood solution, that does not account for selection effects, results in biased posteriors, particularly for the cosmological parameters which are 5$\sigma$ away from the true values. Figure~\ref{fig:toy_post_full} shows the full posterior over all 11 global parameters, including those describing the colour and stretch population distributions and their correlations, again showing near-perfect agreement between the \textit{FlowSN} and analytical likelihood solution inferences.

To further investigate the reliability of our \textit{FlowSN} methodology, we repeat the simplified forward model experiment on $N_{\text{SN}}^{\text{obs}} \approx$~2000 selected SNe one hundred times. The random realisations have the same true global parameters, but have different random seeds in the generative forward model, resulting in varying latent variables and observed summary statistics. Table \ref{tab:toy_results} shows the average posterior prediction from the hundred random realisations along with the standard error of the predictions. The results show \textit{FlowSN} consistently recovers the true global parameters within 1\% on average. It also shows close agreement with the analytical solution agreeing by less than two standard errors for most parameters. With \textit{FlowSN}, we see almost exact agreement for the non-cosmological SN population parameter inferences when constraining different cosmologies. This consistency of SN population parameter inferences between models demonstrates a clear advantage for our method, as this reproducibility cannot be guaranteed in other SBI approaches where the neural network is retrained for each cosmological model. Finally, Figure~\ref{fig:toy_calib} assesses the empirical frequentist coverage of our Bayesian credible regions, demonstrating that all 11 global parameters, including cosmology, are adequately calibrated within the 95\% Monte Carlo confidence band. The calibration checks are important when using SBI, as some methods have been found to be overconfident \citep{hermans2021}.

These convincing results on the simplified forward model, where we have an analytical likelihood solution, give us additional confidence in the \textit{FlowSN} methodology when moving to regimes where the analytical solution is intractable.

\section{Experiments with SNANA Simulations}
\label{sec:SNANA}
In this section we train and perform inference on SNANA simulations \citep{kessler2009}. Since these are non-linear complex forward models, there is no analytical likelihood solution, so we instead compare with BBC \citep{kessler2017}.

\subsection{SNANA Forward Model}
\label{sec:snana_model}
We simulate realistic LSST \citep{lsst2019} Deep Drilling Fields (DDFs) SNe using the SNANA simulator \citep{kessler2009}. The sample is constructed to represent spectroscopically confirmed Time-Domain Extragalactic Survey  \citetext{TiDES; \citealt{tides}} SNe, observed with the 4-metre Multi-Object Spectroscopic Telescope spectrograph \citetext{4MOST; \citealt{4most}}. Since this is a spectroscopically confirmed SN Ia sample, we assume there are no contaminants, i.e. misclassified supernovae.

We test the methods on three survey simulations using three different sets of true cosmological parameters, defined in Table~\ref{tab:snana_results}. SNe are simulated from a power-law redshift distribution between $z_{\text{low}} = 0.05$ and $z_{\text{high}} = 1$ with $\xi_{\text{rate}} = 1.5$, assuming cosmology $\bm{C}$.  We adopt a peculiar velocity dispersion of $\sigma_{\text{pec}} = 300$~$\text{km s}^{-1}$. Observed heliocentric redshifts are computed using each SN’s simulated sky coordinates relative to the CMB, assuming a small spectroscopic redshift uncertainty of $\sigma_{z,s} = 1 \times 10^{-5}$. For this initial demonstration of the \textit{FlowSN} methodology, the SNe are generated using the SALT3.K21 \citep{kenworthy2021} trained model with the \cite{guy2010} scatter model, and the residual scatter is set to $\sigma_{\text{res}} = 0.1$. Future iterations of \textit{FlowSN} will include more complex scatter models including explicit dust treatment. Population asymmetric colour and stretch distributions follow \cite{scolnic2016} and are redefined in Appendix~\ref{app:cxpop}. No correlation is assumed between the simulated colour and stretch. SN distances are computed according to Eq.~\eqref{eq:dl} and Eq.~\eqref{eq:mu_func}, given $\bm{C}$. The SNe follow the simple \cite{tripp1998} relation defined in Eq.~\eqref{eq:tripp} with $\alpha = -0.14$ and $\beta = 3.1$. The simulations assume a \cite{fitzpatrick1999} dust law and use a \cite{schlafly2011} dust map. 

There are three layers of selection in the forward model:
\begin{itemize}
    \item We simulate LSST $ugrizy$ light curve data using realistic cadences and observation plans, using the \cite{mitra23} high-z DDF sample as a template. Each band has its own efficiency that impacts light curve quality \citep{sanchez2022}. Only SNe with at least five observations proceed to the light curve fitting stage.
    \item Contrary to \cite{mitra23}, we select only SNe Ia that have been confirmed through spectroscopic follow-up. To model this, we use the mock 4MOST spectroscopic confirmation selection efficiency from the PLAsTiCC simulations \citep{plasticc}, which mimics the criteria for triggering spectroscopic follow-up. This efficiency is defined by the latent peak apparent magnitude in the $i$ band \citep{kessler2019} and is the dominant selection function over the LSST photometric selection function.
    \item There are standard cuts based on quality and fit of the light curve summary statistics \citep{betoule2014} that we reproduce in Appendix~\ref{app:SNANA_cuts}.
\end{itemize}
Figure~\ref{fig:SNANA_data} shows the influence that the selection effects have on our observed SN distributions and Hubble diagram. 
The SALT3 model \citep{kenworthy2021} is used to fit the light curves to obtain the observed summary statistics. We produce several test datasets of varying sizes from $N_{\text{SN}}^{\text{obs}} \approx$~2,000 selected SNe to $N_{\text{SN}}^{\text{obs}} \approx$~200,000 selected SNe.

\begin{figure}
\centering
	\includegraphics[width=1.0\columnwidth]{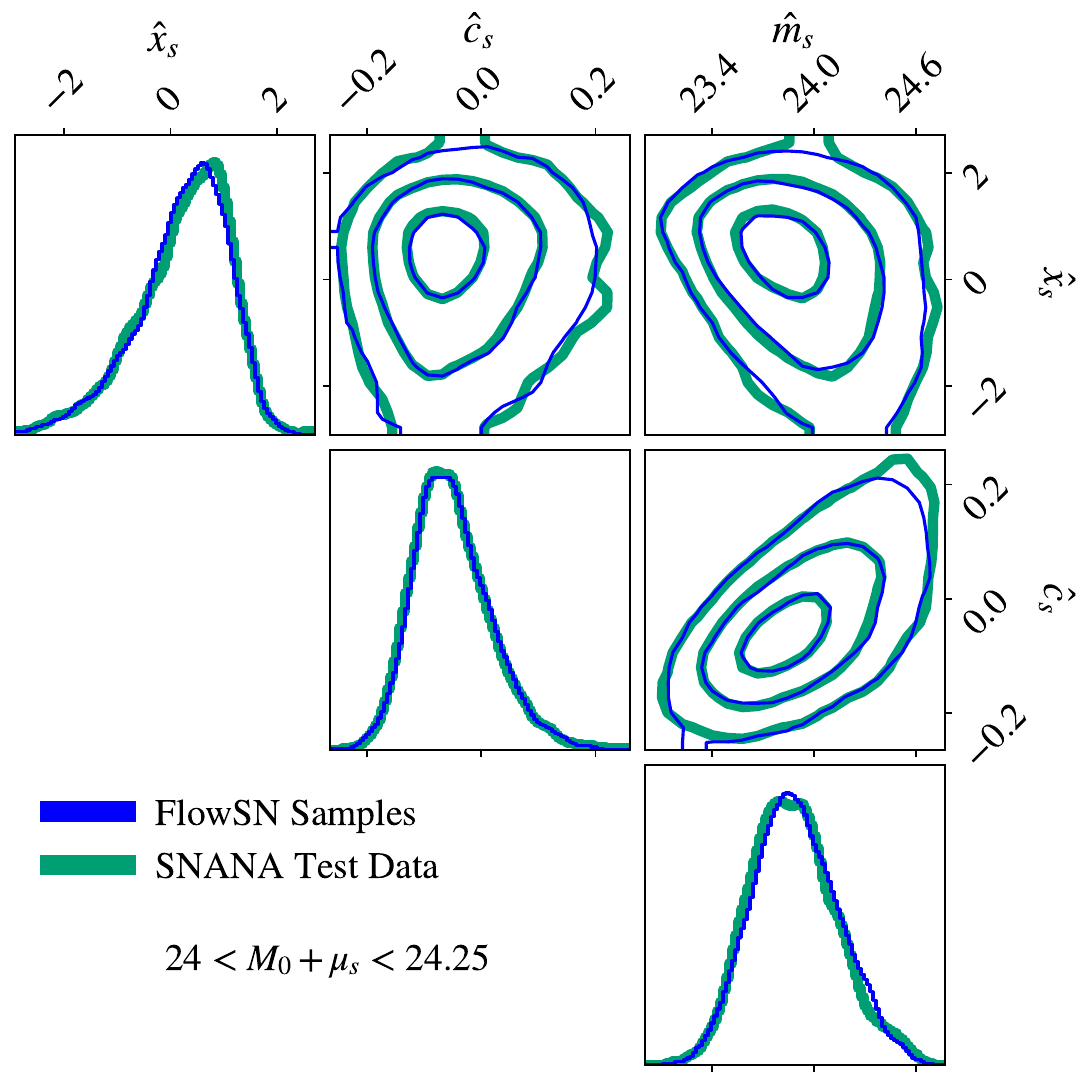}

    \caption{\textit{FlowSN} normalising flow samples conditioned on SNe with dim latent apparent magnitudes $m_0^s = M_0 + \mu_s$ between 24 and 24.5, where Malmquist bias is strongest. We also plot the underlying data distribution from the held-out SNANA test set.}
    \label{fig:SNANA_samples}
\end{figure}
\begin{figure*}
\centering
	\includegraphics[width=0.95\textwidth]{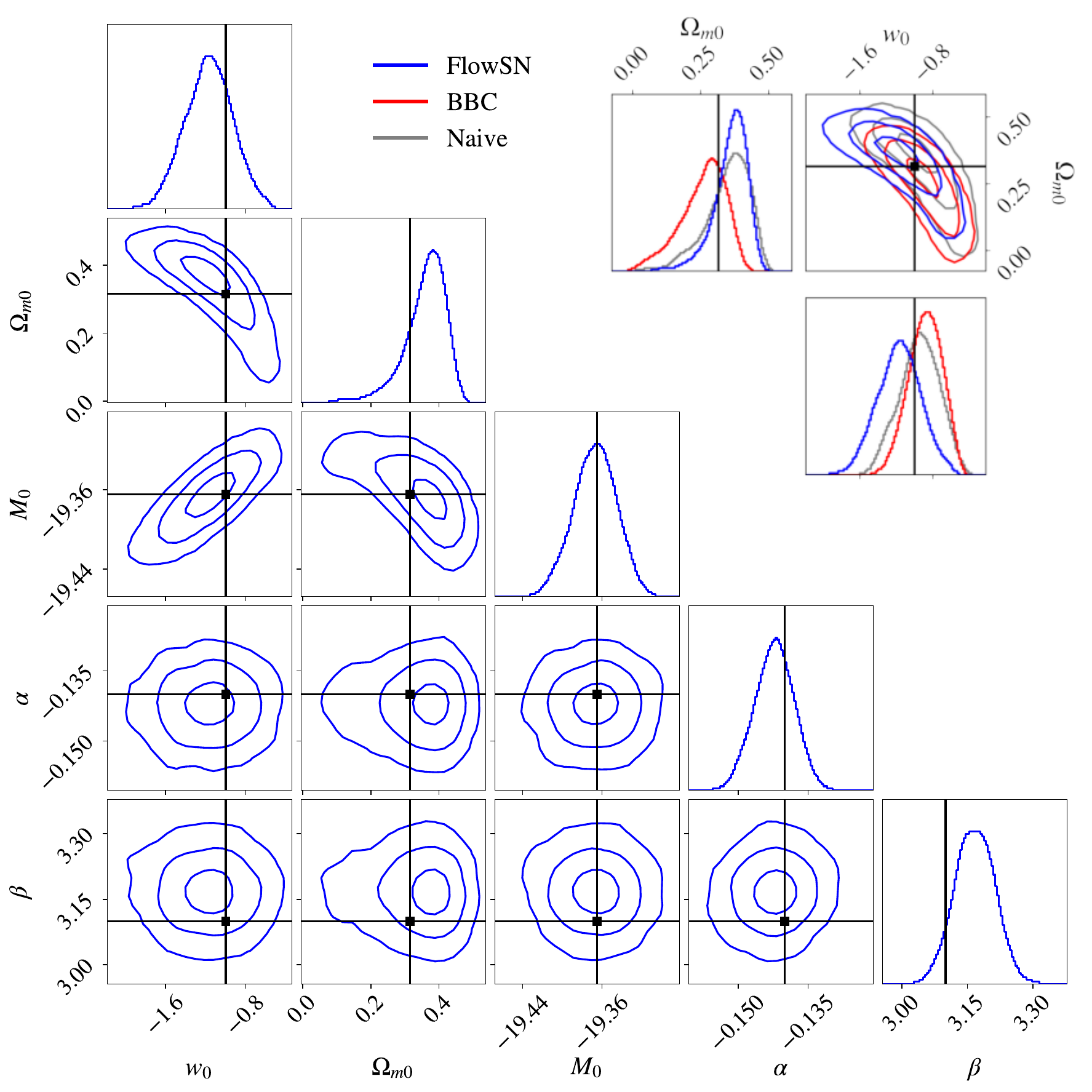}

    \caption{\textit{FlowSN} posteriors on flat $w$CDM cosmological parameters and supernova population parameters using 2000 simulated SNANA spectroscopic SNe. True parameters used in simulations are illustrated by the straight black lines. In the top right, we additionally include posterior inferences on cosmological parameters for BBC and a naive approach. The naive approach uses the BBC pipeline, but does not make any bias corrections $\Delta \mu = 0$. We see that all three posteriors assign probabilities to different regions of cosmological parameter space. Note that these results are without a low redshift anchor sample and without a CMB prior. }
    \label{fig:snana_post}
\end{figure*}

\subsection{\textit{FlowSN} Training and Global Parameter Inference}
Within SNANA, the population distributions cannot be modified for each SN event, so it is not possible to infer $\sigma_{\text{res}}$, $\bm{\theta}_x$, $\bm{\theta}_c$ and $\alpha_c$ without doing multiple computationally expensive survey simulations. This limitation arises from the simulation software rather than the inference method. As shown in Section~\ref{sec:toy}, the simplified forward model allows joint inference of these population parameters together with cosmology. For comparison with BBC using SNANA simulations, we therefore focus on inferring only the cosmological parameters and the basic \cite{tripp1998} population parameters $\bm{\Theta} = (\bm{C},M_0,\alpha,\beta)$ in Eq.~\eqref{eq:tripp}. The priors used for inference of the global parameters are listed in Table~\ref{tab:snana_priors}. We present results with and without using CMB prior \citetext{\citealt{komatsu2009}; defined in Appendix~\ref{app:cmb_prior}} with $\sigma^{\text{CMB}}_{\text{R}} = 0.007$ which further constrains the cosmological parameters in lieu of a low redshift anchor sample.

To train the \textit{FlowSN} normalising flow we generate a training sample using the same forward model as in Section~\ref{sec:snana_model}, but modify the population sampling distributions to preferentially simulate SNe that are more strongly impacted by selection effects. For 90\% of the SNe we increase the Eq.\eqref{eq:samp_z} redshift power law rate to $\xi_{\text{rate}} = 7$. For the remaining 10\% of the training set we sample a uniform redshift distribution $z_t \sim U(0,1)$ to ensure coverage of the full redshift range. We uniformly sample $\alpha^t \sim U(-0.2,-0.1)$ and $\beta^t \sim U(2.5,3.5)$ for each SN training example.  Our training set uses a fixed fiducial $\Lambda$CDM cosmology $\bm{C}_1 = ( \Omega_{m 0} = 0.315, \Omega_{\Lambda 0}  = 0.685, w_0 = -1, w_a = 0)$. To correctly capture the residual impact of the observed heliocentric redshift on the SALT fit, we must sample pairs of $\hat{z}_t^\text{hel}$ and $m_0^t$ associated with different cosmological parameters $\bm{C}^t$. As a workaround in SNANA, where the cosmology is fixed for each survey simulation, we emulate this effect by introducing a distance smearing term $\Delta \mu_t$ for each training example as described in Appendix~\ref{app:SNANA_sims}. We generate examples continuously until the training set reaches $N^{\text{train}} = 10$~million SNe that satisfy the selection criteria. This trained normalising flow is used in all the SNANA experiments. 

\subsection{BBC Settings}
To compare with \textit{FlowSN}, we use the SNANA pipeline's implementation of BBC \citep{kessler2017}, with the same settings as \cite{mitra23}. The BBC 5D method used corrects distances on a five-dimensional grid of observed redshift, colour, stretch, $\alpha$ and $\beta$. Redshift is divided into 14 logarithmically spaced bins. The stretch and colour parameters are each divided into twelve equal-width bins over the ranges $-3.0 < \hat{x}_s < 3.0$ and $-0.3 < \hat{c}_s < 0.3$ respectively. The $\alpha$ and $\beta$ parameters are each split into two equal bins spanning $-0.16 < \alpha < -0.08$ and $2.5 < \beta < 3.6$ respectively. In the bias correction sample, we use a fiducial flat $\Lambda$CDM cosmology $\bm{C}_1 = ( \Omega_{m 0} = 0.315, \Omega_{\Lambda 0}  = 0.685, w_0 = -1, w_a = 0)$ and follow the same forward model as described in Section \ref{sec:snana_model}. We generate 480,000 light curves across 40 random realisations for a total of 19.2 million simulated SNe, of which approximately 674,000 are observed and pass the selection cuts for use in the binning. We use the default binned \texttt{wfit} \citep{kessler2009} settings when doing $\chi^2$ fits for cosmological parameters, again following \cite{mitra23}.

\subsection{Results and Discussion}

\begin{table*}
	\centering
	\caption{Average of posterior estimates over 100 repeats and the standard error on this average using SNANA survey simulations with 2000 spectroscopic SNe. Due to their non-Gaussian nature, posterior estimates for the SNe only results were taken to be the posterior median.  For the Gaussian posteriors that used a CMB prior, we use the posterior mean as the estimate. \textit{FlowSN} was trained and BBC bins were populated using simulations with assumed fiducial cosmology $\bm{C}_1$. We see BBC combined with the CMB prior is biased when testing on different cosmologies $\bm{C}_2$ and $\bm{C}_3$. Note uncertainties are the standard error on the average so are a factor of 10 smaller than those in individual experiments. The experiments do not include a low redshift anchor sample. The $\sigma$ values are calculated using unrounded averages and standard errors.}

	\label{tab:snana_results}

\begin{tabular}{lcccccc}
\hline
\multicolumn{6}{c}{Cosmology $\bm{C}_1$} \\
\hline
Test &  &  \multicolumn{2}{c}{SN Only} &  \multicolumn{2}{c}{SN + CMB Prior}\\
\hline
Global Parameter & Truth  & \textit{FlowSN} & BBC & \textit{FlowSN} & BBC \\
\hline
$w_0$&-1&-1.025 $\pm$ 0.022 (-1.2$\sigma$)&-1.02 $\pm$ 0.02 (-1.0$\sigma$)&-1.0 $\pm$ 0.003 (0.2$\sigma$)&-1.004 $\pm$ 0.003 (-1.4$\sigma$)\\
$\Omega_{m0}$&0.315&0.307 $\pm$ 0.007 (-1.1$\sigma$)&0.309 $\pm$ 0.006 (-1.0$\sigma$)&0.315 $\pm$ 0.001 (0.5$\sigma$)&0.317 $\pm$ 0.001 (1.7$\sigma$)\\
$M_0$&-19.365&-19.367 $\pm$ 0.002 (-1.0$\sigma$)&- &-19.364 $\pm$ 0.001 (0.9$\sigma$)&- \\
$\alpha$&-0.14&-0.139 $\pm$ 0.0004 (2.5$\sigma$)&-0.14 $\pm$ 0.0003 (-0.1$\sigma$)&-0.139 $\pm$ 0.0004 (2.6$\sigma$)&-0.14 $\pm$ 0.0003 (-0.1$\sigma$)\\
$\beta$&3.1&3.112 $\pm$ 0.004 (2.7$\sigma$)&3.106 $\pm$ 0.004 (1.5$\sigma$)&3.112 $\pm$ 0.004 (2.8$\sigma$)&3.106 $\pm$ 0.004 (1.5$\sigma$)\\
\hline
\multicolumn{6}{c}{Cosmology $\bm{C}_2$} \\
\hline
Test &  &  \multicolumn{2}{c}{SN Only} &  \multicolumn{2}{c}{SN + CMB Prior}\\
\hline
Global Parameter & Truth  & \textit{FlowSN} & BBC & \textit{FlowSN} & BBC \\
\hline
$w_0$&-0.8&-0.842 $\pm$ 0.023 (-1.8$\sigma$)&-0.824 $\pm$ 0.021 (-1.1$\sigma$)&-0.801 $\pm$ 0.002 (-0.6$\sigma$)&-0.813 $\pm$ 0.002 (-5.6$\sigma$)\\
$\Omega_{m0}$&0.35&0.34 $\pm$ 0.009 (-1.1$\sigma$)&0.331 $\pm$ 0.008 (-2.2$\sigma$)&0.35 $\pm$ 0.001 (-0.3$\sigma$)&0.349 $\pm$ 0.001 (-0.7$\sigma$)\\
$M_0$&-19.365&-19.369 $\pm$ 0.002 (-2.1$\sigma$)&- &-19.365 $\pm$ 0.001 (0.3$\sigma$)&- \\
$\alpha$&-0.14&-0.1402 $\pm$ 0.0004 (-0.4$\sigma$)&-0.1411 $\pm$ 0.0002 (-4.4$\sigma$)&-0.1402 $\pm$ 0.0004 (-0.6$\sigma$)&-0.1411 $\pm$ 0.0002 (-4.4$\sigma$)\\
$\beta$&3.1&3.107 $\pm$ 0.004 (1.6$\sigma$)&3.117 $\pm$ 0.004 (4.1$\sigma$)&3.108 $\pm$ 0.005 (1.7$\sigma$)&3.117 $\pm$ 0.004 (4.1$\sigma$)\\
\hline 
\multicolumn{6}{c}{Cosmology $\bm{C}_3$} \\
\hline
Test &  &  \multicolumn{2}{c}{SN Only} &  \multicolumn{2}{c}{SN + CMB Prior}\\
\hline
Global Parameter & Truth  & \textit{FlowSN} & BBC & \textit{FlowSN} & BBC \\
\hline
$w_0$&-1.2&-1.237 $\pm$ 0.017 (-2.1$\sigma$)&-1.22 $\pm$ 0.017 (-1.2$\sigma$)&-1.197 $\pm$ 0.003 (0.9$\sigma$)&-1.186 $\pm$ 0.003 (4.3$\sigma$)\\
$\Omega_{m0}$&0.25&0.255 $\pm$ 0.004 (1.1$\sigma$)&0.256 $\pm$ 0.004 (1.7$\sigma$)&0.251 $\pm$ 0.001 (1.8$\sigma$)&0.254 $\pm$ 0.001 (5.0$\sigma$)\\
$M_0$&-19.365&-19.366 $\pm$ 0.002 (-0.6$\sigma$)&- &-19.362 $\pm$ 0.001 (3.0$\sigma$)&- \\
$\alpha$&-0.14&-0.1402 $\pm$ 0.0004 (-0.5$\sigma$)&-0.1396 $\pm$ 0.0003 (1.2$\sigma$)&-0.1402 $\pm$ 0.0004 (-0.5$\sigma$)&-0.1396 $\pm$ 0.0003 (1.2$\sigma$)\\
$\beta$&3.1&3.107 $\pm$ 0.005 (1.6$\sigma$)&3.097 $\pm$ 0.004 (-0.8$\sigma$)&3.108 $\pm$ 0.005 (1.7$\sigma$)&3.097 $\pm$ 0.004 (-0.8$\sigma$)\\
\hline
\end{tabular}

\end{table*}
\begin{figure}
\centering
	\includegraphics[width=0.95\columnwidth]{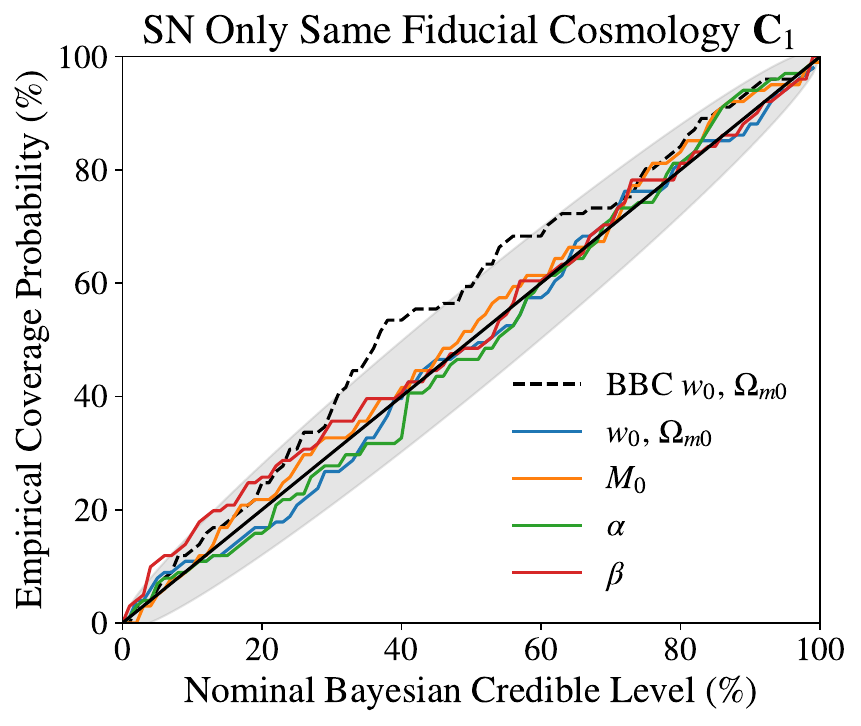}
    \caption{\textit{FlowSN} and BBC frequentist calibration of posteriors after 100 repeated experiments on SNANA survey simulations with 2000  spectroscopic SNe. The coloured lines correspond to \textit{FlowSN} coverage, while the dashed black line corresponds to BBC coverage. The grey shaded region shows the 95\% Monte Carlo confidence band for the empirical coverage under perfect calibration. These results are using simulations with the same fiducial flat $\Lambda$CDM cosmology for training and testing the methods. Note these results are without a CMB prior and low redshift anchor sample.}
    \label{fig:calib}
\end{figure}

\begin{figure*}
\centering
	\includegraphics[width=0.95\textwidth]{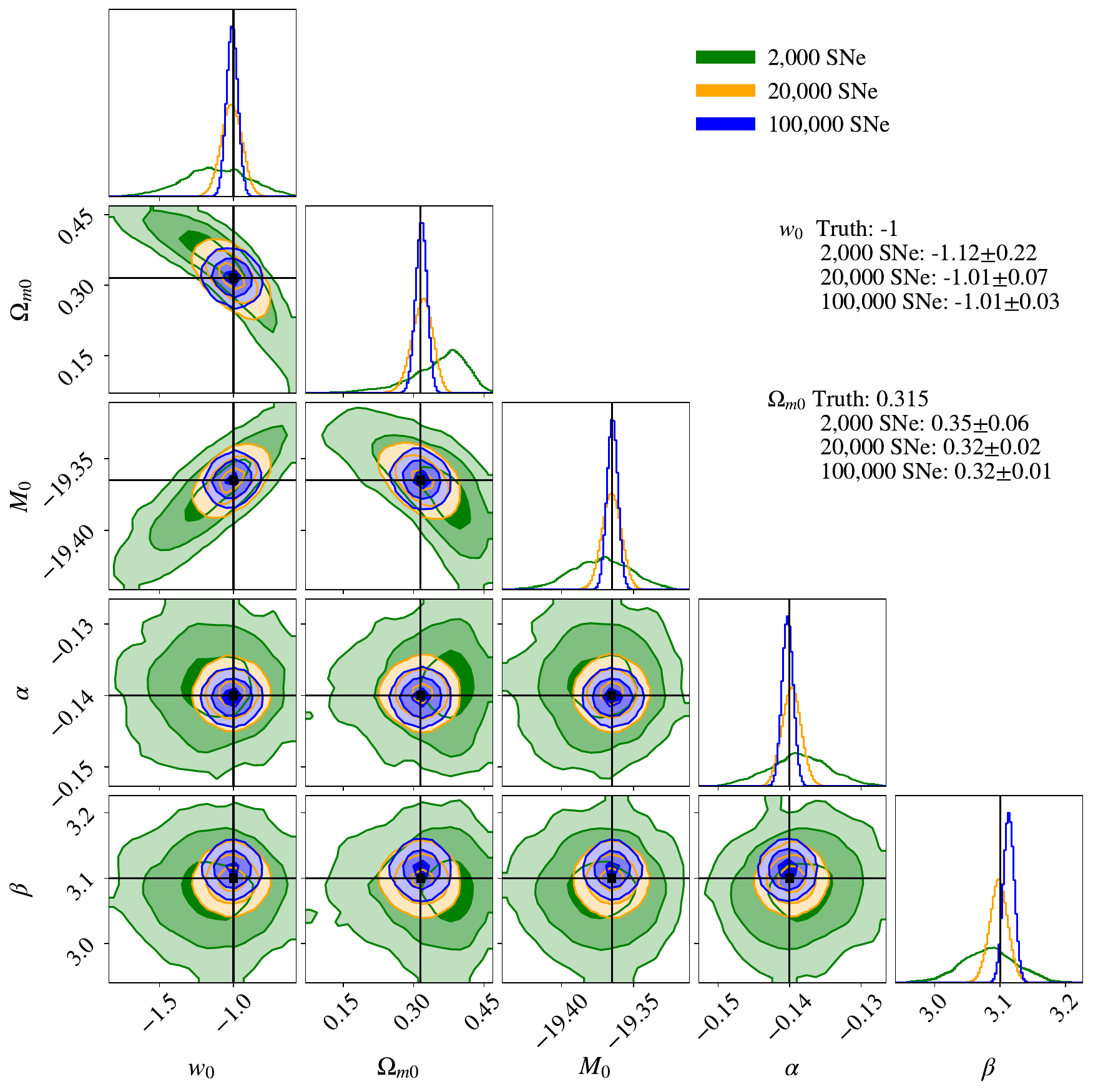}
    \caption{\textit{FlowSN} posteriors using SNANA spectroscopic SN simulations, demonstrating scalability with increasing sample size. The straight line shows the true parameters used in the simulations. The cosmological parameter results on the right are posterior mean averages and posterior standard deviations. Note these constraints are without a low redshift anchor sample and without a CMB prior.}
    \label{fig:scale}
\end{figure*}

To begin, we train the \textit{FlowSN} normalising flow using the training set described in Section~\ref{sec:snana_model}, following the procedure in Section \ref{sec:flowsn}. Since the SNANA simulations involve non-Gaussian population distributions, as well as complex selection criteria, we are unable to derive an analytical likelihood solution for comparison. Instead, we plot the data distribution of a held-out test set of SNANA simulations and samples from our normalising flow likelihood approximation conditioned on the same parameters in Figure \ref{fig:SNANA_samples}. We monitor how the distributions change in bins of latent apparent magnitude $m_0^s$, paying special attention to the dimmer ranges where Malmquist bias is strongest. This test confirms the normalising flow likelihood approximation learns the SNANA summary statistic data distributions and correlations sufficiently across the three dimensions.

We then use the learnt normalising flow likelihood approximation to sample the posterior on the global SNANA parameters. To begin, the inference is performed on $N_{\text{SN}}^{\text{obs}} \approx$~2,000 selected SNe, which is a similar size to our largest present day spectroscopic samples \citep{scolnic2022}. For comparison we also use the SNANA pipeline to perform BBC \citep{kessler2017} on the same dataset and then do cosmology fits with \texttt{wfit} \citep{kessler2009}. To demonstrate that both \textit{FlowSN} and BBC approaches significantly impact results, we can add a third comparison using naive constraints that ignore selection effects. In practice we do this using the SNANA pipeline assuming the bias correction is zero in each bin $\Delta \mu=0$. This baseline highlights the specific influence each procedure has on cosmological parameter inference. The posteriors on a single random realisation of the survey are included in Figure \ref{fig:snana_post}. We first see that \textit{FlowSN} is able to recover both the true global parameters within 2$\sigma$. Both \textit{FlowSN} and BBC recover the true underlying cosmology within 1$\sigma$, while the naive approach still recovers the true cosmology within 2$\sigma$. \textit{FlowSN} infers $w_0 = -1.16 \pm 0.24$ and $\Omega_{m0}= 0.37 \pm 0.06 $, whereas BBC infers  $w_0 = -0.87 \pm 0.27$ and $\Omega_{m0}= 0.27 \pm 0.09 $. Although, \textit{FlowSN} and BBC still recover the truth, their posterior densities occupy significantly different regions of parameter space. This finding implies that if we reanalysed previous works that made use of BBC \citep{brout2022,des5yr} with our alternative technique instead, we may find differing constraints on cosmology.

To look further into the differences between our \textit{FlowSN} inferences and those using BBC, we repeat the experiments on $N_{\text{SN}}^{\text{obs}} \approx$~2,000 selected SNe over one hundred SNANA random realisations. Table~\ref{tab:snana_results} shows the mean average posterior predictions over the one hundred realisations along with the standard error of the predictions. We show results using both SNe alone and with the inclusion of a CMB prior (defined in Appendix~\ref{app:cmb_prior}) for additional constraining power. We repeat both experiments for simulations with three different underlying cosmological parameters. The $\bm{C}_1 = (w_0 = -1, \Omega_{m0} =0.315)$ cosmology is the same flat $\Lambda$CDM fiducial cosmology that \textit{FlowSN} and BBC are trained using. To test effects of this assumed fiducial cosmology, we repeat the experiment on test sets that follow $\bm{C}_2 = (w_0 = -0.8, \Omega_{m0} =0.35)$ and $\bm{C}_3 = (w_0 = -1.2, \Omega_{m0} =0.25)$, representing cosmologies that lie below and above a flat $\Lambda$CDM model on the Hubble diagram respectively. 

We find that \textit{FlowSN} consistently recovers the true underlying cosmology with and without the CMB prior. We also find that both \textit{FlowSN} and BBC are able to infer the true $\alpha$ and $\beta$ parameters for $\bm{C}_1$ and $\bm{C}_3$. These results are largely consistent with the findings of \cite{camilleri2024}, that finds weak dependency of the fiducial cosmology on the $\alpha$ and $\beta$ inference. We do find that BBC yields a 4$\sigma$ biased inference of $\alpha$ and $\beta$ on cosmology $\bm{C}_2$, though one could argue the magnitude of this bias is negligible in comparison to uncertainties from individual realisations. 

Both \textit{FlowSN} and BBC are able to infer the true underlying cosmology within two standard errors when analysing the SNe alone, however, the posteriors are often centred in different regions of parameter space as we see in Figure~\ref{fig:snana_post}.  The most notable difference between the methods is found when combining the SNe Ia with a CMB prior. We find that both methods recover the true cosmological parameters when they are equal to the assumed fiducial cosmology used in the training set. However, we importantly find that when the true cosmology used in the test simulations is significantly different from the fiducial training cosmology, the BBC constraints with the CMB prior are biased by over four standard errors in the direction of the fiducial $w_0=-1$. These BBC biases are similar to those found in \cite{camilleri2024}, though we find the biases to be in the direction of the fiducial cosmology rather than along the $w_0-\Omega_{m0}$ degeneracy plane. Our interpretation is that this bias is always present, though adding strong constraints such as the CMB prior amplify its influence as other uncertainties shrink. The \textit{FlowSN} approach on the other hand results in $w_0$ biases that are less than one standard error. This demonstrates that our method is insensitive to the assumed cosmology used in training, whereas BBC is sensitive to this when combined with other strong cosmological probes. While the magnitude of the BBC biases are small, it is important to understand these effects as our sample sizes grow and our modelling of other systematic uncertainties improve.

We further investigate the quality of the \textit{FlowSN} posteriors by plotting each parameter's Bayesian credibility level against its frequentist empirical coverage in Figure~\ref{fig:calib}. We see that for both the cosmological parameters and the SN population parameters, the credible levels lie well within the 95\% Monte Carlo confidence band, confirming the posteriors are well-calibrated. We find the BBC posterior when using SNe only to be under-confident, where part of its calibration curve lies outside the 95\% confidence band. We find when the SNe are combined with the strong CMB prior, the BBC posteriors are better calibrated, however, it is important to understand these properties in isolation. 

To demonstrate our \textit{FlowSN} methodology scales to future SN sample sizes we show posteriors when analysing both $N_{\text{SN}}^{\text{obs}} \approx$~20,000 and $N_{\text{SN}}^{\text{obs}} \approx$~100,000 selected spectroscopic SNe simulated with SNANA. In these experiments we use the same normalising flow as before, without retraining. Figure~\ref{fig:scale} demonstrates that the approach is able to recover the true underlying cosmology and SN population parameters to within 2$\sigma$ for both of the larger datasets. One might expect compounding errors for each of the $N_{\text{SN}}^{\text{obs}}$ times likelihood approximation is evaluated, however, Figure~\ref{fig:scale} provides strong evidence the normalising flow does in fact learn the density to the required level of detailed numerical accuracy. Furthermore, the approach still only takes 20 minutes to sample the posterior despite the $N_{\text{SN}}^{\text{obs}}$ latent apparent magnitudes $m^s_0$ also being sampled. These results give us confidence that the \textit{FlowSN} methodology is robust enough to scale to the demands of next generation surveys, such as LSST \citep{lsst2019} and TiDES \citep{tides}.

As a final experiment, we apply the trained normalising flow to a $w_0 w_a$CDM model inference using three SNANA simulations, one from each cosmology ($\bm{C}_1$ to $\bm{C}_3$). Figure~\ref{fig:w0wa_plot} demonstrates that despite being trained exclusively on $\Lambda$CDM, \textit{FlowSN} provides reliable uncertainties in the $w_0-w_a$ plane, recovering the true cosmological parameter values with 2$\sigma$. This is possible because the normalising flow is conditioned on the latent apparent magnitude $m^s_0 = \mu(z_s;\,\bm{C}) + M_0$, rather than the cosmological parameters $\bm{C}$ directly. This represents a distinct advantage of the \textit{FlowSN} methodology, whereas other SBI approaches typically require retraining for each new cosmological model. Additional details on this experiment are provided in Appendix~\ref{app:w0wa}. We save further repeated experiments with varying values of $w_a$ and comparisons with BBC for future work.

\section{Conclusions and Future Work}
\label{sec:conc}

This work demonstrates that the \textit{FlowSN} methodology provides a reliable and interpretable solution to account for selection effects in SN Ia cosmology. Our approach is lightweight and fast, comparable with BBC \citep{kessler2017} timescales. The normalising flow training takes 20 minutes and inference on 2,000 SNe takes as little as 5 minutes on a GPU. The simplified forward model verifies that the learnt normalising flow density closely matches the analytical likelihood solution. Sampling the posterior that uses the normalising flow allows for well-calibrated inferences over both cosmological parameters and an additional nine SN population parameters. 

We followed this with the first application of SBI techniques to realistic SNANA survey simulations \citep{kessler2009} that are used in current cosmology analyses. The interpretable properties of our technique allow us to verify that our normalising flow has learnt the correct likelihood approximation by plotting samples along with held-out data distributions, without needing to rerun the simulator. In posterior sampling, we found that the \textit{FlowSN} technique allows for inferences that have better frequentist calibration and are less biased than the traditional BBC technique \citep{kessler2017} when combined with a CMB prior. We show BBC biases that are in the direction of the fiducial $w_0=-1$ used in training, whereas the \textit{FlowSN} approach remains insensitive to this assumption. Importantly, we found that BBC and \textit{FlowSN} posteriors often assign probability densities to different regions of cosmological parameter space, motivating the need to reanalyse existing datasets using an alternative method for handling selection effects.

The modular approach of \textit{FlowSN} allows us to use the same trained normalising flow to constrain different sets of cosmological parameters, without retraining. This is a distinct advantage over other SBI methods that would require retraining on each cosmology and additional verification that the model has learnt SN population parameter distributions to the same level of detail. The modular structure of our methodology allows us to make an arbitrary number of model changes to the SN parameters that are not learnt by the normalising flow. For example, without retraining we could extend the framework to infer broken $\alpha$ \citep{ginolin2025a} or $\beta$ relationships, as well as the inclusion of galaxy host mass relationships \citep{kelly2010,sullivan2010,lampeitl2010,childress2013}. Further versions of this work will introduce extrinsic dust parameters and their influence on the observed summary statistic data distributions, building on the Simple-BayeSN model \citep{mandel2017}. The achromatic host galaxy mass-step \citep{kelly2010,sullivan2010} can be incorporated into the current version of \textit{FlowSN} by introducing an additional constant, $\gamma$, to the latent apparent magnitude ($m^s_0 \rightarrow m^s_0 + \gamma$) for SNe in high-mass galaxies. We can then infer $\gamma$ as an additional population parameter in the hierarchical model. More complex host-galaxy correlations will require conditioning the normalising flow on key host parameters, such as stellar age or metallicity, then learning their influence from simulations. These various modelling decisions motivate the need for model comparison using the Bayesian evidence \citep{karchev2023b}. A further benefit of the \textit{FlowSN} approach is that we are able to use the same trained normalising flow in evidence calculations, as well as posterior inference, using nested sampling techniques which we plan to exploit in future work.

Our current formulation of the \textit{FlowSN} methodology is compatible with present day spectroscopically confirmed SN Ia samples. For compilations such as Pantheon+ \citep{scolnic2022,brout2022}, this would involve training a normalising flow for each survey to correctly handle their selection effects. Similarly, to model spectroscopic SNe from both the LSST Deep Drilling Fields and Wide Fast Deep fields \citep{lsst2019}, we would require a normalising flow for each. We show our work is able to scale to sample sizes of 100,000 selected SNe, still taking a matter of 20 minutes at inference time. This shows us that the methodology is robust to next-generation survey sample sizes. To model photometric SN Ia samples, such as the Dark Energy Survey (DES) 5-year sample \citep{des5yr} and the majority of SNe Ia observed with LSST \citep{lsst2019}, we will need to extend our methodology. For SNe Ia that are only classified using photometry, we will need to include a well-calibrated Ia classification probability and also consider models for non-Ia contaminants. When considering photometric redshifts, we will need to reframe the approach as a joint problem, where we include a photometric redshift likelihood, as well as including the additional cosmological constraining power one can gain from redshift distributions alone. This joint modelling approach will make the methodology useful to areas outside SN Ia cosmology, particularly those that require careful rate calculations \citep{schmidt1968,kauffmann2003,petigura2013,hu2023,grayling2023}.

Throughout this work, we have shown that \textit{FlowSN} can effectively infer different cosmological models, a capability that will be vital for addressing ongoing $w_0$--$w_a$ debates in the LSST-era. However, as with other SBI techniques, our methodology may remain vulnerable to other types of model misspecification, for example, if differences exist between the simulated training data and reality regarding SNe~Ia standardisation or selection effects. To quantify any potential biases induced by such misspecifications, we will need to perform standardised tests specifically developed for SBI techniques \citep{montel2025,ocallaghan2025}.

The current \textit{FlowSN} formulation assumes that the selection effect is stationary and does not vary with time. If LSST introduces time-dependent changes in selection, there are several ways we could address this. If there are distinct shifts in survey selection between the beginning and end of the survey, we could train a normalising flow for each phase, effectively treating them as independent surveys. Alternatively, if the change is more gradual, we could explore conditioning the flow on the time of observation for each SN, assuming we are able to simulate this dependence. A third option is to leave the change in selection unmodelled and instead monitor its influence using model misspecification tests.

Further systematics to consider include correlations between SN observations, which could be implemented as additional conditionals in the normalising flow, similar to the treatment of covariances in the SALT fit. Regarding photometric calibration, we could infer these systematics as global variables by varying them in simulation, allowing the flow to learn their influence directly. Alternatively, as is standard in traditional analysis, we could repeatedly perform inference on simulations with different calibration systematics and average their \textit{FlowSN} posteriors to determine the resulting systematic uncertainty.

Our understanding and handling of systematics, including selection effects, is playing an ever more central role in constraining the nature of dark energy and the expansion of the Universe \citep{desi2025,efstathiou2024,vincenzi2025,dhawan2025,wiseman2026}. \textit{FlowSN} provides a principled and computationally efficient likelihood-based framework that combines the flexibility of SBI with the transparency and reusability of an explicit likelihood approximation, bridging the gap between traditional analytical approaches and black-box neural methods. As forthcoming surveys push SN Ia cosmology into the systematics-dominated regime, robust and well-calibrated treatments of selection effects will be essential, with \textit{FlowSN} providing a promising solution for this next era of precision cosmology.

\section*{Acknowledgements}
We thank Roberto Trotta, Konstantin Karchev, and Lucas Makinen for  useful discussions.

BMB is supported by the Cambridge Centre for Doctoral Training in Data-Intensive Science funded by the UK Science and Technology Facilities Council (STFC) and two G-Research Early Career Researcher grants used for equipment and travel. BMB also acknowledges travel support provided by Christ's College Cambridge's Clayton Fund, LSST DESC supported by DOE funds administered by SLAC and STFC for UK participation in LSST through grant ST/S006206/1. KSM, MG, MA, AD, MG, MOC, and this research are supported by the European Union’s Horizon 2020 research and innovation programmes under European Research Council Grant Agreement No 101002652 (BayeSN) and Marie Skłodowska-Curie Grant Agreement No 873089 (ASTROSTAT-II). This research was supported by the Munich Institute for Astro-, Particle and BioPhysics (MIAPbP) which is funded by the Deutsche Forschungsgemeinschaft (DFG, German Research Foundation) under Germany's Excellence Strategy – EXC-2094 – 390783311.  ST has been supported by funding from the European Research Council (ERC) under the European Union's Horizon 2020 research and innovation programmes (grant agreement no.\ 101018897 CosmicExplorer). LK acknowledges support for an Early Career Fellowship from the Leverhulme Trust through grant ECF-2024-054 and the Isaac Newton Trust through grant 24.08(w).

\section*{Data Availability}

The implementation of \textit{FlowSN}, as well as the simplified forward model, is 
publicly available at:~\url{https://github.com/bayesn/flowsn}.
The SNANA training and testing data used in this work will be shared on reasonable request to the corresponding author.



\bibliographystyle{mnras}
\bibliography{references} 

@ARTICLE{carr2022,
       author = {{Carr}, Anthony and {Davis}, Tamara M. and {Scolnic}, Dan and {Said}, Khaled and {Brout}, Dillon and {Peterson}, Erik R. and {Kessler}, Richard},
        title = "{The Pantheon+ analysis: Improving the redshifts and peculiar velocities of Type Ia supernovae used in cosmological analyses}",
      journal = {\pasa},
         year = 2022,
        month = oct,
       volume = {39},
          eid = {e046},
        pages = {e046},
          doi = {10.1017/pasa.2022.41},
archivePrefix = {arXiv},
       eprint = {2112.01471},
 primaryClass = {astro-ph.CO},
       adsurl = {https://ui.adsabs.harvard.edu/abs/2022PASA...39...46C}
}

@software{jax2018github,
  author = {James Bradbury and Roy Frostig and Peter Hawkins and Matthew James Johnson and Chris Leary and Dougal Maclaurin and George Necula and Adam Paszke and Jake Vander{P}las and Skye Wanderman-{M}ilne and Qiao Zhang},
  title = {{JAX}: composable transformations of {P}ython+{N}um{P}y programs},
  url = {http://github.com/google/jax},
  version = {0.3.13},
  year = {2018},
}

@article{tabak2010density,
  title={Density estimation by dual ascent of the log-likelihood},
  author={Tabak, Esteban G and Vanden-Eijnden, Eric},
  journal={Comm. Math. Sci.},
  volume={8},
  number={1},
  pages={217--233},
  year={2010},
  publisher={International Press of Boston}
}

@inproceedings{papamakarios2017masked,
 author = {Papamakarios, George and Pavlakou, Theo and Murray, Iain},
 booktitle = {Advances in Neural Information Processing Systems 30},
 editor = {I. Guyon and U. Von Luxburg and S. Bengio and H. Wallach and R. Fergus and S. Vishwanathan and R. Garnett},
 pages = {2335-2344},
 publisher = {Curran Associates, Inc.},
 title = {Masked Autoregressive Flow for Density Estimation},
 url = {https://proceedings.neurips.cc/paper_files/paper/2017/file/6c1da886822c67822bcf3679d04369fa-Paper.pdf},
 year = {2017},
 archivePrefix = {arXiv},
 eprint = {1705.07057}
}

@InProceedings{germain2015made,
  title = 	 {MADE: Masked Autoencoder for Distribution Estimation},
  author = 	 {Germain, Mathieu and Gregor, Karol and Murray, Iain and Larochelle, Hugo},
  booktitle = 	 {Proceedings of the 32nd International Conference on Machine Learning},
  pages = 	 {881--889},
  year = 	 {2015},
  editor = 	 {Bach, Francis and Blei, David},
  volume = 	 {37},
  series = 	 {Proceedings of Machine Learning Research},
  address = 	 {Lille, France},
  month = 	 {07--09 Jul},
  publisher =    {PMLR},
  url = 	 {https://proceedings.mlr.press/v37/germain15.html},
}

@ARTICLE{karchev2024,
       author = {{Karchev}, Konstantin and {Grayling}, Matthew and {Boyd}, Benjamin M. and {Trotta}, Roberto and {Mandel}, Kaisey S. and {Weniger}, Christoph},
        title = "{SIDE-real: Supernova Ia Dust Extinction with truncated marginal neural ratio estimation applied to real data}",
      journal = {\mnras},
         year = 2024,
        month = jun,
       volume = {530},
       number = {4},
        pages = {3881-3896},
          doi = {10.1093/mnras/stae995},
archivePrefix = {arXiv},
       eprint = {2403.07871},
 primaryClass = {astro-ph.CO},
       adsurl = {https://ui.adsabs.harvard.edu/abs/2024MNRAS.530.3881K}
}

@ARTICLE{scolnic2022,
       author = {{Scolnic}, Dan and {Brout}, Dillon and {Carr}, Anthony and {Riess}, Adam G. and {Davis}, Tamara M. and {Dwomoh}, Arianna and {Jones}, David O. and {Ali}, Noor and {Charvu}, Pranav and {Chen}, Rebecca and {Peterson}, Erik R. and {Popovic}, Brodie and {Rose}, Benjamin M. and {Wood}, Charlotte M. and {Brown}, Peter J. and {Chambers}, Ken and {Coulter}, David A. and {Dettman}, Kyle G. and {Dimitriadis}, Georgios and {Filippenko}, Alexei V. and {Foley}, Ryan J. and {Jha}, Saurabh W. and {Kilpatrick}, Charles D. and {Kirshner}, Robert P. and {Pan}, Yen-Chen and {Rest}, Armin and {Rojas-Bravo}, Cesar and {Siebert}, Matthew R. and {Stahl}, Benjamin E. and {Zheng}, WeiKang},
        title = "{The Pantheon+ Analysis: The Full Data Set and Light-curve Release}",
      journal = {\apj},
         year = 2022,
        month = oct,
       volume = {938},
       number = {2},
          eid = {113},
        pages = {113},
          doi = {10.3847/1538-4357/ac8b7a},
archivePrefix = {arXiv},
       eprint = {2112.03863},
 primaryClass = {astro-ph.CO},
       adsurl = {https://ui.adsabs.harvard.edu/abs/2022ApJ...938..113S}
}

@ARTICLE{rubin2023,
       author = {{Rubin}, David and {Aldering}, Greg and {Betoule}, Marc and {Fruchter}, Andy and {Huang}, Xiaosheng and {Kim}, Alex G. and {Lidman}, Chris and {Linder}, Eric and {Perlmutter}, Saul and {Ruiz-Lapuente}, Pilar and {Suzuki}, Nao},
        title = "{Union through UNITY: Cosmology with 2000 SNe Using a Unified Bayesian Framework}",
      journal = {\apj},
         year = 2025,
        month = jun,
       volume = {986},
       number = {2},
          eid = {231},
        pages = {231},
          doi = {10.3847/1538-4357/adc0a5},
archivePrefix = {arXiv},
       eprint = {2311.12098},
 primaryClass = {astro-ph.CO},
       adsurl = {https://ui.adsabs.harvard.edu/abs/2025ApJ...986..231R}
}

@article{duane1987,
title = {Hybrid Monte Carlo},
journal = {Phys.\ Lett.\ B},
volume = {195},
number = {2},
pages = {216-222},
year = {1987},
issn = {0370-2693},
doi = {10.1016/0370-2693(87)91197-X},
author = {Simon Duane and A.D. Kennedy and Brian J. Pendleton and Duncan Roweth}
}

@ARTICLE{karchev2023b,
       author = {{Karchev}, Konstantin and {Trotta}, Roberto and {Weniger}, Christoph},
        title = "{SimSIMS: Simulation-based Supernova Ia Model Selection with thousands of latent variables}",
      journal = {ArXiv e-prints},
         year = 2023,
        month = nov,
          eid = {arXiv:2311.15650},
          doi = {10.48550/arXiv.2311.15650},
archivePrefix = {arXiv},
       eprint = {2311.15650},
 primaryClass = {astro-ph.CO},
       adsurl = {https://ui.adsabs.harvard.edu/abs/2023arXiv231115650K}
}

@article{tabak2013family,
  title={A family of nonparametric density estimation algorithms},
  author={Tabak, Esteban G and Turner, Cristina V},
  journal={Comm. Pure Applied Math.},
  volume={66},
  number={2},
  pages={145--164},
  year={2013},
  publisher={Wiley Online Library},
  doi={10.1002/cpa.21423}
}

@ARTICLE{tripp1998,
       author = {{Tripp}, Robert},
        title = "{A two-parameter luminosity correction for Type IA supernovae}",
      journal = {\aap},
         year = 1998,
        month = mar,
       volume = {331},
        pages = {815-820},
       adsurl = {https://ui.adsabs.harvard.edu/abs/1998A&A...331..815T}
}

@ARTICLE{kelly2010,
       author = {{Kelly}, Patrick L. and {Hicken}, Malcolm and {Burke}, David L. and {Mandel}, Kaisey S. and {Kirshner}, Robert P.},
        title = "{Hubble Residuals of Nearby Type Ia Supernovae are Correlated with Host Galaxy Masses}",
      journal = {\apj},
         year = 2010,
        month = jun,
       volume = {715},
       number = {2},
        pages = {743-756},
          doi = {10.1088/0004-637X/715/2/743},
archivePrefix = {arXiv},
       eprint = {0912.0929},
 primaryClass = {astro-ph.CO},
       adsurl = {https://ui.adsabs.harvard.edu/abs/2010ApJ...715..743K}
}

@ARTICLE{sullivan2010,
       author = {{Sullivan}, M. and {Conley}, A. and {Howell}, D.~A. and {Neill}, J.~D. and {Astier}, P. and {Balland}, C. and {Basa}, S. and {Carlberg}, R.~G. and {Fouchez}, D. and {Guy}, J. and {Hardin}, D. and {Hook}, I.~M. and {Pain}, R. and {Palanque-Delabrouille}, N. and {Perrett}, K.~M. and {Pritchet}, C.~J. and {Regnault}, N. and {Rich}, J. and {Ruhlmann-Kleider}, V. and {Baumont}, S. and {Hsiao}, E. and {Kronborg}, T. and {Lidman}, C. and {Perlmutter}, S. and {Walker}, E.~S.},
        title = "{The dependence of Type Ia Supernovae luminosities on their host galaxies}",
      journal = {\mnras},
         year = 2010,
        month = aug,
       volume = {406},
       number = {2},
        pages = {782-802},
          doi = {10.1111/j.1365-2966.2010.16731.x},
archivePrefix = {arXiv},
       eprint = {1003.5119},
 primaryClass = {astro-ph.CO},
       adsurl = {https://ui.adsabs.harvard.edu/abs/2010MNRAS.406..782S}
}

@ARTICLE{lampeitl2010,
       author = {{Lampeitl}, Hubert and {Smith}, Mathew and {Nichol}, Robert C. and {Bassett}, Bruce and {Cinabro}, David and {Dilday}, Benjamin and {Foley}, Ryan J. and {Frieman}, Joshua A. and {Garnavich}, Peter M. and {Goobar}, Ariel and {Im}, Myungshin and {Jha}, Saurabh W. and {Marriner}, John and {Miquel}, Ramon and {Nordin}, Jakob and {{\"O}stman}, Linda and {Riess}, Adam G. and {Sako}, Masao and {Schneider}, Donald P. and {Sollerman}, Jesper and {Stritzinger}, Maximilian},
        title = "{The Effect of Host Galaxies on Type Ia Supernovae in the SDSS-II Supernova Survey}",
      journal = {\apj},
         year = 2010,
        month = oct,
       volume = {722},
       number = {1},
        pages = {566-576},
          doi = {10.1088/0004-637X/722/1/566},
archivePrefix = {arXiv},
       eprint = {1005.4687},
 primaryClass = {astro-ph.CO},
       adsurl = {https://ui.adsabs.harvard.edu/abs/2010ApJ...722..566L}
}

@ARTICLE{childress2013,
       author = {{Childress}, M. and {Aldering}, G. and {Antilogus}, P. and {Aragon}, C. and {Bailey}, S. and {Baltay}, C. and {Bongard}, S. and {Buton}, C. and {Canto}, A. and {Cellier-Holzem}, F. and {Chotard}, N. and {Copin}, Y. and {Fakhouri}, H.~K. and {Gangler}, E. and {Guy}, J. and {Hsiao}, E.~Y. and {Kerschhaggl}, M. and {Kim}, A.~G. and {Kowalski}, M. and {Loken}, S. and {Nugent}, P. and {Paech}, K. and {Pain}, R. and {Pecontal}, E. and {Pereira}, R. and {Perlmutter}, S. and {Rabinowitz}, D. and {Rigault}, M. and {Runge}, K. and {Scalzo}, R. and {Smadja}, G. and {Tao}, C. and {Thomas}, R.~C. and {Weaver}, B.~A. and {Wu}, C.},
        title = "{Host Galaxy Properties and Hubble Residuals of Type Ia Supernovae from the Nearby Supernova Factory}",
      journal = {\apj},
         year = 2013,
        month = jun,
       volume = {770},
       number = {2},
          eid = {108},
        pages = {108},
          doi = {10.1088/0004-637X/770/2/108},
archivePrefix = {arXiv},
       eprint = {1304.4720},
 primaryClass = {astro-ph.CO},
       adsurl = {https://ui.adsabs.harvard.edu/abs/2013ApJ...770..108C}
}

@ARTICLE{betoule2014,
       author = {{Betoule}, M. and {Kessler}, R. and {Guy}, J. and {Mosher}, J. and {Hardin}, D. and {Biswas}, R. and {Astier}, P. and {El-Hage}, P. and {Konig}, M. and {Kuhlmann}, S. and {Marriner}, J. and {Pain}, R. and {Regnault}, N. and {Balland}, C. and {Bassett}, B.~A. and {Brown}, P.~J. and {Campbell}, H. and {Carlberg}, R.~G. and {Cellier-Holzem}, F. and {Cinabro}, D. and {Conley}, A. and {D'Andrea}, C.~B. and {DePoy}, D.~L. and {Doi}, M. and {Ellis}, R.~S. and {Fabbro}, S. and {Filippenko}, A.~V. and {Foley}, R.~J. and {Frieman}, J.~A. and {Fouchez}, D. and {Galbany}, L. and {Goobar}, A. and {Gupta}, R.~R. and {Hill}, G.~J. and {Hlozek}, R. and {Hogan}, C.~J. and {Hook}, I.~M. and {Howell}, D.~A. and {Jha}, S.~W. and {Le Guillou}, L. and {Leloudas}, G. and {Lidman}, C. and {Marshall}, J.~L. and {M{\"o}ller}, A. and {Mour{\~a}o}, A.~M. and {Neveu}, J. and {Nichol}, R. and {Olmstead}, M.~D. and {Palanque-Delabrouille}, N. and {Perlmutter}, S. and {Prieto}, J.~L. and {Pritchet}, C.~J. and {Richmond}, M. and {Riess}, A.~G. and {Ruhlmann-Kleider}, V. and {Sako}, M. and {Schahmaneche}, K. and {Schneider}, D.~P. and {Smith}, M. and {Sollerman}, J. and {Sullivan}, M. and {Walton}, N.~A. and {Wheeler}, C.~J.},
        title = "{Improved cosmological constraints from a joint analysis of the SDSS-II and SNLS supernova samples}",
      journal = {\aap},
         year = 2014,
        month = aug,
       volume = {568},
          eid = {A22},
        pages = {A22},
          doi = {10.1051/0004-6361/201423413},
archivePrefix = {arXiv},
       eprint = {1401.4064},
 primaryClass = {astro-ph.CO},
       adsurl = {https://ui.adsabs.harvard.edu/abs/2014A&A...568A..22B}
}

@ARTICLE{guy2007,
       author = {{Guy}, J. and {Astier}, P. and {Baumont}, S. and {Hardin}, D. and {Pain}, R. and {Regnault}, N. and {Basa}, S. and {Carlberg}, R.~G. and {Conley}, A. and {Fabbro}, S. and {Fouchez}, D. and {Hook}, I.~M. and {Howell}, D.~A. and {Perrett}, K. and {Pritchet}, C.~J. and {Rich}, J. and {Sullivan}, M. and {Antilogus}, P. and {Aubourg}, E. and {Bazin}, G. and {Bronder}, J. and {Filiol}, M. and {Palanque-Delabrouille}, N. and {Ripoche}, P. and {Ruhlmann-Kleider}, V.},
        title = "{SALT2: using distant supernovae to improve the use of type Ia supernovae as distance indicators}",
      journal = {\aap},
         year = 2007,
        month = apr,
       volume = {466},
       number = {1},
        pages = {11-21},
          doi = {10.1051/0004-6361:20066930},
archivePrefix = {arXiv},
       eprint = {astro-ph/0701828},
 primaryClass = {astro-ph},
       adsurl = {https://ui.adsabs.harvard.edu/abs/2007A&A...466...11G}
}

@ARTICLE{guy2010,
       author = {{Guy}, J. and {Sullivan}, M. and {Conley}, A. and {Regnault}, N. and {Astier}, P. and {Balland}, C. and {Basa}, S. and {Carlberg}, R.~G. and {Fouchez}, D. and {Hardin}, D. and {Hook}, I.~M. and {Howell}, D.~A. and {Pain}, R. and {Palanque-Delabrouille}, N. and {Perrett}, K.~M. and {Pritchet}, C.~J. and {Rich}, J. and {Ruhlmann-Kleider}, V. and {Balam}, D. and {Baumont}, S. and {Ellis}, R.~S. and {Fabbro}, S. and {Fakhouri}, H.~K. and {Fourmanoit}, N. and {Gonz{\'a}lez-Gait{\'a}n}, S. and {Graham}, M.~L. and {Hsiao}, E. and {Kronborg}, T. and {Lidman}, C. and {Mourao}, A.~M. and {Perlmutter}, S. and {Ripoche}, P. and {Suzuki}, N. and {Walker}, E.~S.},
        title = "{The Supernova Legacy Survey 3-year sample: Type Ia supernovae photometric distances and cosmological constraints}",
      journal = {\aap},
         year = 2010,
        month = nov,
       volume = {523},
          eid = {A7},
        pages = {A7},
          doi = {10.1051/0004-6361/201014468},
archivePrefix = {arXiv},
       eprint = {1010.4743},
 primaryClass = {astro-ph.CO},
       adsurl = {https://ui.adsabs.harvard.edu/abs/2010A&A...523A...7G}
}

@ARTICLE{kessler2019,
       author = {{Kessler}, R. and {Brout}, D. and {D'Andrea}, C.~B. and {Davis}, T.~M. and {Hinton}, S.~R. and {Kim}, A.~G. and {Lasker}, J. and {Lidman}, C. and {Macaulay}, E. and {M{\"o}ller}, A. and {Sako}, M. and {Scolnic}, D. and {Smith}, M. and {Sullivan}, M. and {Zhang}, B. and {Andersen}, P. and {Asorey}, J. and {Avelino}, A. and {Calcino}, J. and {Carollo}, D. and {Challis}, P. and {Childress}, M. and {Clocchiatti}, A. and {Crawford}, S. and {Filippenko}, A.~V. and {Foley}, R.~J. and {Glazebrook}, K. and {Hoormann}, J.~K. and {Kasai}, E. and {Kirshner}, R.~P. and {Lewis}, G.~F. and {Mandel}, K.~S. and {March}, M. and {Morganson}, E. and {Muthukrishna}, D. and {Nugent}, P. and {Pan}, Y. -C. and {Sommer}, N.~E. and {Swann}, E. and {Thomas}, R.~C. and {Tucker}, B.~E. and {Uddin}, S.~A. and {Abbott}, T.~M.~C. and {Allam}, S. and {Annis}, J. and {Avila}, S. and {Banerji}, M. and {Bechtol}, K. and {Bertin}, E. and {Brooks}, D. and {Buckley-Geer}, E. and {Burke}, D.~L. and {Carnero Rosell}, A. and {Carrasco Kind}, M. and {Carretero}, J. and {Castander}, F.~J. and {Crocce}, M. and {da Costa}, L.~N. and {Davis}, C. and {De Vicente}, J. and {Desai}, S. and {Diehl}, H.~T. and {Doel}, P. and {Eifler}, T.~F. and {Flaugher}, B. and {Fosalba}, P. and {Frieman}, J. and {Garc{\'\i}a-Bellido}, J. and {Gaztanaga}, E. and {Gerdes}, D.~W. and {Gruen}, D. and {Gruendl}, R.~A. and {Gutierrez}, G. and {Hartley}, W.~G. and {Hollowood}, D.~L. and {Honscheid}, K. and {James}, D.~J. and {Johnson}, M.~W.~G. and {Johnson}, M.~D. and {Krause}, E. and {Kuehn}, K. and {Kuropatkin}, N. and {Lahav}, O. and {Li}, T.~S. and {Lima}, M. and {Marshall}, J.~L. and {Martini}, P. and {Menanteau}, F. and {Miller}, C.~J. and {Miquel}, R. and {Nord}, B. and {Plazas}, A.~A. and {Roodman}, A. and {Sanchez}, E. and {Scarpine}, V. and {Schindler}, R. and {Schubnell}, M. and {Serrano}, S. and {Sevilla-Noarbe}, I. and {Soares-Santos}, M. and {Sobreira}, F. and {Suchyta}, E. and {Tarle}, G. and {Thomas}, D. and {Walker}, A.~R. and {Zhang}, Y. and {DES Collaboration}},
        title = "{First cosmology results using Type Ia supernova from the Dark Energy Survey: simulations to correct supernova distance biases}",
      journal = {\mnras},
         year = 2019,
        month = may,
       volume = {485},
       number = {1},
        pages = {1171-1187},
          doi = {10.1093/mnras/stz463},
archivePrefix = {arXiv},
       eprint = {1811.02379},
 primaryClass = {astro-ph.CO},
       adsurl = {https://ui.adsabs.harvard.edu/abs/2019MNRAS.485.1171K}
}

@ARTICLE{kessler2009,
       author = {{Kessler}, Richard and {Bernstein}, Joseph P. and {Cinabro}, David and {Dilday}, Benjamin and {Frieman}, Joshua A. and {Jha}, Saurabh and {Kuhlmann}, Stephen and {Miknaitis}, Gajus and {Sako}, Masao and {Taylor}, Matt and {Vanderplas}, Jake},
        title = "{SNANA: A Public Software Package for Supernova Analysis}",
      journal = {\pasp},
         year = 2009,
        month = sep,
       volume = {121},
       number = {883},
        pages = {1028},
          doi = {10.1086/605984},
archivePrefix = {arXiv},
       eprint = {0908.4280},
 primaryClass = {astro-ph.CO},
       adsurl = {https://ui.adsabs.harvard.edu/abs/2009PASP..121.1028K}
}

@ARTICLE{riess1998,
       author = {{Riess}, Adam G. and {Filippenko}, Alexei V. and {Challis}, Peter and {Clocchiatti}, Alejandro and {Diercks}, Alan and {Garnavich}, Peter M. and {Gilliland}, Ron L. and {Hogan}, Craig J. and {Jha}, Saurabh and {Kirshner}, Robert P. and {Leibundgut}, B. and {Phillips}, M.~M. and {Reiss}, David and {Schmidt}, Brian P. and {Schommer}, Robert A. and {Smith}, R. Chris and {Spyromilio}, J. and {Stubbs}, Christopher and {Suntzeff}, Nicholas B. and {Tonry}, John},
        title = "{Observational Evidence from Supernovae for an Accelerating Universe and a Cosmological Constant}",
      journal = {\aj},
         year = 1998,
        month = sep,
       volume = {116},
       number = {3},
        pages = {1009-1038},
          doi = {10.1086/300499},
archivePrefix = {arXiv},
       eprint = {astro-ph/9805201},
 primaryClass = {astro-ph},
       adsurl = {https://ui.adsabs.harvard.edu/abs/1998AJ....116.1009R}
}

@ARTICLE{hu2023,
       author = {{Hu}, Jian-Ping and {Wang}, Fa-Yin},
        title = "{Hubble Tension: The Evidence of New Physics}",
      journal = {Universe},
         year = 2023,
        month = feb,
       volume = {9},
       number = {2},
        pages = {94},
          doi = {10.3390/universe9020094},
archivePrefix = {arXiv},
       eprint = {2302.05709},
 primaryClass = {astro-ph.CO},
       adsurl = {https://ui.adsabs.harvard.edu/abs/2023Univ....9...94H}
}

@ARTICLE{brout2021,
       author = {{Brout}, Dillon and {Scolnic}, Daniel},
        title = "{It's Dust: Solving the Mysteries of the Intrinsic Scatter and Host-galaxy Dependence of Standardized Type Ia Supernova Brightnesses}",
      journal = {\apj},
         year = 2021,
        month = mar,
       volume = {909},
       number = {1},
          eid = {26},
        pages = {26},
          doi = {10.3847/1538-4357/abd69b},
archivePrefix = {arXiv},
       eprint = {2004.10206},
 primaryClass = {astro-ph.CO},
       adsurl = {https://ui.adsabs.harvard.edu/abs/2021ApJ...909...26B}
}

@ARTICLE{brout2022,
       author = {{Brout}, Dillon and {Scolnic}, Dan and {Popovic}, Brodie and {Riess}, Adam G. and {Carr}, Anthony and {Zuntz}, Joe and {Kessler}, Rick and {Davis}, Tamara M. and {Hinton}, Samuel and {Jones}, David and {Kenworthy}, W. D'Arcy and {Peterson}, Erik R. and {Said}, Khaled and {Taylor}, Georgie and {Ali}, Noor and {Armstrong}, Patrick and {Charvu}, Pranav and {Dwomoh}, Arianna and {Meldorf}, Cole and {Palmese}, Antonella and {Qu}, Helen and {Rose}, Benjamin M. and {Sanchez}, Bruno and {Stubbs}, Christopher W. and {Vincenzi}, Maria and {Wood}, Charlotte M. and {Brown}, Peter J. and {Chen}, Rebecca and {Chambers}, Ken and {Coulter}, David A. and {Dai}, Mi and {Dimitriadis}, Georgios and {Filippenko}, Alexei V. and {Foley}, Ryan J. and {Jha}, Saurabh W. and {Kelsey}, Lisa and {Kirshner}, Robert P. and {M{\"o}ller}, Anais and {Muir}, Jessie and {Nadathur}, Seshadri and {Pan}, Yen-Chen and {Rest}, Armin and {Rojas-Bravo}, Cesar and {Sako}, Masao and {Siebert}, Matthew R. and {Smith}, Mat and {Stahl}, Benjamin E. and {Wiseman}, Phil},
        title = "{The Pantheon+ Analysis: Cosmological Constraints}",
      journal = {\apj},
         year = 2022,
        month = oct,
       volume = {938},
       number = {2},
          eid = {110},
        pages = {110},
          doi = {10.3847/1538-4357/ac8e04},
archivePrefix = {arXiv},
       eprint = {2202.04077},
 primaryClass = {astro-ph.CO},
       adsurl = {https://ui.adsabs.harvard.edu/abs/2022ApJ...938..110B}
}

@ARTICLE{mandel2017,
       author = {{Mandel}, Kaisey S. and {Scolnic}, Daniel M. and {Shariff}, Hikmatali and {Foley}, Ryan J. and {Kirshner}, Robert P.},
        title = "{The Type Ia Supernova Color-Magnitude Relation and Host Galaxy Dust: A Simple Hierarchical Bayesian Model}",
      journal = {\apj},
         year = 2017,
        month = jun,
       volume = {842},
       number = {2},
          eid = {93},
        pages = {93},
          doi = {10.3847/1538-4357/aa6038},
archivePrefix = {arXiv},
       eprint = {1609.04470},
 primaryClass = {astro-ph.CO},
       adsurl = {https://ui.adsabs.harvard.edu/abs/2017ApJ...842...93M}
}

@ARTICLE{uzsoy2024,
       author = {{Uzsoy}, Ana Sof{\'\i}a M. and {Thorp}, Stephen and {Grayling}, Matthew and {Mandel}, Kaisey S.},
        title = "{Variational inference for acceleration of SN Ia photometric distance estimation with BayeSN}",
      journal = {\mnras},
         year = 2024,
        month = dec,
       volume = {535},
       number = {3},
        pages = {2306-2321},
          doi = {10.1093/mnras/stae2465},
archivePrefix = {arXiv},
       eprint = {2405.06013},
 primaryClass = {astro-ph.IM},
       adsurl = {https://ui.adsabs.harvard.edu/abs/2024MNRAS.535.2306U}
}

@ARTICLE{march2011,
       author = {{March}, M.~C. and {Trotta}, R. and {Berkes}, P. and {Starkman}, G.~D. and {Vaudrevange}, P.~M.},
        title = "{Improved constraints on cosmological parameters from Type Ia supernova data}",
      journal = {\mnras},
         year = 2011,
        month = dec,
       volume = {418},
       number = {4},
        pages = {2308-2329},
          doi = {10.1111/j.1365-2966.2011.19584.x},
archivePrefix = {arXiv},
       eprint = {1102.3237},
 primaryClass = {astro-ph.CO},
       adsurl = {https://ui.adsabs.harvard.edu/abs/2011MNRAS.418.2308M}
}

@ARTICLE{rubin2015,
       author = {{Rubin}, D. and {Aldering}, G. and {Barbary}, K. and {Boone}, K. and {Chappell}, G. and {Currie}, M. and {Deustua}, S. and {Fagrelius}, P. and {Fruchter}, A. and {Hayden}, B. and {Lidman}, C. and {Nordin}, J. and {Perlmutter}, S. and {Saunders}, C. and {Sofiatti}, C. and {Supernova Cosmology Project}, The},
        title = "{UNITY: Confronting Supernova Cosmology's Statistical and Systematic Uncertainties in a Unified Bayesian Framework}",
      journal = {\apj},
         year = 2015,
        month = nov,
       volume = {813},
       number = {2},
          eid = {137},
        pages = {137},
          doi = {10.1088/0004-637X/813/2/137},
archivePrefix = {arXiv},
       eprint = {1507.01602},
 primaryClass = {astro-ph.CO},
       adsurl = {https://ui.adsabs.harvard.edu/abs/2015ApJ...813..137R}
}

@ARTICLE{shariff2016,
       author = {{Shariff}, Hikmatali and {Jiao}, Xiyun and {Trotta}, Roberto and {van Dyk}, David A.},
        title = "{BAHAMAS: New Analysis of Type Ia Supernovae Reveals Inconsistencies with Standard Cosmology}",
      journal = {\apj},
         year = 2016,
        month = aug,
       volume = {827},
       number = {1},
          eid = {1},
        pages = {1},
          doi = {10.3847/0004-637X/827/1/1},
archivePrefix = {arXiv},
       eprint = {1510.05954},
 primaryClass = {astro-ph.CO},
       adsurl = {https://ui.adsabs.harvard.edu/abs/2016ApJ...827....1S}
}

@ARTICLE{mandel2009,
       author = {{Mandel}, Kaisey S. and {Wood-Vasey}, W. Michael and {Friedman}, Andrew S. and {Kirshner}, Robert P.},
        title = "{Type Ia Supernova Light-Curve Inference: Hierarchical Bayesian Analysis in the Near-Infrared}",
      journal = {\apj},
         year = 2009,
        month = oct,
       volume = {704},
       number = {1},
        pages = {629-651},
          doi = {10.1088/0004-637X/704/1/629},
archivePrefix = {arXiv},
       eprint = {0908.0536},
 primaryClass = {astro-ph.CO},
       adsurl = {https://ui.adsabs.harvard.edu/abs/2009ApJ...704..629M}
}

@ARTICLE{mandel2022,
       author = {{Mandel}, Kaisey S. and {Thorp}, Stephen and {Narayan}, Gautham and {Friedman}, Andrew S. and {Avelino}, Arturo},
        title = "{A hierarchical Bayesian SED model for Type Ia supernovae in the optical to near-infrared}",
      journal = {\mnras},
         year = 2022,
        month = mar,
       volume = {510},
       number = {3},
        pages = {3939-3966},
          doi = {10.1093/mnras/stab3496},
archivePrefix = {arXiv},
       eprint = {2008.07538},
 primaryClass = {astro-ph.CO},
       adsurl = {https://ui.adsabs.harvard.edu/abs/2022MNRAS.510.3939M}
}

@ARTICLE{thorp2021,
       author = {{Thorp}, Stephen and {Mandel}, Kaisey S. and {Jones}, David O. and {Ward}, Sam M. and {Narayan}, Gautham},
        title = "{Testing the consistency of dust laws in SN Ia host galaxies: a BAYESN examination of Foundation DR1}",
      journal = {\mnras},
         year = 2021,
        month = dec,
       volume = {508},
       number = {3},
        pages = {4310-4331},
          doi = {10.1093/mnras/stab2849},
archivePrefix = {arXiv},
       eprint = {2102.05678},
 primaryClass = {astro-ph.GA},
       adsurl = {https://ui.adsabs.harvard.edu/abs/2021MNRAS.508.4310T}
}

@ARTICLE{mandel2011,
       author = {{Mandel}, Kaisey S. and {Narayan}, Gautham and {Kirshner}, Robert P.},
        title = "{Type Ia Supernova Light Curve Inference: Hierarchical Models in the Optical and Near-infrared}",
      journal = {\apj},
         year = 2011,
        month = apr,
       volume = {731},
       number = {2},
          eid = {120},
        pages = {120},
          doi = {10.1088/0004-637X/731/2/120},
archivePrefix = {arXiv},
       eprint = {1011.5910},
 primaryClass = {astro-ph.CO},
       adsurl = {https://ui.adsabs.harvard.edu/abs/2011ApJ...731..120M}
}

@ARTICLE{thorp2022,
       author = {{Thorp}, Stephen and {Mandel}, Kaisey S.},
        title = "{Constraining the SN Ia host galaxy dust law distribution and mass step: hierarchical BAYESN analysis of optical and near-infrared light curves}",
      journal = {\mnras},
         year = 2022,
        month = dec,
       volume = {517},
       number = {2},
        pages = {2360-2382},
          doi = {10.1093/mnras/stac2714},
archivePrefix = {arXiv},
       eprint = {2209.10552},
 primaryClass = {astro-ph.CO},
       adsurl = {https://ui.adsabs.harvard.edu/abs/2022MNRAS.517.2360T}
}

@ARTICLE{thorp2024,
       author = {{Thorp}, Stephen and {Mandel}, Kaisey S. and {Jones}, David O. and {Kirshner}, Robert P. and {Challis}, Peter M.},
        title = "{Using rest-frame optical and NIR data from the RAISIN survey to explore the redshift evolution of dust laws in SN Ia host galaxies}",
      journal = {\mnras},
         year = 2024,
        month = jun,
       volume = {530},
       number = {4},
        pages = {4016-4031},
          doi = {10.1093/mnras/stae1111},
archivePrefix = {arXiv},
       eprint = {2402.18624},
 primaryClass = {astro-ph.CO},
       adsurl = {https://ui.adsabs.harvard.edu/abs/2024MNRAS.530.4016T}
}

@ARTICLE{burns2011,
       author = {{Burns}, Christopher R. and {Stritzinger}, Maximilian and {Phillips}, M.~M. and {Kattner}, ShiAnne and {Persson}, S.~E. and {Madore}, Barry F. and {Freedman}, Wendy L. and {Boldt}, Luis and {Campillay}, Abdo and {Contreras}, Carlos and {Folatelli}, Gaston and {Gonzalez}, Sergio and {Krzeminski}, Wojtek and {Morrell}, Nidia and {Salgado}, Francisco and {Suntzeff}, Nicholas B.},
        title = "{The Carnegie Supernova Project: Light-curve Fitting with SNooPy}",
      journal = {\aj},
         year = 2011,
        month = jan,
       volume = {141},
       number = {1},
          eid = {19},
        pages = {19},
          doi = {10.1088/0004-6256/141/1/19},
archivePrefix = {arXiv},
       eprint = {1010.4040},
 primaryClass = {astro-ph.CO},
       adsurl = {https://ui.adsabs.harvard.edu/abs/2011AJ....141...19B}
}

@ARTICLE{burns2014,
       author = {{Burns}, Christopher R. and {Stritzinger}, Maximilian and {Phillips}, M.~M. and {Hsiao}, E.~Y. and {Contreras}, Carlos and {Persson}, S.~E. and {Folatelli}, Gaston and {Boldt}, Luis and {Campillay}, Abdo and {Castell{\'o}n}, Sergio and {Freedman}, Wendy L. and {Madore}, Barry F. and {Morrell}, Nidia and {Salgado}, Francisco and {Suntzeff}, Nicholas B.},
        title = "{The Carnegie Supernova Project: Intrinsic Colors of Type Ia Supernovae}",
      journal = {\apj},
         year = 2014,
        month = jul,
       volume = {789},
       number = {1},
          eid = {32},
        pages = {32},
          doi = {10.1088/0004-637X/789/1/32},
archivePrefix = {arXiv},
       eprint = {1405.3934},
 primaryClass = {astro-ph.CO},
       adsurl = {https://ui.adsabs.harvard.edu/abs/2014ApJ...789...32B}
}

@ARTICLE{lsst2019,
       author = {{Ivezi{\'c}}, {\v{Z}}eljko and {Kahn}, Steven M. and {Tyson}, J. Anthony and {Abel}, Bob and {Acosta}, Emily and {Allsman}, Robyn and {Alonso}, David and {AlSayyad}, Yusra and {Anderson}, Scott F. and {Andrew}, John and {Angel}, James Roger P. and {Angeli}, George Z. and {Ansari}, Reza and {Antilogus}, Pierre and {Araujo}, Constanza and {Armstrong}, Robert and {Arndt}, Kirk T. and {Astier}, Pierre and {Aubourg}, {\'E}ric and {Auza}, Nicole and {Axelrod}, Tim S. and {Bard}, Deborah J. and {Barr}, Jeff D. and {Barrau}, Aurelian and {Bartlett}, James G. and {Bauer}, Amanda E. and {Bauman}, Brian J. and {Baumont}, Sylvain and {Bechtol}, Ellen and {Bechtol}, Keith and {Becker}, Andrew C. and {Becla}, Jacek and {Beldica}, Cristina and {Bellavia}, Steve and {Bianco}, Federica B. and {Biswas}, Rahul and {Blanc}, Guillaume and {Blazek}, Jonathan and {Blandford}, Roger D. and {Bloom}, Josh S. and {Bogart}, Joanne and {Bond}, Tim W. and {Booth}, Michael T. and {Borgland}, Anders W. and {Borne}, Kirk and {Bosch}, James F. and {Boutigny}, Dominique and {Brackett}, Craig A. and {Bradshaw}, Andrew and {Brandt}, William Nielsen and {Brown}, Michael E. and {Bullock}, James S. and {Burchat}, Patricia and {Burke}, David L. and {Cagnoli}, Gianpietro and {Calabrese}, Daniel and {Callahan}, Shawn and {Callen}, Alice L. and {Carlin}, Jeffrey L. and {Carlson}, Erin L. and {Chandrasekharan}, Srinivasan and {Charles-Emerson}, Glenaver and {Chesley}, Steve and {Cheu}, Elliott C. and {Chiang}, Hsin-Fang and {Chiang}, James and {Chirino}, Carol and {Chow}, Derek and {Ciardi}, David R. and {Claver}, Charles F. and {Cohen-Tanugi}, Johann and {Cockrum}, Joseph J. and {Coles}, Rebecca and {Connolly}, Andrew J. and {Cook}, Kem H. and {Cooray}, Asantha and {Covey}, Kevin R. and {Cribbs}, Chris and {Cui}, Wei and {Cutri}, Roc and {Daly}, Philip N. and {Daniel}, Scott F. and {Daruich}, Felipe and {Daubard}, Guillaume and {Daues}, Greg and {Dawson}, William and {Delgado}, Francisco and {Dellapenna}, Alfred and {de Peyster}, Robert and {de Val-Borro}, Miguel and {Digel}, Seth W. and {Doherty}, Peter and {Dubois}, Richard and {Dubois-Felsmann}, Gregory P. and {Durech}, Josef and {Economou}, Frossie and {Eifler}, Tim and {Eracleous}, Michael and {Emmons}, Benjamin L. and {Fausti Neto}, Angelo and {Ferguson}, Henry and {Figueroa}, Enrique and {Fisher-Levine}, Merlin and {Focke}, Warren and {Foss}, Michael D. and {Frank}, James and {Freemon}, Michael D. and {Gangler}, Emmanuel and {Gawiser}, Eric and {Geary}, John C. and {Gee}, Perry and {Geha}, Marla and {Gessner}, Charles J.~B. and {Gibson}, Robert R. and {Gilmore}, D. Kirk and {Glanzman}, Thomas and {Glick}, William and {Goldina}, Tatiana and {Goldstein}, Daniel A. and {Goodenow}, Iain and {Graham}, Melissa L. and {Gressler}, William J. and {Gris}, Philippe and {Guy}, Leanne P. and {Guyonnet}, Augustin and {Haller}, Gunther and {Harris}, Ron and {Hascall}, Patrick A. and {Haupt}, Justine and {Hernandez}, Fabio and {Herrmann}, Sven and {Hileman}, Edward and {Hoblitt}, Joshua and {Hodgson}, John A. and {Hogan}, Craig and {Howard}, James D. and {Huang}, Dajun and {Huffer}, Michael E. and {Ingraham}, Patrick and {Innes}, Walter R. and {Jacoby}, Suzanne H. and {Jain}, Bhuvnesh and {Jammes}, Fabrice and {Jee}, M. James and {Jenness}, Tim and {Jernigan}, Garrett and {Jevremovi{\'c}}, Darko and {Johns}, Kenneth and {Johnson}, Anthony S. and {Johnson}, Margaret W.~G. and {Jones}, R. Lynne and {Juramy-Gilles}, Claire and {Juri{\'c}}, Mario and {Kalirai}, Jason S. and {Kallivayalil}, Nitya J. and {Kalmbach}, Bryce and {Kantor}, Jeffrey P. and {Karst}, Pierre and {Kasliwal}, Mansi M. and {Kelly}, Heather and {Kessler}, Richard and {Kinnison}, Veronica and {Kirkby}, David and {Knox}, Lloyd and {Kotov}, Ivan V. and {Krabbendam}, Victor L. and {Krughoff}, K. Simon and {Kub{\'a}nek}, Petr and {Kuczewski}, John and {Kulkarni}, Shri and {Ku}, John and {Kurita}, Nadine R. and {Lage}, Craig S. and {Lambert}, Ron and {Lange}, Travis and {Langton}, J. Brian and {Le Guillou}, Laurent and {Levine}, Deborah and {Liang}, Ming and {Lim}, Kian-Tat and {Lintott}, Chris J. and {Long}, Kevin E. and {Lopez}, Margaux and {Lotz}, Paul J. and {Lupton}, Robert H. and {Lust}, Nate B. and {MacArthur}, Lauren A. and {Mahabal}, Ashish and {Mandelbaum}, Rachel and {Markiewicz}, Thomas W. and {Marsh}, Darren S. and {Marshall}, Philip J. and {Marshall}, Stuart and {May}, Morgan and {McKercher}, Robert and {McQueen}, Michelle and {Meyers}, Joshua and {Migliore}, Myriam and {Miller}, Michelle and {Mills}, David J. and {Miraval}, Connor and {Moeyens}, Joachim and {Moolekamp}, Fred E. and {Monet}, David G. and {Moniez}, Marc and {Monkewitz}, Serge and {Montgomery}, Christopher and {Morrison}, Christopher B. and {Mueller}, Fritz and {Muller}, Gary P. and {Mu{\~n}oz Arancibia}, Freddy and {Neill}, Douglas R. and {Newbry}, Scott P. and {Nief}, Jean-Yves and {Nomerotski}, Andrei and {Nordby}, Martin and {O'Connor}, Paul and {Oliver}, John and {Olivier}, Scot S. and {Olsen}, Knut and {O'Mullane}, William and {Ortiz}, Sandra and {Osier}, Shawn and {Owen}, Russell E. and {Pain}, Reynald and {Palecek}, Paul E. and {Parejko}, John K. and {Parsons}, James B. and {Pease}, Nathan M. and {Peterson}, J. Matt and {Peterson}, John R. and {Petravick}, Donald L. and {Libby Petrick}, M.~E. and {Petry}, Cathy E. and {Pierfederici}, Francesco and {Pietrowicz}, Stephen and {Pike}, Rob and {Pinto}, Philip A. and {Plante}, Raymond and {Plate}, Stephen and {Plutchak}, Joel P. and {Price}, Paul A. and {Prouza}, Michael and {Radeka}, Veljko and {Rajagopal}, Jayadev and {Rasmussen}, Andrew P. and {Regnault}, Nicolas and {Reil}, Kevin A. and {Reiss}, David J. and {Reuter}, Michael A. and {Ridgway}, Stephen T. and {Riot}, Vincent J. and {Ritz}, Steve and {Robinson}, Sean and {Roby}, William and {Roodman}, Aaron and {Rosing}, Wayne and {Roucelle}, Cecille and {Rumore}, Matthew R. and {Russo}, Stefano and {Saha}, Abhijit and {Sassolas}, Benoit and {Schalk}, Terry L. and {Schellart}, Pim and {Schindler}, Rafe H. and {Schmidt}, Samuel and {Schneider}, Donald P. and {Schneider}, Michael D. and {Schoening}, William and {Schumacher}, German and {Schwamb}, Megan E. and {Sebag}, Jacques and {Selvy}, Brian and {Sembroski}, Glenn H. and {Seppala}, Lynn G. and {Serio}, Andrew and {Serrano}, Eduardo and {Shaw}, Richard A. and {Shipsey}, Ian and {Sick}, Jonathan and {Silvestri}, Nicole and {Slater}, Colin T. and {Smith}, J. Allyn and {Smith}, R. Chris and {Sobhani}, Shahram and {Soldahl}, Christine and {Storrie-Lombardi}, Lisa and {Stover}, Edward and {Strauss}, Michael A. and {Street}, Rachel A. and {Stubbs}, Christopher W. and {Sullivan}, Ian S. and {Sweeney}, Donald and {Swinbank}, John D. and {Szalay}, Alexander and {Takacs}, Peter and {Tether}, Stephen A. and {Thaler}, Jon J. and {Thayer}, John Gregg and {Thomas}, Sandrine and {Thornton}, Adam J. and {Thukral}, Vaikunth and {Tice}, Jeffrey and {Trilling}, David E. and {Turri}, Max and {Van Berg}, Richard and {Vanden Berk}, Daniel and {Vetter}, Kurt and {Virieux}, Francoise and {Vucina}, Tomislav and {Wahl}, William and {Walkowicz}, Lucianne and {Walsh}, Brian and {Walter}, Christopher W. and {Wang}, Daniel L. and {Wang}, Shin-Yawn and {Warner}, Michael and {Wiecha}, Oliver and {Willman}, Beth and {Winters}, Scott E. and {Wittman}, David and {Wolff}, Sidney C. and {Wood-Vasey}, W. Michael and {Wu}, Xiuqin and {Xin}, Bo and {Yoachim}, Peter and {Zhan}, Hu},
        title = "{LSST: From Science Drivers to Reference Design and Anticipated Data Products}",
      journal = {\apj},
         year = 2019,
        month = mar,
       volume = {873},
       number = {2},
          eid = {111},
        pages = {111},
          doi = {10.3847/1538-4357/ab042c},
archivePrefix = {arXiv},
       eprint = {0805.2366},
 primaryClass = {astro-ph},
       adsurl = {https://ui.adsabs.harvard.edu/abs/2019ApJ...873..111I}
}

@ARTICLE{narayan2019,
       author = {{Narayan}, Gautham and {Matheson}, Thomas and {Saha}, Abhijit and {Axelrod}, Tim and {Calamida}, Annalisa and {Olszewski}, Edward and {Claver}, Jenna and {Mandel}, Kaisey S. and {Bohlin}, Ralph C. and {Holberg}, Jay B. and {Deustua}, Susana and {Rest}, Armin and {Stubbs}, Christopher W. and {Shanahan}, Clare E. and {Vaz}, Amali L. and {Zenteno}, Alfredo and {Strampelli}, Giovanni and {Hubeny}, Ivan and {Points}, Sean and {Sabbi}, Elena and {Mackenty}, John},
        title = "{Subpercent Photometry: Faint DA White Dwarf Spectrophotometric Standards for Astrophysical Observatories}",
      journal = {\apjs},
         year = 2019,
        month = apr,
       volume = {241},
       number = {2},
          eid = {20},
        pages = {20},
          doi = {10.3847/1538-4365/ab0557},
archivePrefix = {arXiv},
       eprint = {1811.12534},
 primaryClass = {astro-ph.IM},
       adsurl = {https://ui.adsabs.harvard.edu/abs/2019ApJS..241...20N}
}

@ARTICLE{numpyro,
       author = {{Phan}, Du and {Pradhan}, Neeraj and {Jankowiak}, Martin},
        title = "{Composable Effects for Flexible and Accelerated Probabilistic Programming in NumPyro}",
      journal = {ArXiv e-prints},
         year = 2019,
        month = dec,
          eid = {arXiv:1912.11554},
          doi = {10.48550/arXiv.1912.11554},
archivePrefix = {arXiv},
       eprint = {1912.11554},
 primaryClass = {stat.ML},
       adsurl = {https://ui.adsabs.harvard.edu/abs/2019arXiv191211554P}
}

@ARTICLE{ward2023,
       author = {{Ward}, Sam M. and {Thorp}, Stephen and {Mandel}, Kaisey S. and {Dhawan}, Suhail and {Jones}, David O. and {Taggart}, Kirsty and {Foley}, Ryan J. and {Narayan}, Gautham and {Chambers}, Kenneth C. and {Coulter}, David A. and {Davis}, Kyle W. and {de Boer}, Thomas and {de Soto}, Kaylee and {Earl}, Nicholas and {Gagliano}, Alex and {Gao}, Hua and {Hjorth}, Jens and {Huber}, Mark E. and {Izzo}, Luca and {Langeroodi}, Danial and {Magnier}, Eugene A. and {McGill}, Peter and {Rest}, Armin and {Rojas-Bravo}, C{\'e}sar and {Wojtak}, Rados{\l}aw and {Young Supernova Experiment}},
        title = "{Relative Intrinsic Scatter in Hierarchical Type Ia Supernova Sibling Analyses: Application to SNe 2021hpr, 1997bq, and 2008fv in NGC 3147}",
      journal = {\apj},
         year = 2023,
        month = oct,
       volume = {956},
       number = {2},
          eid = {111},
        pages = {111},
          doi = {10.3847/1538-4357/acf7bb},
archivePrefix = {arXiv},
       eprint = {2209.10558},
 primaryClass = {astro-ph.CO},
       adsurl = {https://ui.adsabs.harvard.edu/abs/2023ApJ...956..111W}
}

@ARTICLE{malmquist1922,
       author = {{Malmquist}, K.~G.},
        title = "{On some relations in stellar statistics}",
      journal = {Medd.\ Lunds Astron.\ Obser.\ Serie I},
         year = 1922,
        month = mar,
       volume = {100},
        pages = {1-52},
       adsurl = {https://ui.adsabs.harvard.edu/abs/1922MeLuF.100....1M}
}

@ARTICLE{kessler2017,
       author = {{Kessler}, R. and {Scolnic}, D.},
        title = "{Correcting Type Ia Supernova Distances for Selection Biases and Contamination in Photometrically Identified Samples}",
      journal = {\apj},
         year = 2017,
        month = feb,
       volume = {836},
       number = {1},
          eid = {56},
        pages = {56},
          doi = {10.3847/1538-4357/836/1/56},
archivePrefix = {arXiv},
       eprint = {1610.04677},
 primaryClass = {astro-ph.CO},
       adsurl = {https://ui.adsabs.harvard.edu/abs/2017ApJ...836...56K}
}

@ARTICLE{march2018,
       author = {{March}, M C and {Wolf}, R C and {Sako}, M and {D'Andrea}, C and {Brout}, D},
        title = "{A Bayesian approach to truncated data sets: An application to Malmquist bias in Supernova Cosmology}",
      journal = {ArXiv e-prints},
         year = 2018,
        month = apr,
          eid = {arXiv:1804.02474},
          doi = {10.48550/arXiv.1804.02474},
archivePrefix = {arXiv},
       eprint = {1804.02474},
 primaryClass = {astro-ph.CO},
       adsurl = {https://ui.adsabs.harvard.edu/abs/2018arXiv180402474M}
}

@techreport{bishop1994,
  title={Mixture density networks},
  author={Bishop, Christopher M},
  year={1994},
  number = {NCRG/94/004},
  institution={Aston University},
  address = {\url{https://publications.aston.ac.uk/id/eprint/373/1/NCRG_94_004.pdf}}
}

@InProceedings{rezende2015,
  title = 	 {Variational Inference with Normalizing Flows},
  author = 	 {Rezende, Danilo and Mohamed, Shakir},
  booktitle = 	 {Proceedings of the 32nd International Conference on Machine Learning},
  pages = 	 {1530--1538},
  year = 	 {2015},
  editor = 	 {Bach, Francis and Blei, David},
  volume = 	 {37},
  series = 	 {Proceedings of Machine Learning Research},
  address = 	 {Lille, France},
  month = 	 {07--09 Jul},
  publisher =    {PMLR},
  url = 	 {https://proceedings.mlr.press/v37/rezende15.html},
}

@InProceedings{hermans2020,
  title = 	 {Likelihood-free {MCMC} with Amortized Approximate Ratio Estimators},
  author =       {Hermans, Joeri and Begy, Volodimir and Louppe, Gilles},
  booktitle = 	 {Proceedings of the 37th International Conference on Machine Learning},
  pages = 	 {4239--4248},
  year = 	 {2020},
  editor = 	 {Daum\'{e} III, Hal and Singh, Aarti},
  volume = 	 {119},
  series = 	 {Proceedings of Machine Learning Research},
  month = 	 {13--18 Jul},
  publisher =    {PMLR},
  url = 	 {https://proceedings.mlr.press/v119/hermans20a.html}
}

@ARTICLE{karchev2023,
       author = {{Karchev}, Konstantin and {Trotta}, Roberto and {Weniger}, Christoph},
        title = "{SICRET: Supernova Ia Cosmology with truncated marginal neural Ratio EsTimation}",
      journal = {\mnras},
         year = 2023,
        month = mar,
       volume = {520},
       number = {1},
        pages = {1056-1072},
          doi = {10.1093/mnras/stac3785},
archivePrefix = {arXiv},
       eprint = {2209.06733},
 primaryClass = {astro-ph.CO},
       adsurl = {https://ui.adsabs.harvard.edu/abs/2023MNRAS.520.1056K}
}

@ARTICLE{perlmutter1999,
       author = {{Perlmutter}, S. and {Aldering}, G. and {Goldhaber}, G. and {Knop}, R.~A. and {Nugent}, P. and {Castro}, P.~G. and {Deustua}, S. and {Fabbro}, S. and {Goobar}, A. and {Groom}, D.~E. and {Hook}, I.~M. and {Kim}, A.~G. and {Kim}, M.~Y. and {Lee}, J.~C. and {Nunes}, N.~J. and {Pain}, R. and {Pennypacker}, C.~R. and {Quimby}, R. and {Lidman}, C. and {Ellis}, R.~S. and {Irwin}, M. and {McMahon}, R.~G. and {Ruiz-Lapuente}, P. and {Walton}, N. and {Schaefer}, B. and {Boyle}, B.~J. and {Filippenko}, A.~V. and {Matheson}, T. and {Fruchter}, A.~S. and {Panagia}, N. and {Newberg}, H.~J.~M. and {Couch}, W.~J. and {Project}, The Supernova Cosmology},
        title = "{Measurements of {\ensuremath{\Omega}} and {\ensuremath{\Lambda}} from 42 High-Redshift Supernovae}",
      journal = {\apj},
         year = 1999,
        month = jun,
       volume = {517},
       number = {2},
        pages = {565-586},
          doi = {10.1086/307221},
       adsurl = {https://ui.adsabs.harvard.edu/abs/1999ApJ...517..565P}
}

@ARTICLE{alsing2019,
       author = {{Alsing}, Justin and {Charnock}, Tom and {Feeney}, Stephen and {Wandelt}, Benjamin},
        title = "{Fast likelihood-free cosmology with neural density estimators and active learning}",
      journal = {\mnras},
         year = 2019,
        month = sep,
       volume = {488},
       number = {3},
        pages = {4440-4458},
          doi = {10.1093/mnras/stz1960},
archivePrefix = {arXiv},
       eprint = {1903.00007},
 primaryClass = {astro-ph.CO},
       adsurl = {https://ui.adsabs.harvard.edu/abs/2019MNRAS.488.4440A}
}

@article{papamakarios2021,
  author  = {George Papamakarios and Eric Nalisnick and Danilo Jimenez Rezende and Shakir Mohamed and Balaji Lakshminarayanan},
  title   = {Normalizing Flows for Probabilistic Modeling and Inference},
  journal = {J.\ Machine Learning Res.},
  year    = {2021},
  volume  = {22},
  number  = {57},
  pages   = {1--64},
  url     = {http://jmlr.org/papers/v22/19-1028.html}
}

@article{pyro,
  author    = {Eli Bingham and
               Jonathan P. Chen and
               Martin Jankowiak and
               Fritz Obermeyer and
               Neeraj Pradhan and
               Theofanis Karaletsos and
               Rohit Singh and
               Paul A. Szerlip and
               Paul Horsfall and
               Noah D. Goodman},
  title     = {Pyro: Deep Universal Probabilistic Programming},
  journal   = {J. Machine Learning Res.},
  volume    = {20},
  pages     = {1-6},
  year      = {2019},
  url       = {http://jmlr.org/papers/v20/18-403.html},
  adsurl = {https://ui.adsabs.harvard.edu/abs/2018arXiv181009538B}
}

@ARTICLE{davis2011,
       author = {{Davis}, Tamara M. and {Hui}, Lam and {Frieman}, Joshua A. and {Haugb{\o}lle}, Troels and {Kessler}, Richard and {Sinclair}, Benjamin and {Sollerman}, Jesper and {Bassett}, Bruce and {Marriner}, John and {M{\"o}rtsell}, Edvard and {Nichol}, Robert C. and {Richmond}, Michael W. and {Sako}, Masao and {Schneider}, Donald P. and {Smith}, Mathew},
        title = "{The Effect of Peculiar Velocities on Supernova Cosmology}",
      journal = {\apj},
         year = 2011,
        month = nov,
       volume = {741},
       number = {1},
          eid = {67},
        pages = {67},
          doi = {10.1088/0004-637X/741/1/67},
archivePrefix = {arXiv},
       eprint = {1012.2912},
 primaryClass = {astro-ph.CO},
       adsurl = {https://ui.adsabs.harvard.edu/abs/2011ApJ...741...67D}
}

@ARTICLE{nicolas2021,
       author = {{Nicolas}, N. and {Rigault}, M. and {Copin}, Y. and {Graziani}, R. and {Aldering}, G. and {Briday}, M. and {Kim}, Y. -L. and {Nordin}, J. and {Perlmutter}, S. and {Smith}, M.},
        title = "{Redshift evolution of the underlying type Ia supernova stretch distribution}",
      journal = {\aap},
         year = 2021,
        month = may,
       volume = {649},
          eid = {A74},
        pages = {A74},
          doi = {10.1051/0004-6361/202038447},
archivePrefix = {arXiv},
       eprint = {2005.09441},
 primaryClass = {astro-ph.CO},
       adsurl = {https://ui.adsabs.harvard.edu/abs/2021A&A...649A..74N}
}

@ARTICLE{dimitriadis2025,
       author = {{Dimitriadis}, G. and {Burgaz}, U. and {Deckers}, M. and {Maguire}, K. and {Johansson}, J. and {Smith}, M. and {Rigault}, M. and {Frohmaier}, C. and {Sollerman}, J. and {Galbany}, L. and {Kim}, Y. -L. and {Liu}, C. and {Miller}, A.~A. and {Nugent}, P.~E. and {Alburai}, A. and {Chen}, P. and {Dhawan}, S. and {Ginolin}, M. and {Goobar}, A. and {Groom}, S.~L. and {Harvey}, L. and {Kenworthy}, W.~D. and {Kulkarni}, S.~R. and {Phan}, K. and {Popovic}, B. and {Riddle}, R.~L. and {Rusholme}, B. and {M{\"u}ller-Bravo}, T.~E. and {Nordin}, J. and {Terwel}, J.~H. and {Townsend}, A.},
        title = "{ZTF SN Ia DR2: The diversity and relative rates of the thermonuclear supernova population}",
      journal = {\aap},
         year = 2025,
        month = feb,
       volume = {694},
          eid = {A10},
        pages = {A10},
          doi = {10.1051/0004-6361/202451852},
archivePrefix = {arXiv},
       eprint = {2406.01434},
 primaryClass = {astro-ph.CO},
       adsurl = {https://ui.adsabs.harvard.edu/abs/2025A&A...694A..10D}
}

@ARTICLE{ginolin2025a,
       author = {{Ginolin}, M. and {Rigault}, M. and {Smith}, M. and {Copin}, Y. and {Ruppin}, F. and {Dimitriadis}, G. and {Goobar}, A. and {Johansson}, J. and {Maguire}, K. and {Nordin}, J. and {Amenouche}, M. and {Aubert}, M. and {Barjou-Delayre}, C. and {Betoule}, M. and {Burgaz}, U. and {Carreres}, B. and {Deckers}, M. and {Dhawan}, S. and {Feinstein}, F. and {Fouchez}, D. and {Galbany}, L. and {Ganot}, C. and {Harvey}, L. and {de Jaeger}, T. and {Kenworthy}, W.~D. and {Kim}, Y. -L. and {Kowalski}, M. and {Kuhn}, D. and {Lacroix}, L. and {M{\"u}ller-Bravo}, T.~E. and {Nugent}, P. and {Popovic}, B. and {Racine}, B. and {Rosnet}, P. and {Rosselli}, D. and {Sollerman}, J. and {Terwel}, J.~H. and {Townsend}, A. and {Brugger}, J. and {Bellm}, E.~C. and {Kasliwal}, M.~M. and {Kulkarni}, S. and {Laher}, R.~R. and {Masci}, F.~J. and {Riddle}, R.~L. and {Sharma}, Y.},
        title = "{ZTF SN Ia DR2: Environmental dependencies of stretch and luminosity for a volume-limited sample of 1000 type Ia supernovae}",
      journal = {\aap},
         year = 2025,
        month = mar,
       volume = {695},
          eid = {A140},
        pages = {A140},
          doi = {10.1051/0004-6361/202450378},
archivePrefix = {arXiv},
       eprint = {2405.20965},
 primaryClass = {astro-ph.CO},
       adsurl = {https://ui.adsabs.harvard.edu/abs/2025A&A...695A.140G}
}

@ARTICLE{ginolin2025b,
       author = {{Ginolin}, M. and {Rigault}, M. and {Copin}, Y. and {Popovic}, B. and {Dimitriadis}, G. and {Goobar}, A. and {Johansson}, J. and {Maguire}, K. and {Nordin}, J. and {Smith}, M. and {Aubert}, M. and {Barjou-Delayre}, C. and {Burgaz}, U. and {Carreres}, B. and {Dhawan}, S. and {Deckers}, M. and {Feinstein}, F. and {Fouchez}, D. and {Galbany}, L. and {Ganot}, C. and {de Jaeger}, T. and {Kim}, Y. -L. and {Kuhn}, D. and {Lacroix}, L. and {M{\"u}ller-Bravo}, T.~E. and {Nugent}, P. and {Racine}, B. and {Rosnet}, P. and {Rosselli}, D. and {Ruppin}, F. and {Sollerman}, J. and {Terwel}, J.~H. and {Townsend}, A. and {Dekany}, R. and {Graham}, M. and {Kasliwal}, M. and {Groom}, S.~L. and {Purdum}, J. and {Rusholme}, B. and {van der Walt}, S.},
        title = "{ZTF SN Ia DR2: Colour standardisation of type Ia supernovae and its dependence on the environment}",
      journal = {\aap},
         year = 2025,
        month = feb,
       volume = {694},
          eid = {A4},
        pages = {A4},
          doi = {10.1051/0004-6361/202450943},
archivePrefix = {arXiv},
       eprint = {2406.02072},
 primaryClass = {astro-ph.CO},
       adsurl = {https://ui.adsabs.harvard.edu/abs/2025A&A...694A...4G}
}

@ARTICLE{popovic2021,
       author = {{Popovic}, Brodie and {Brout}, Dillon and {Kessler}, Richard and {Scolnic}, Daniel},
        title = "{The Pantheon+ Analysis: Forward Modeling the Dust and Intrinsic Color Distributions of Type Ia Supernovae, and Quantifying Their Impact on Cosmological Inferences}",
      journal = {\apj},
         year = 2023,
        month = mar,
       volume = {945},
       number = {1},
          eid = {84},
        pages = {84},
          doi = {10.3847/1538-4357/aca273},
archivePrefix = {arXiv},
       eprint = {2112.04456},
 primaryClass = {astro-ph.CO},
       adsurl = {https://ui.adsabs.harvard.edu/abs/2023ApJ...945...84P}
}

@ARTICLE{grayling2024,
       author = {{Grayling}, Matthew and {Thorp}, Stephen and {Mandel}, Kaisey S. and {Dhawan}, Suhail and {Uzsoy}, Ana Sofia M. and {Boyd}, Benjamin M. and {Hayes}, Erin E. and {Ward}, Sam M.},
        title = "{Scalable hierarchical BayeSN inference: investigating dependence of SN Ia host galaxy dust properties on stellar mass and redshift}",
      journal = {\mnras},
         year = 2024,
        month = jun,
       volume = {531},
       number = {1},
        pages = {953-976},
          doi = {10.1093/mnras/stae1202},
archivePrefix = {arXiv},
       eprint = {2401.08755},
 primaryClass = {astro-ph.CO},
       adsurl = {https://ui.adsabs.harvard.edu/abs/2024MNRAS.531..953G}
}

@ARTICLE{phillips,
       author = {{Phillips}, M.~M.},
        title = "{The Absolute Magnitudes of Type IA Supernovae}",
      journal = {\apjl},
         year = 1993,
        month = aug,
       volume = {413},
        pages = {L105},
          doi = {10.1086/186970},
       adsurl = {https://ui.adsabs.harvard.edu/abs/1993ApJ...413L.105P}
}

@ARTICLE{kenworthy2021,
       author = {{Kenworthy}, W.~D. and {Jones}, D.~O. and {Dai}, M. and {Kessler}, R. and {Scolnic}, D. and {Brout}, D. and {Siebert}, M.~R. and {Pierel}, J.~D.~R. and {Dettman}, K.~G. and {Dimitriadis}, G. and {Foley}, R.~J. and {Jha}, S.~W. and {Pan}, Y. -C. and {Riess}, A. and {Rodney}, S. and {Rojas-Bravo}, C.},
        title = "{SALT3: An Improved Type Ia Supernova Model for Measuring Cosmic Distances}",
      journal = {\apj},
         year = 2021,
        month = dec,
       volume = {923},
       number = {2},
          eid = {265},
        pages = {265},
          doi = {10.3847/1538-4357/ac30d8},
archivePrefix = {arXiv},
       eprint = {2104.07795},
 primaryClass = {astro-ph.CO},
       adsurl = {https://ui.adsabs.harvard.edu/abs/2021ApJ...923..265K}
}

@ARTICLE{boyd2024,
       author = {{Boyd}, Benjamin M. and {Grayling}, Matthew and {Thorp}, Stephen and {Mandel}, Kaisey S.},
        title = "{Accounting for Selection Effects in Supernova Cosmology with Simulation-Based Inference and Hierarchical Bayesian Modelling}",
      journal = {ArXiv e-prints},
         year = 2024,
        month = jul,
          eid = {arXiv:2407.15923},
          doi = {10.48550/arXiv.2407.15923},
archivePrefix = {arXiv},
       eprint = {2407.15923},
 primaryClass = {astro-ph.CO},
       adsurl = {https://ui.adsabs.harvard.edu/abs/2024arXiv240715923B}
}

@ARTICLE{karchev2025,
       author = {{Karchev}, Konstantin and {Trotta}, Roberto},
        title = "{STAR NRE: solving supernova selection effects with set-based truncated auto-regressive neural ratio estimation}",
      journal = {\jcap},
         year = 2025,
        month = jul,
       volume = {2025},
       number = {7},
          eid = {031},
        pages = {031},
          doi = {10.1088/1475-7516/2025/07/031},
archivePrefix = {arXiv},
       eprint = {2409.03837},
 primaryClass = {astro-ph.CO},
       adsurl = {https://ui.adsabs.harvard.edu/abs/2025JCAP...07..031K}
}

@article{gelman1992,
author = {Andrew Gelman and Donald B. Rubin},
title = {{Inference from Iterative Simulation Using Multiple Sequences}},
volume = {7},
journal = {Statistical Sci.},
number = {4},
publisher = {Institute of Mathematical Statistics},
pages = {457 -- 472},
year = {1992},
doi = {10.1214/ss/1177011136},
URL = {https://doi.org/10.1214/ss/1177011136}}

@ARTICLE{vehtari2021,
       author = {{Vehtari}, Aki and {Gelman}, Andrew and {Simpson}, Daniel and {Carpenter}, Bob and {B{\"u}rkner}, Paul-Christian},
        title = "{Rank-normalization, folding, and localization: An improved R-hat for assessing convergence of MCMC (with Discussion)}",
      journal = {Bayesian Analysis},
         year = 2021,
        month = jun,
       volume = {16},
       number = {2},
        pages = {667-718},
          doi = {10.1214/20-BA1221},
archivePrefix = {arXiv},
       eprint = {1903.08008},
 primaryClass = {stat.CO},
       adsurl = {https://ui.adsabs.harvard.edu/abs/2021BayAn..16..667V}
}

@ARTICLE{betancourt2014,
       author = {{Betancourt}, M.~J. and {Byrne}, Simon and {Girolami}, Mark},
        title = "{Optimizing The Integrator Step Size for Hamiltonian Monte Carlo}",
      journal = {ArXiv e-prints},
         year = 2014,
        month = nov,
          eid = {arXiv:1411.6669},
          doi = {10.48550/arXiv.1411.6669},
archivePrefix = {arXiv},
       eprint = {1411.6669},
 primaryClass = {stat.ME},
       adsurl = {https://ui.adsabs.harvard.edu/abs/2014arXiv1411.6669B}
}

@ARTICLE{Betancourt2017,
       author = {{Betancourt}, Michael},
        title = "{A Conceptual Introduction to Hamiltonian Monte Carlo}",
      journal = {ArXiv e-prints},
         year = 2017,
        month = jan,
          eid = {arXiv:1701.02434},
          doi = {10.48550/arXiv.1701.02434},
archivePrefix = {arXiv},
       eprint = {1701.02434},
 primaryClass = {stat.ME},
       adsurl = {https://ui.adsabs.harvard.edu/abs/2017arXiv170102434B}
}

@ARTICLE{kessler2023,
       author = {{Kessler}, R. and {Vincenzi}, M. and {Armstrong}, P.},
        title = "{Binning is Sinning: Redemption for Hubble Diagram Using Photometrically Classified Type Ia Supernovae}",
      journal = {\apjl},
         year = 2023,
        month = jul,
       volume = {952},
       number = {1},
          eid = {L8},
        pages = {L8},
          doi = {10.3847/2041-8213/ace34d},
archivePrefix = {arXiv},
       eprint = {2306.05819},
 primaryClass = {astro-ph.CO},
       adsurl = {https://ui.adsabs.harvard.edu/abs/2023ApJ...952L...8K}
}

@ARTICLE{brout2021bin,
       author = {{Brout}, Dillon and {Hinton}, Samuel R. and {Scolnic}, Dan},
        title = "{Binning is Sinning (Supernova Version): The Impact of Self-calibration in Cosmological Analyses with Type Ia Supernovae}",
      journal = {\apjl},
         year = 2021,
        month = may,
       volume = {912},
       number = {2},
          eid = {L26},
        pages = {L26},
          doi = {10.3847/2041-8213/abf4db},
archivePrefix = {arXiv},
       eprint = {2012.05900},
 primaryClass = {astro-ph.CO},
       adsurl = {https://ui.adsabs.harvard.edu/abs/2021ApJ...912L..26B}
}

@ARTICLE{karchev2025cigars,
       author = {{Karchev}, Konstantin and {Trotta}, Roberto and {Jim{\'e}nez}, Ra{\'u}l},
        title = "{CIGaRS I: combined simulation-based inference from type Ia supernovae and host photometry}",
      journal = {Nature Astronomy},
         year = 2026,
        month = may,
          doi = {10.1038/s41550-026-02842-5},
archivePrefix = {arXiv},
       eprint = {2508.15899},
 primaryClass = {astro-ph.CO},
       adsurl = {https://ui.adsabs.harvard.edu/abs/2026NatAs.tmp...91K}
}

@inproceedings{
heinrich2023,
title={Hierarchical Neural Simulation-Based Inference Over Event Ensembles},
author={Lukas Heinrich and Siddharth Mishra-Sharma and Chris Pollard and Philipp Windischhofer},
booktitle={Transactions on Machine Learning Research},
issn={2835-8856},
year={2024},
url={https://openreview.net/forum?id=Jy2IgzjoFH},
note={},
archivePrefix = {arXiv},
eprint = {2306.12584}
}

@ARTICLE{jha2007,
       author = {{Jha}, Saurabh and {Riess}, Adam G. and {Kirshner}, Robert P.},
        title = "{Improved Distances to Type Ia Supernovae with Multicolor Light-Curve Shapes: MLCS2k2}",
      journal = {\apj},
         year = 2007,
        month = apr,
       volume = {659},
       number = {1},
        pages = {122-148},
          doi = {10.1086/512054},
archivePrefix = {arXiv},
       eprint = {astro-ph/0612666},
 primaryClass = {astro-ph},
       adsurl = {https://ui.adsabs.harvard.edu/abs/2007ApJ...659..122J}
}

@ARTICLE{hogg1999,
       author = {{Hogg}, David W.},
        title = "{Distance measures in cosmology}",
      journal = {ArXiv e-prints},
         year = 1999,
        month = may,
          eid = {astro-ph/9905116},
          doi = {10.48550/arXiv.astro-ph/9905116},
archivePrefix = {arXiv},
       eprint = {astro-ph/9905116},
 primaryClass = {astro-ph},
       adsurl = {https://ui.adsabs.harvard.edu/abs/1999astro.ph..5116H}
}

@ARTICLE{dilday2008,
       author = {{Dilday}, Benjamin and {Kessler}, Richard and {Frieman}, Joshua A. and {Holtzman}, Jon and {Marriner}, John and {Miknaitis}, Gajus and {Nichol}, Robert C. and {Romani}, Roger and {Sako}, Masao and {Bassett}, Bruce and {Becker}, Andrew and {Cinabro}, David and {DeJongh}, Fritz and {Depoy}, Darren L. and {Doi}, Mamoru and {Garnavich}, Peter M. and {Hogan}, Craig J. and {Jha}, Saurabh and {Konishi}, Kohki and {Lampeitl}, Hubert and {Marshall}, Jennifer L. and {McGinnis}, David and {Prieto}, Jose Luis and {Riess}, Adam G. and {Richmond}, Michael W. and {Schneider}, Donald P. and {Smith}, Mathew and {Takanashi}, Naohiro and {Tokita}, Kouichi and {van der Heyden}, Kurt and {Yasuda}, Naoki and {Zheng}, Chen and {Barentine}, John and {Brewington}, Howard and {Choi}, Changsu and {Crotts}, Arlin and {Dembicky}, Jack and {Harvanek}, Michael and {Im}, Myunshin and {Ketzeback}, William and {Kleinman}, Scott J. and {Krzesi{\'n}ski}, Jurek and {Long}, Daniel C. and {Malanushenko}, Elena and {Malanushenko}, Viktor and {McMillan}, Russet J. and {Nitta}, Atsuko and {Pan}, Kaike and {Saurage}, Gabrelle and {Snedden}, Stephanie A. and {Watters}, Shannon and {Wheeler}, J. Craig and {York}, Donald},
        title = "{A Measurement of the Rate of Type Ia Supernovae at Redshift z {\ensuremath{\approx}} 0.1 from the First Season of the SDSS-II Supernova Survey}",
      journal = {\apj},
         year = 2008,
        month = jul,
       volume = {682},
       number = {1},
        pages = {262-282},
          doi = {10.1086/587733},
archivePrefix = {arXiv},
       eprint = {0801.3297},
 primaryClass = {astro-ph},
       adsurl = {https://ui.adsabs.harvard.edu/abs/2008ApJ...682..262D}
}

@ARTICLE{damodel,
       author = {{Boyd}, Benjamin M. and {Narayan}, Gautham and {Mandel}, Kaisey S. and {Grayling}, Matthew and {Saha}, Abhijit and {Axelrod}, Tim and {Matheson}, Thomas and {Olszewski}, Edward W. and {Calamida}, Annalisa and {Do}, Aaron and {Bohlin}, Ralph C. and {Holberg}, Jay B. and {Hubeny}, Ivan and {Deustua}, Susana and {Rest}, Armin and {Stubbs}, Christopher W. and {Berres}, Aidan and {Li}, Mai and {Mackenty}, John W. and {Sabbi}, Elena},
        title = "{DAmodel: hierarchical Bayesian modelling of DA white dwarfs for spectrophotometric calibration}",
      journal = {\mnras},
         year = 2025,
        month = jun,
       volume = {540},
       number = {1},
        pages = {385-415},
          doi = {10.1093/mnras/staf629},
archivePrefix = {arXiv},
       eprint = {2412.08809},
 primaryClass = {astro-ph.IM},
       adsurl = {https://ui.adsabs.harvard.edu/abs/2025MNRAS.540..385B}
}

@ARTICLE{mandel2014,
       author = {{Mandel}, Kaisey S. and {Foley}, Ryan J. and {Kirshner}, Robert P.},
        title = "{Type Ia Supernova Colors and Ejecta Velocities: Hierarchical Bayesian Regression with Non-Gaussian Distributions}",
      journal = {\apj},
         year = 2014,
        month = dec,
       volume = {797},
       number = {2},
          eid = {75},
        pages = {75},
          doi = {10.1088/0004-637X/797/2/75},
archivePrefix = {arXiv},
       eprint = {1402.7079},
 primaryClass = {astro-ph.CO},
       adsurl = {https://ui.adsabs.harvard.edu/abs/2014ApJ...797...75M}
}

@ARTICLE{desmond2025,
       author = {{Desmond}, Harry and {Stiskalek}, Richard and {Najera}, Jose Antonio and {Banik}, Indranil},
        title = "{The subtle statistics of the distance ladder: On the distance prior and selection effects}",
      journal = {ArXiv e-prints},
         year = 2025,
        month = nov,
          eid = {arXiv:2511.03394},
          doi = {10.48550/arXiv.2511.03394},
archivePrefix = {arXiv},
       eprint = {2511.03394},
 primaryClass = {astro-ph.CO},
       adsurl = {https://ui.adsabs.harvard.edu/abs/2025arXiv251103394D}
}

@ARTICLE{4most,
       author = {{\VAN{Jong}{De}{de} Jong}, R.~S. and {Agertz}, O. and {Berbel}, A.~A. and {Aird}, J. and {Alexander}, D.~A. and {Amarsi}, A. and {Anders}, F. and {Andrae}, R. and {Ansarinejad}, B. and {Ansorge}, W. and {Antilogus}, P. and {Anwand-Heerwart}, H. and {Arentsen}, A. and {Arnadottir}, A. and {Asplund}, M. and {Auger}, M. and {Azais}, N. and {Baade}, D. and {Baker}, G. and {Baker}, S. and {Balbinot}, E. and {Baldry}, I.~K. and {Banerji}, M. and {Barden}, S. and {Barklem}, P. and {Barth{\'e}l{\'e}my-Mazot}, E. and {Battistini}, C. and {Bauer}, S. and {Bell}, C.~P.~M. and {Bellido-Tirado}, O. and {Bellstedt}, S. and {Belokurov}, V. and {Bensby}, T. and {Bergemann}, M. and {Bestenlehner}, J.~M. and {Bielby}, R. and {Bilicki}, M. and {Blake}, C. and {Bland-Hawthorn}, J. and {Boeche}, C. and {Boland}, W. and {Boller}, T. and {Bongard}, S. and {Bongiorno}, A. and {Bonifacio}, P. and {Boudon}, D. and {Brooks}, D. and {Brown}, M.~J.~I. and {Brown}, R. and {Br{\"u}ggen}, M. and {Brynnel}, J. and {Brzeski}, J. and {Buchert}, T. and {Buschkamp}, P. and {Caffau}, E. and {Caillier}, P. and {Carrick}, J. and {Casagrande}, L. and {Case}, S. and {Casey}, A. and {Cesarini}, I. and {Cescutti}, G. and {Chapuis}, D. and {Chiappini}, C. and {Childress}, M. and {Christlieb}, N. and {Church}, R. and {Cioni}, M.-R.~L. and {Cluver}, M. and {Colless}, M. and {Collett}, T. and {Comparat}, J. and {Cooper}, A. and {Couch}, W. and {Courbin}, F. and {Croom}, S. and {Croton}, D. and {Daguis{\'e}}, E. and {Dalton}, G. and {Davies}, L.~J.~M. and {Davis}, T. and {de Laverny}, P. and {Deason}, A. and {Dionies}, F. and {Disseau}, K. and {Doel}, P. and {D{\"o}scher}, D. and {Driver}, S.~P. and {Dwelly}, T. and {Eckert}, D. and {Edge}, A. and {Edvardsson}, B. and {Youssoufi}, D.~E. and {Elhaddad}, A. and {Enke}, H. and {Erfanianfar}, G. and {Farrell}, T. and {Fechner}, T. and {Feiz}, C. and {Feltzing}, S. and {Ferreras}, I. and {Feuerstein}, D. and {Feuillet}, D. and {Finoguenov}, A. and {Ford}, D. and {Fotopoulou}, S. and {Fouesneau}, M. and {Frenk}, C. and {Frey}, S. and {Gaessler}, W. and {Geier}, S. and {Gentile Fusillo}, N. and {Gerhard}, O. and {Giannantonio}, T. and {Giannone}, D. and {Gibson}, B. and {Gillingham}, P. and {Gonz{\'a}lez-Fern{\'a}ndez}, C. and {Gonzalez-Solares}, E. and {Gottloeber}, S. and {Gould}, A. and {Grebel}, E.~K. and {Gueguen}, A. and {Guiglion}, G. and {Haehnelt}, M. and {Hahn}, T. and {Hansen}, C.~J. and {Hartman}, H. and {Hauptner}, K. and {Hawkins}, K. and {Haynes}, D. and {Haynes}, R. and {Heiter}, U. and {Helmi}, A. and {Aguayo}, C.~H. and {Hewett}, P. and {Hinton}, S. and {Hobbs}, D. and {Hoenig}, S. and {Hofman}, D. and {Hook}, I. and {Hopgood}, J. and {Hopkins}, A. and {Hourihane}, A. and {Howes}, L. and {Howlett}, C. and {Huet}, T. and {Irwin}, M. and {Iwert}, O. and {Jablonka}, P. and {Jahn}, T. and {Jahnke}, K. and {Jarno}, A. and {Jin}, S. and {Jofre}, P. and {Johl}, D. and {Jones}, D. and {J{\"o}nsson}, H. and {Jordan}, C. and {Karovicova}, I. and {Khalatyan}, A. and {Kelz}, A. and {Kennicutt}, R. and {King}, D. and {Kitaura}, F. and {Klar}, J. and {Klauser}, U. and {Kneib}, J.-P. and {Koch}, A. and {Koposov}, S. and {Kordopatis}, G. and {Korn}, A. and {Kosmalski}, J. and {Kotak}, R. and {Kovalev}, M. and {Kreckel}, K. and {Kripak}, Y. and {Krumpe}, M. and {Kuijken}, K. and {Kunder}, A. and {Kushniruk}, I. and {Lam}, M.~I. and {Lamer}, G. and {Laurent}, F. and {Lawrence}, J. and {Lehmitz}, M. and {Lemasle}, B. and {Lewis}, J. and {Li}, B. and {Lidman}, C. and {Lind}, K. and {Liske}, J. and {Lizon}, J.-L. and {Loveday}, J. and {Ludwig}, H.-G. and {McDermid}, R.~M. and {Maguire}, K. and {Mainieri}, V. and {Mali}, S. and {Mandel}, H.},
        title = "{4MOST: Project overview and information for the First Call for Proposals}",
      journal = {The Messenger},
         year = 2019,
        month = mar,
       volume = {175},
        pages = {3-11},
          doi = {10.18727/0722-6691/5117},
archivePrefix = {arXiv},
       eprint = {1903.02464},
 primaryClass = {astro-ph.IM},
       adsurl = {https://ui.adsabs.harvard.edu/abs/2019Msngr.175....3D}
}

@ARTICLE{tides,
       author = {{Frohmaier}, C. and {Vincenzi}, M. and {Sullivan}, M. and {H{\"o}nig}, S.~F. and {Smith}, M. and {Addison}, H. and {Collett}, T. and {Dimitriadis}, G. and {Ellis}, R.~S. and {Gandhi}, P. and {Graur}, O. and {Hook}, I. and {Kelsey}, L. and {Kim}, Y.-L. and {Lidman}, C. and {Maguire}, K. and {Makrygianni}, L. and {Martin}, B. and {M{\"o}ller}, A. and {Nichol}, R.~C. and {Nicholl}, M. and {Schady}, P. and {Simmons}, B.~D. and {Smartt}, S.~J. and {Tempel}, E. and {Wiseman}, P. and {the LSST Dark Energy Science Collaboration}},
        title = "{TiDES: The 4MOST Time Domain Extragalactic Survey}",
      journal = {\apj},
         year = 2025,
        month = oct,
       volume = {992},
       number = {1},
          eid = {158},
        pages = {158},
          doi = {10.3847/1538-4357/adff4e},
archivePrefix = {arXiv},
       eprint = {2501.16311},
 primaryClass = {astro-ph.HE},
       adsurl = {https://ui.adsabs.harvard.edu/abs/2025ApJ...992..158F}
}

@ARTICLE{mitra23,
       author = {{Mitra}, Ayan and {Kessler}, Richard and {More}, Surhud and {Hlozek}, Renee and {LSST Dark Energy Science Collaboration}},
        title = "{Using Host Galaxy Photometric Redshifts to Improve Cosmological Constraints with Type Ia Supernovae in the LSST Era}",
      journal = {\apj},
         year = 2023,
        month = feb,
       volume = {944},
       number = {2},
          eid = {212},
        pages = {212},
          doi = {10.3847/1538-4357/acb057},
archivePrefix = {arXiv},
       eprint = {2210.07560},
 primaryClass = {astro-ph.CO},
       adsurl = {https://ui.adsabs.harvard.edu/abs/2023ApJ...944..212M}
}

@ARTICLE{plasticc,
       author = {{Kessler}, R. and {Narayan}, G. and {Avelino}, A. and {Bachelet}, E. and {Biswas}, R. and {Brown}, P.~J. and {Chernoff}, D.~F. and {Connolly}, A.~J. and {Dai}, M. and {Daniel}, S. and {Di Stefano}, R. and {Drout}, M.~R. and {Galbany}, L. and {Gonz{\'a}lez-Gait{\'a}n}, S. and {Graham}, M.~L. and {Hlo{\v{z}}ek}, R. and {Ishida}, E.~E.~O. and {Guillochon}, J. and {Jha}, S.~W. and {Jones}, D.~O. and {Mandel}, K.~S. and {Muthukrishna}, D. and {O'Grady}, A. and {Peters}, C.~M. and {Pierel}, J.~R. and {Ponder}, K.~A. and {Pr{\v{s}}a}, A. and {Rodney}, S. and {Villar}, V.~A. and {LSST Dark Energy Science Collaboration} and {Transient and Variable Stars Science Collaboration}},
        title = "{Models and Simulations for the Photometric LSST Astronomical Time Series Classification Challenge (PLAsTiCC)}",
      journal = {\pasp},
         year = 2019,
        month = sep,
       volume = {131},
       number = {1003},
        pages = {094501},
          doi = {10.1088/1538-3873/ab26f1},
archivePrefix = {arXiv},
       eprint = {1903.11756},
 primaryClass = {astro-ph.HE},
       adsurl = {https://ui.adsabs.harvard.edu/abs/2019PASP..131i4501K}
}

@ARTICLE{scolnic2016,
       author = {{Scolnic}, D. and {Kessler}, R.},
        title = "{Measuring Type Ia Supernova Populations of Stretch and Color and Predicting Distance Biases}",
      journal = {\apjl},
         year = 2016,
        month = may,
       volume = {822},
       number = {2},
          eid = {L35},
        pages = {L35},
          doi = {10.3847/2041-8205/822/2/L35},
archivePrefix = {arXiv},
       eprint = {1603.01559},
 primaryClass = {astro-ph.CO},
       adsurl = {https://ui.adsabs.harvard.edu/abs/2016ApJ...822L..35S}
}

@ARTICLE{fitzpatrick1999,
       author = {{Fitzpatrick}, Edward L.},
        title = "{Correcting for the Effects of Interstellar Extinction}",
      journal = {\pasp},
         year = 1999,
        month = jan,
       volume = {111},
       number = {755},
        pages = {63-75},
          doi = {10.1086/316293},
archivePrefix = {arXiv},
       eprint = {astro-ph/9809387},
 primaryClass = {astro-ph},
       adsurl = {https://ui.adsabs.harvard.edu/abs/1999PASP..111...63F}
}

@ARTICLE{schlafly2011,
       author = {{Schlafly}, Edward F. and {Finkbeiner}, Douglas P.},
        title = "{Measuring Reddening with Sloan Digital Sky Survey Stellar Spectra and Recalibrating SFD}",
      journal = {\apj},
         year = 2011,
        month = aug,
       volume = {737},
       number = {2},
          eid = {103},
        pages = {103},
          doi = {10.1088/0004-637X/737/2/103},
archivePrefix = {arXiv},
       eprint = {1012.4804},
 primaryClass = {astro-ph.GA},
       adsurl = {https://ui.adsabs.harvard.edu/abs/2011ApJ...737..103S}
}

@ARTICLE{komatsu2009,
       author = {{Komatsu}, E. and {Dunkley}, J. and {Nolta}, M.~R. and {Bennett}, C.~L. and {Gold}, B. and {Hinshaw}, G. and {Jarosik}, N. and {Larson}, D. and {Limon}, M. and {Page}, L. and {Spergel}, D.~N. and {Halpern}, M. and {Hill}, R.~S. and {Kogut}, A. and {Meyer}, S.~S. and {Tucker}, G.~S. and {Weiland}, J.~L. and {Wollack}, E. and {Wright}, E.~L.},
        title = "{Five-Year Wilkinson Microwave Anisotropy Probe Observations: Cosmological Interpretation}",
      journal = {\apjs},
         year = 2009,
        month = feb,
       volume = {180},
       number = {2},
        pages = {330-376},
          doi = {10.1088/0067-0049/180/2/330},
archivePrefix = {arXiv},
       eprint = {0803.0547},
 primaryClass = {astro-ph},
       adsurl = {https://ui.adsabs.harvard.edu/abs/2009ApJS..180..330K}
}

@article{hermans2021,
   author = {{Hermans}, Joeri and {Delaunoy}, Arnaud and {Rozet}, Fran{\c{c}}ois and {Wehenkel}, Antoine and {Begy}, Volodimir and {Louppe}, Gilles},
        title = "{A Trust Crisis In Simulation-Based Inference? Your Posterior Approximations Can Be Unfaithful}",
      journal = {ArXiv e-prints},
         year = 2021,
        month = oct,
          eid = {arXiv:2110.06581},
          doi = {10.48550/arXiv.2110.06581},
archivePrefix = {arXiv},
       eprint = {2110.06581},
 primaryClass = {stat.ML},
       adsurl = {https://ui.adsabs.harvard.edu/abs/2021arXiv211006581H}
}

@ARTICLE{des5yr,
       author = {{DES Collaboration} and {Abbott}, T.~M.~C. and {Acevedo}, M. and {Aguena}, M. and {Alarcon}, A. and {Allam}, S. and {Alves}, O. and {Amon}, A. and {Andrade-Oliveira}, F. and {Annis}, J. and {Armstrong}, P. and {Asorey}, J. and {Avila}, S. and {Bacon}, D. and {Bassett}, B.~A. and {Bechtol}, K. and {Bernardinelli}, P.~H. and {Bernstein}, G.~M. and {Bertin}, E. and {Blazek}, J. and {Bocquet}, S. and {Brooks}, D. and {Brout}, D. and {Buckley-Geer}, E. and {Burke}, D.~L. and {Camacho}, H. and {Camilleri}, R. and {Campos}, A. and {Carnero Rosell}, A. and {Carollo}, D. and {Carr}, A. and {Carretero}, J. and {Castander}, F.~J. and {Cawthon}, R. and {Chang}, C. and {Chen}, R. and {Choi}, A. and {Conselice}, C. and {Costanzi}, M. and {da Costa}, L.~N. and {Crocce}, M. and {Davis}, T.~M. and {DePoy}, D.~L. and {Desai}, S. and {Diehl}, H.~T. and {Dixon}, M. and {Dodelson}, S. and {Doel}, P. and {Doux}, C. and {Drlica-Wagner}, A. and {Elvin-Poole}, J. and {Everett}, S. and {Ferrero}, I. and {Fert{\'e}}, A. and {Flaugher}, B. and {Foley}, R.~J. and {Fosalba}, P. and {Friedel}, D. and {Frieman}, J. and {Frohmaier}, C. and {Galbany}, L. and {Garc{\'\i}a-Bellido}, J. and {Gatti}, M. and {Gaztanaga}, E. and {Giannini}, G. and {Glazebrook}, K. and {Graur}, O. and {Gruen}, D. and {Gruendl}, R.~A. and {Gutierrez}, G. and {Hartley}, W.~G. and {Herner}, K. and {Hinton}, S.~R. and {Hollowood}, D.~L. and {Honscheid}, K. and {Huterer}, D. and {Jain}, B. and {James}, D.~J. and {Jeffrey}, N. and {Kasai}, E. and {Kelsey}, L. and {Kent}, S. and {Kessler}, R. and {Kim}, A.~G. and {Kirshner}, R.~P. and {Kovacs}, E. and {Kuehn}, K. and {Lahav}, O. and {Lee}, J. and {Lee}, S. and {Lewis}, G.~F. and {Li}, T.~S. and {Lidman}, C. and {Lin}, H. and {Malik}, U. and {Marshall}, J.~L. and {Martini}, P. and {Mena-Fern{\'a}ndez}, J. and {Menanteau}, F. and {Miquel}, R. and {Mohr}, J.~J. and {Mould}, J. and {Muir}, J. and {M{\"o}ller}, A. and {Neilsen}, E. and {Nichol}, R.~C. and {Nugent}, P. and {Ogando}, R.~L.~C. and {Palmese}, A. and {Pan}, Y.-C. and {Paterno}, M. and {Percival}, W.~J. and {Pereira}, M.~E.~S. and {Pieres}, A. and {Malag{\'o}n}, A.~A. Plazas and {Popovic}, B. and {Porredon}, A. and {Prat}, J. and {Qu}, H. and {Raveri}, M. and {Rodr{\'\i}guez-Monroy}, M. and {Romer}, A.~K. and {Roodman}, A. and {Rose}, B. and {Sako}, M. and {Sanchez}, E. and {Sanchez Cid}, D. and {Schubnell}, M. and {Scolnic}, D. and {Sevilla-Noarbe}, I. and {Shah}, P. and {Smith}, J. Allyn. and {Smith}, M. and {Soares-Santos}, M. and {Suchyta}, E. and {Sullivan}, M. and {Suntzeff}, N. and {Swanson}, M.~E.~C. and {S{\'a}nchez}, B.~O. and {Tarle}, G. and {Taylor}, G. and {Thomas}, D. and {To}, C. and {Toy}, M. and {Troxel}, M.~A. and {Tucker}, B.~E. and {Tucker}, D.~L. and {Uddin}, S.~A. and {Vincenzi}, M. and {Walker}, A.~R. and {Weaverdyck}, N. and {Wechsler}, R.~H. and {Weller}, J. and {Wester}, W. and {Wiseman}, P. and {Yamamoto}, M. and {Yuan}, F. and {Zhang}, B. and {Zhang}, Y.},
        title = "{The Dark Energy Survey: Cosmology Results with {\ensuremath{\sim}}1500 New High-redshift Type Ia Supernovae Using the Full 5 yr Data Set}",
      journal = {\apjl},
         year = 2024,
        month = sep,
       volume = {973},
       number = {1},
          eid = {L14},
        pages = {L14},
          doi = {10.3847/2041-8213/ad6f9f},
archivePrefix = {arXiv},
       eprint = {2401.02929},
 primaryClass = {astro-ph.CO},
       adsurl = {https://ui.adsabs.harvard.edu/abs/2024ApJ...973L..14D}
}

@ARTICLE{efstathiou2024,
       author = {{Efstathiou}, George},
        title = "{Evolving dark energy or supernovae systematics?}",
      journal = {\mnras},
         year = 2025,
        month = apr,
       volume = {538},
       number = {2},
        pages = {875-882},
          doi = {10.1093/mnras/staf301},
archivePrefix = {arXiv},
       eprint = {2408.07175},
 primaryClass = {astro-ph.CO},
       adsurl = {https://ui.adsabs.harvard.edu/abs/2025MNRAS.538..875E}
}

@ARTICLE{dhawan2025,
       author = {{Dhawan}, Suhail and {Popovic}, Brodie and {Goobar}, Ariel},
        title = "{The axis of systematic bias in SN Ia cosmology and implications for DESI 2024 results}",
      journal = {\mnras},
         year = 2025,
        month = jun,
       volume = {540},
       number = {2},
        pages = {1626-1634},
          doi = {10.1093/mnras/staf779},
archivePrefix = {arXiv},
       eprint = {2409.18668},
 primaryClass = {astro-ph.CO},
       adsurl = {https://ui.adsabs.harvard.edu/abs/2025MNRAS.540.1626D}
}

@ARTICLE{wiseman2026,
       author = {{Wiseman}, Phil and {Popovic}, Brodie and {Sullivan}, Mark and {Riess}, Adam G. and {Scolnic}, Dan and {Chen}, Rebecca C. and {Davis}, Tamara M. and {Galbany}, Llu{\'\i}s and {Hook}, Isobel M. and {Jha}, Saurabh W. and {Kelsey}, Lisa and {Murakami}, Yukei S. and {Rigault}, Micka{\"e}l and {Rose}, Benjamin M. and {Schmidt}, Brian and {Smith}, Mat and {Vincenzi}, Maria},
        title = "{Still Accelerating: Type Ia supernova cosmology is robust to host galaxy age evolution}",
      journal = {ArXiv e-prints},
         year = 2026,
        month = jan,
          eid = {arXiv:2601.13785},
          doi = {10.48550/arXiv.2601.13785},
archivePrefix = {arXiv},
       eprint = {2601.13785},
 primaryClass = {astro-ph.CO},
       adsurl = {https://ui.adsabs.harvard.edu/abs/2026arXiv260113785W}
}

@ARTICLE{sanchez2022,
       author = {{S{\'a}nchez}, B.~O. and {Kessler}, R. and {Scolnic}, D. and {Armstrong}, R. and {Biswas}, R. and {Bogart}, J. and {Chiang}, J. and {Cohen-Tanugi}, J. and {Fouchez}, D. and {Gris}, Ph. and {Heitmann}, K. and {Hlo{\v{z}}ek}, R. and {Jha}, S. and {Kelly}, H. and {Liu}, S. and {Narayan}, G. and {Racine}, B. and {Rykoff}, E. and {Sullivan}, M. and {Walter}, C.~W. and {Wood-Vasey}, W.~M. and {LSST Dark Energy Science Collaboration (DESC)}},
        title = "{SNIa Cosmology Analysis Results from Simulated LSST Images: From Difference Imaging to Constraints on Dark Energy}",
      journal = {\apj},
         year = 2022,
        month = aug,
       volume = {934},
       number = {2},
          eid = {96},
        pages = {96},
          doi = {10.3847/1538-4357/ac7a37},
archivePrefix = {arXiv},
       eprint = {2111.06858},
 primaryClass = {astro-ph.CO},
       adsurl = {https://ui.adsabs.harvard.edu/abs/2022ApJ...934...96S}
}

@ARTICLE{kenworthy2025,
       author = {{Kenworthy}, W.~D. and {Goobar}, A. and {Jones}, D.~O. and {Johansson}, J. and {Thorp}, S. and {Kessler}, R. and {Burgaz}, U. and {Dhawan}, S. and {Dimitriadis}, G. and {Galbany}, L. and {Ginolin}, M. and {Kim}, Y.-L. and {Maguire}, K. and {M{\"u}ller-Bravo}, T.~E. and {Nugent}, P. and {Nordin}, J. and {Popovic}, B. and {Pessi}, P.~J. and {Rigault}, M. and {Rosnet}, P. and {Sollerman}, J. and {Terwel}, J.~H. and {Townsend}, A. and {Laher}, R.~R. and {Purdum}, J. and {Rosselli}, D. and {Rusholme}, B.},
        title = "{ZTF SN Ia DR2: Improved SN Ia colors through expanded dimensionality with SALT3+}",
      journal = {\aap},
         year = 2025,
        month = may,
       volume = {697},
          eid = {A125},
        pages = {A125},
          doi = {10.1051/0004-6361/202452578},
archivePrefix = {arXiv},
       eprint = {2502.09713},
 primaryClass = {astro-ph.CO},
       adsurl = {https://ui.adsabs.harvard.edu/abs/2025A&A...697A.125K}
}

@ARTICLE{taylor2021,
       author = {{Taylor}, G. and {Lidman}, C. and {Tucker}, B.~E. and {Brout}, D. and {Hinton}, S.~R. and {Kessler}, R.},
        title = "{A revised SALT2 surface for fitting Type Ia supernova light curves}",
      journal = {\mnras},
         year = 2021,
        month = jul,
       volume = {504},
       number = {3},
        pages = {4111-4122},
          doi = {10.1093/mnras/stab962},
archivePrefix = {arXiv},
       eprint = {2104.00172},
 primaryClass = {astro-ph.CO},
       adsurl = {https://ui.adsabs.harvard.edu/abs/2021MNRAS.504.4111T}
}

@ARTICLE{popovic2025,
       author = {{Popovic}, B. and {Rigault}, M. and {Smith}, M. and {Ginolin}, M. and {Goobar}, A. and {Kenworthy}, W.~D. and {Ganot}, C. and {Ruppin}, F. and {Dimitriadis}, G. and {Johansson}, J. and {Amenouche}, M. and {Aubert}, M. and {Barjou-Delayre}, C. and {Burgaz}, U. and {Carreres}, B. and {Feinstein}, F. and {Fouchez}, D. and {Galbany}, L. and {de Jaeger}, T. and {Kim}, Y.-L. and {Lacroix}, L. and {Nugent}, P.~E. and {Racine}, B. and {Rosselli}, D. and {Rosnet}, P. and {Sollerman}, J. and {Hale}, D. and {Laher}, R. and {M{\"u}ller-Bravo}, T.~E. and {Reed}, R. and {Rusholme}, B. and {Terwel}, J.},
        title = "{ZTF SN Ia DR2: Evidence of changing dust distribution with redshift using type Ia supernovae}",
      journal = {\aap},
         year = 2025,
        month = feb,
       volume = {694},
          eid = {A5},
        pages = {A5},
          doi = {10.1051/0004-6361/202450391},
archivePrefix = {arXiv},
       eprint = {2406.06215},
 primaryClass = {astro-ph.CO},
       adsurl = {https://ui.adsabs.harvard.edu/abs/2025A&A...694A...5P}
}

@ARTICLE{schmidt1968,
       author = {{Schmidt}, Maarten},
        title = "{Space Distribution and Luminosity Functions of Quasi-Stellar Radio Sources}",
      journal = {\apj},
         year = 1968,
        month = feb,
       volume = {151},
        pages = {393},
          doi = {10.1086/149446},
       adsurl = {https://ui.adsabs.harvard.edu/abs/1968ApJ...151..393S}
}

@ARTICLE{abbott2021,
       author = {{Abbott}, R. and {Abbott}, T.~D. and {Abraham}, S. and {Acernese}, F. and {Ackley}, K. and {Adams}, A. and {Adams}, C. and {Adhikari}, R.~X. and {Adya}, V.~B. and {Affeldt}, C. and {Agathos}, M. and {Agatsuma}, K. and {Aggarwal}, N. and {Aguiar}, O.~D. and {Aiello}, L. and {Ain}, A. and {Ajith}, P. and {Allen}, G. and {Allocca}, A. and {Altin}, P.~A. and {Amato}, A. and {Anand}, S. and {Ananyeva}, A. and {Anderson}, S.~B. and {Anderson}, W.~G. and {Angelova}, S.~V. and {Ansoldi}, S. and {Antelis}, J.~M. and {Antier}, S. and {Appert}, S. and {Arai}, K. and {Araya}, M.~C. and {Areeda}, J.~S. and {Ar{\`e}ne}, M. and {Arnaud}, N. and {Aronson}, S.~M. and {Arun}, K.~G. and {Asali}, Y. and {Ascenzi}, S. and {Ashton}, G. and {Aston}, S.~M. and {Astone}, P. and {Aubin}, F. and {Aufmuth}, P. and {AultONeal}, K. and {Austin}, C. and {Avendano}, V. and {Babak}, S. and {Badaracco}, F. and {Bader}, M.~K.~M. and {Bae}, S. and {Baer}, A.~M. and {Bagnasco}, S. and {Baird}, J. and {Ball}, M. and {Ballardin}, G. and {Ballmer}, S.~W. and {Bals}, A. and {Balsamo}, A. and {Baltus}, G. and {Banagiri}, S. and {Bankar}, D. and {Bankar}, R.~S. and {Barayoga}, J.~C. and {Barbieri}, C. and {Barish}, B.~C. and {Barker}, D. and {Barneo}, P. and {Barnum}, S. and {Barone}, F. and {Barr}, B. and {Barsotti}, L. and {Barsuglia}, M. and {Barta}, D. and {Bartlett}, J. and {Bartos}, I. and {Bassiri}, R. and {Basti}, A. and {Bawaj}, M. and {Bayley}, J.~C. and {Bazzan}, M. and {Becher}, B.~R. and {B{\'e}csy}, B. and {Bedakihale}, V.~M. and {Bejger}, M. and {Belahcene}, I. and {Beniwal}, D. and {Benjamin}, M.~G. and {Bennett}, T.~F. and {Bentley}, J.~D. and {Bergamin}, F. and {Berger}, B.~K. and {Bergmann}, G. and {Bernuzzi}, S. and {Berry}, C.~P.~L. and {Bersanetti}, D. and {Bertolini}, A. and {Betzwieser}, J. and {Bhandare}, R. and {Bhandari}, A.~V. and {Bhattacharjee}, D. and {Bidler}, J. and {Bilenko}, I.~A. and {Billingsley}, G. and {Birney}, R. and {Birnholtz}, O. and {Biscans}, S. and {Bischi}, M. and {Biscoveanu}, S. and {Bisht}, A. and {Bitossi}, M. and {Bizouard}, M.-A. and {Blackburn}, J.~K. and {Blackman}, J. and {Blair}, C.~D. and {Blair}, D.~G. and {Blair}, R.~M. and {Blanch}, O. and {Bobba}, F. and {Bode}, N. and {Boer}, M. and {Boetzel}, Y. and {Bogaert}, G. and {Boldrini}, M. and {Bondu}, F. and {Bonilla}, E. and {Bonnand}, R. and {Booker}, P. and {Boom}, B.~A. and {Bork}, R. and {Boschi}, V. and {Bose}, S. and {Bossilkov}, V. and {Boudart}, V. and {Bouffanais}, Y. and {Bozzi}, A. and {Bradaschia}, C. and {Brady}, P.~R. and {Bramley}, A. and {Branchesi}, M. and {Brau}, J.~E. and {Breschi}, M. and {Briant}, T. and {Briggs}, J.~H. and {Brighenti}, F. and {Brillet}, A. and {Brinkmann}, M. and {Brockill}, P. and {Brooks}, A.~F. and {Brooks}, J. and {Brown}, D.~D. and {Brunett}, S. and {Bruno}, G. and {Bruntz}, R. and {Buikema}, A. and {Bulik}, T. and {Bulten}, H.~J. and {Buonanno}, A. and {Buscicchio}, R. and {Buskulic}, D. and {Byer}, R.~L. and {Cabero}, M. and {Cadonati}, L. and {Caesar}, M. and {Cagnoli}, G. and {Cahillane}, C. and {Calder{\'o}n Bustillo}, J. and {Callaghan}, J.~D. and {Callister}, T.~A. and {Calloni}, E. and {Camp}, J.~B. and {Canepa}, M. and {Cannon}, K.~C. and {Cao}, H. and {Cao}, J. and {Carapella}, G. and {Carbognani}, F. and {Carney}, M.~F. and {Carpinelli}, M. and {Carullo}, G. and {Carver}, T.~L. and {Casanueva Diaz}, J. and {Casentini}, C. and {Caudill}, S. and {Cavagli{\`a}}, M. and {Cavalier}, F. and {Cavalieri}, R. and {Cella}, G. and {Cerd{\'a}-Dur{\'a}n}, P. and {Cesarini}, E. and {Chaibi}, W. and {Chakravarti}, K. and {Chan}, C.-L. and {Chan}, C. and {Chandra}, K. and {Chanial}, P. and {Chao}, S. and {Charlton}, P. and {Chase}, E.~A. and {Chassande-Mottin}, E.},
        title = "{Population Properties of Compact Objects from the Second LIGO-Virgo Gravitational-Wave Transient Catalog}",
      journal = {\apjl},
         year = 2021,
        month = may,
       volume = {913},
       number = {1},
          eid = {L7},
        pages = {L7},
          doi = {10.3847/2041-8213/abe949},
archivePrefix = {arXiv},
       eprint = {2010.14533},
 primaryClass = {astro-ph.HE},
       adsurl = {https://ui.adsabs.harvard.edu/abs/2021ApJ...913L...7A}
}

@ARTICLE{kauffmann2003,
       author = {{Kauffmann}, Guinevere and {Heckman}, Timothy M. and {White}, Simon D.~M. and {Charlot}, St{\'e}phane and {Tremonti}, Christy and {Brinchmann}, Jarle and {Bruzual}, Gustavo and {Peng}, Eric W. and {Seibert}, Mark and {Bernardi}, Mariangela and {Blanton}, Michael and {Brinkmann}, Jon and {Castander}, Francisco and {Cs{\'a}bai}, Istvan and {Fukugita}, Masataka and {Ivezic}, Zeljko and {Munn}, Jeffrey A. and {Nichol}, Robert C. and {Padmanabhan}, Nikhil and {Thakar}, Aniruddha R. and {Weinberg}, David H. and {York}, Donald},
        title = "{Stellar masses and star formation histories for {}10$^{5}$ galaxies from the Sloan Digital Sky Survey}",
      journal = {\mnras},
         year = 2003,
        month = may,
       volume = {341},
       number = {1},
        pages = {33-53},
          doi = {10.1046/j.1365-8711.2003.06291.x},
archivePrefix = {arXiv},
       eprint = {astro-ph/0204055},
 primaryClass = {astro-ph},
       adsurl = {https://ui.adsabs.harvard.edu/abs/2003MNRAS.341...33K}
}

@ARTICLE{petigura2013,
       author = {{Petigura}, Erik A. and {Howard}, Andrew W. and {Marcy}, Geoffrey W.},
        title = "{Prevalence of Earth-size planets orbiting Sun-like stars}",
      journal = {Proc.\ Natl.\ Acad.\ Sci.},
         year = 2013,
        month = nov,
       volume = {110},
       number = {48},
        pages = {19273-19278},
          doi = {10.1073/pnas.1319909110},
archivePrefix = {arXiv},
       eprint = {1311.6806},
 primaryClass = {astro-ph.EP},
       adsurl = {https://ui.adsabs.harvard.edu/abs/2013PNAS..11019273P}
}

@ARTICLE{scalzo2014,
       author = {{Scalzo}, R.~A. and {Ruiter}, A.~J. and {Sim}, S.~A.},
        title = "{The ejected mass distribution of Type Ia supernovae: a significant rate of non-Chandrasekhar-mass progenitors}",
      journal = {\mnras},
         year = 2014,
        month = dec,
       volume = {445},
       number = {3},
        pages = {2535-2544},
          doi = {10.1093/mnras/stu1808},
archivePrefix = {arXiv},
       eprint = {1408.6601},
 primaryClass = {astro-ph.HE},
       adsurl = {https://ui.adsabs.harvard.edu/abs/2014MNRAS.445.2535S}
}

@ARTICLE{sarin2026,
       author = {{Sarin}, Nikhil and {Lindsj{\"o}}, Ellen and {Kelsey}, Lisa and {Grayling}, Matthew and {Sollerman}, Jesper and {Schulze}, Steve and {Miller}, Adam and {Ginolin}, Madeleine and {Hayes}, Erin and {Omand}, Conor and {Mandel}, Kaisey and {Do}, Aaron and {Dhawan}, Suhail and {Johansson}, Joel},
        title = "{Lightcurve Modelling of 2,205 ZTF DR2 Type\raisebox{-0.5ex}\textasciitildeIa Supernovae: Implications for SN Ia Physics and Cosmology}",
      journal = {ArXiv e-prints},
         year = 2026,
        month = feb,
          eid = {arXiv:2602.02677},
          doi = {10.48550/arXiv.2602.02677},
archivePrefix = {arXiv},
       eprint = {2602.02677},
 primaryClass = {astro-ph.HE},
       adsurl = {https://ui.adsabs.harvard.edu/abs/2026arXiv260202677S}
}

@ARTICLE{rubin2026,
       author = {{Rubin}, David and {Hoyt}, Taylor and {Aldering}, Greg and {Perlmutter}, Saul},
        title = "{Banana Split: Improved Cosmological Constraints with Two Light-Curve-Shape and Color Populations Using Union3.1+UNITY1.8}",
      journal = {ArXiv e-prints},
         year = 2026,
        month = jan,
          eid = {arXiv:2601.19854},
          doi = {10.48550/arXiv.2601.19854},
archivePrefix = {arXiv},
       eprint = {2601.19854},
 primaryClass = {astro-ph.CO},
       adsurl = {https://ui.adsabs.harvard.edu/abs/2026arXiv260119854R}
}

@ARTICLE{hoyt2026,
       author = {{Hoyt}, Taylor J. and {Rubin}, David and {Aldering}, Greg and {Perlmutter}, Saul and {Cuceu}, Andrei and {Gupta}, Ravi},
        title = "{Union3.1: Self-consistent Measurements of Host Galaxy Properties for 2000 Type Ia Supernovae}",
      journal = {ArXiv e-prints},
         year = 2026,
        month = jan,
          eid = {arXiv:2601.19424},
          doi = {10.48550/arXiv.2601.19424},
archivePrefix = {arXiv},
       eprint = {2601.19424},
 primaryClass = {astro-ph.CO},
       adsurl = {https://ui.adsabs.harvard.edu/abs/2026arXiv260119424H}
}

@ARTICLE{cranmer2020,
       author = {{Cranmer}, Kyle and {Brehmer}, Johann and {Louppe}, Gilles},
        title = "{The frontier of simulation-based inference}",
      journal = {Proc.\ Natl.\ Acad.\ Sci.},
         year = 2020,
        month = dec,
       volume = {117},
       number = {48},
        pages = {30055-30062},
          doi = {10.1073/pnas.1912789117},
archivePrefix = {arXiv},
       eprint = {1911.01429},
 primaryClass = {stat.ML},
       adsurl = {https://ui.adsabs.harvard.edu/abs/2020PNAS..11730055C}
}

@ARTICLE{ocallaghan2026,
       author = {{O'Callaghan}, Matthew and {Mandel}, Kaisey S. and {Gilmore}, Gerry},
        title = "{Data-driven dust inference at mid-to-high Galactic latitudes using probabilistic machine learning}",
      journal = {\mnras},
         year = 2026,
        month = feb,
       volume = {546},
       number = {2},
          eid = {staf2147},
        pages = {staf2147},
          doi = {10.1093/mnras/staf2147},
archivePrefix = {arXiv},
       eprint = {2508.05781},
 primaryClass = {astro-ph.GA},
       adsurl = {https://ui.adsabs.harvard.edu/abs/2026MNRAS.546f2147O}
}

@ARTICLE{ocallaghan2025,
       author = {{O'Callaghan}, Matthew and {Mandel}, Kaisey S. and {Gilmore}, Gerry},
        title = "{Misspecification-robust amortised simulation-based inference using variational methods}",
      journal = {ArXiv e-prints},
         year = 2025,
        month = sep,
          eid = {arXiv:2509.05724},
          doi = {10.48550/arXiv.2509.05724},
archivePrefix = {arXiv},
       eprint = {2509.05724},
 primaryClass = {stat.ML},
       adsurl = {https://ui.adsabs.harvard.edu/abs/2025arXiv250905724O}
}

@ARTICLE{desi2025,
       author = {{Adame}, A.~G. and {Aguilar}, J. and {Ahlen}, S. and {Alam}, S. and {Alexander}, D.~M. and {Alvarez}, M. and {Alves}, O. and {Anand}, A. and {Andrade}, U. and {Armengaud}, E. and {Avila}, S. and {Aviles}, A. and {Awan}, H. and {Bahr-Kalus}, B. and {Bailey}, S. and {Baltay}, C. and {Bault}, A. and {Behera}, J. and {BenZvi}, S. and {Bera}, A. and {Beutler}, F. and {Bianchi}, D. and {Blake}, C. and {Blum}, R. and {Brieden}, S. and {Brodzeller}, A. and {Brooks}, D. and {Buckley-Geer}, E. and {Burtin}, E. and {Calderon}, R. and {Canning}, R. and {Carnero Rosell}, A. and {Cereskaite}, R. and {Cervantes-Cota}, J.~L. and {Chabanier}, S. and {Chaussidon}, E. and {Chaves-Montero}, J. and {Chen}, S. and {Chen}, X. and {Claybaugh}, T. and {Cole}, S. and {Cuceu}, A. and {Davis}, T.~M. and {Dawson}, K. and {de la Macorra}, A. and {de Mattia}, A. and {Deiosso}, N. and {Dey}, A. and {Dey}, B. and {Ding}, Z. and {Doel}, P. and {Edelstein}, J. and {Eftekharzadeh}, S. and {Eisenstein}, D.~J. and {Elliott}, A. and {Fagrelius}, P. and {Fanning}, K. and {Ferraro}, S. and {Ereza}, J. and {Findlay}, N. and {Flaugher}, B. and {Font-Ribera}, A. and {Forero-S{\'a}nchez}, D. and {Forero-Romero}, J.~E. and {Frenk}, C.~S. and {Garcia-Quintero}, C. and {Gazta{\~n}aga}, E. and {Gil-Mar{\'\i}n}, H. and {Gontcho a Gontcho}, S. and {Gonzalez-Morales}, A.~X. and {Gonzalez-Perez}, V. and {Gordon}, C. and {Green}, D. and {Gruen}, D. and {Gsponer}, R. and {Gutierrez}, G. and {Guy}, J. and {Hadzhiyska}, B. and {Hahn}, C. and {Hanif}, M.~M.~S. and {Herrera-Alcantar}, H.~K. and {Honscheid}, K. and {Howlett}, C. and {Huterer}, D. and {Ir{\v{s}}i{\v{c}}}, V. and {Ishak}, M. and {Juneau}, S. and {Kara{\c{c}}ayl{\i}}, N.~G. and {Kehoe}, R. and {Kent}, S. and {Kirkby}, D. and {Kremin}, A. and {Krolewski}, A. and {Lai}, Y. and {Lan}, T.-W. and {Landriau}, M. and {Lang}, D. and {Lasker}, J. and {Le Goff}, J.~M. and {Le Guillou}, L. and {Leauthaud}, A. and {Levi}, M.~E. and {Li}, T.~S. and {Linder}, E. and {Lodha}, K. and {Magneville}, C. and {Manera}, M. and {Margala}, D. and {Martini}, P. and {Maus}, M. and {McDonald}, P. and {Medina-Varela}, L. and {Meisner}, A. and {Mena-Fern{\'a}ndez}, J. and {Miquel}, R. and {Moon}, J. and {Moore}, S. and {Moustakas}, J. and {Mueller}, E. and {Mu{\~n}oz-Guti{\'e}rrez}, A. and {Myers}, A.~D. and {Nadathur}, S. and {Napolitano}, L. and {Neveux}, R. and {Newman}, J.~A. and {Nguyen}, N.~M. and {Nie}, J. and {Niz}, G. and {Noriega}, H.~E. and {Padmanabhan}, N. and {Paillas}, E. and {Palanque-Delabrouille}, N. and {Pan}, J. and {Penmetsa}, S. and {Percival}, W.~J. and {Pieri}, M.~M. and {Pinon}, M. and {Poppett}, C. and {Porredon}, A. and {Prada}, F. and {P{\'e}rez-Fern{\'a}ndez}, A. and {P{\'e}rez-R{\`a}fols}, I. and {Rabinowitz}, D. and {Raichoor}, A. and {Ram{\'\i}rez-P{\'e}rez}, C. and {Ramirez-Solano}, S. and {Rashkovetskyi}, M. and {Ravoux}, C. and {Rezaie}, M. and {Rich}, J. and {Rocher}, A. and {Rockosi}, C. and {Roe}, N.~A. and {Rosado-Marin}, A. and {Ross}, A.~J. and {Rossi}, G. and {Ruggeri}, R. and {Ruhlmann-Kleider}, V. and {Samushia}, L. and {Sanchez}, E. and {Saulder}, C. and {Schlafly}, E.~F. and {Schlegel}, D. and {Schubnell}, M. and {Seo}, H. and {Shafieloo}, A. and {Sharples}, R. and {Silber}, J. and {Slosar}, A. and {Smith}, A. and {Sprayberry}, D. and {Tan}, T. and {Tarl{\'e}}, G. and {Taylor}, P. and {Trusov}, S. and {Ure{\~n}a-L{\'o}pez}, L.~A. and {Vaisakh}, R. and {Valcin}, D. and {Valdes}, F. and {Vargas-Maga{\~n}a}, M. and {Verde}, L. and {Walther}, M. and {Wang}, B. and {Wang}, M.~S. and {Weaver}, B.~A. and {Weaverdyck}, N. and {Wechsler}, R.~H. and {Weinberg}, D.~H. and {White}, M. and {Yu}, J. and {Yu}, Y. and {Yuan}, S. and {Y{\`e}che}, C. and {Zaborowski}, E.~A. and {Zarrouk}, P. and {Zhang}, H. and {Zhao}, C. and {Zhao}, R. and {Zhou}, R. and {Zhuang}, T.},
        title = "{DESI 2024 VI: cosmological constraints from the measurements of baryon acoustic oscillations}",
      journal = {\jcap},
         year = 2025,
        month = feb,
       volume = {2025},
       number = {2},
          eid = {021},
        pages = {021},
          doi = {10.1088/1475-7516/2025/02/021},
archivePrefix = {arXiv},
       eprint = {2404.03002},
 primaryClass = {astro-ph.CO},
       adsurl = {https://ui.adsabs.harvard.edu/abs/2025JCAP...02..021A}
}

@ARTICLE{vincenzi2025,
       author = {{Vincenzi}, M. and {Kessler}, R. and {Shah}, P. and {Lee}, J. and {Davis}, T.~M. and {Scolnic}, D. and {Armstrong}, P. and {Brout}, D. and {Camilleri}, R. and {Chen}, R. and {Galbany}, L. and {Lidman}, C. and {M{\"o}ller}, A. and {Popovic}, B. and {Rose}, B. and {Sako}, M. and {S{\'a}nchez}, B.~O. and {Smith}, M. and {Sullivan}, M. and {Wiseman}, P. and {Abbott}, T.~M.~C. and {Aguena}, M. and {Allam}, S. and {Andrade-Oliveira}, F. and {Bocquet}, S. and {Brooks}, D. and {Carnero Rosell}, A. and {Carretero}, J. and {da Costa}, L.~N. and {Pereira}, M.~E.~S. and {Diehl}, H.~T. and {Doel}, P. and {Everett}, S. and {Flaugher}, B. and {Frieman}, J. and {Garc{\'\i}a-Bellido}, J. and {Gaztanaga}, E. and {Gruen}, D. and {Gruendl}, R.~A. and {Gutierrez}, G. and {Hinton}, S.~R. and {Hollowood}, D.~L. and {Honscheid}, K. and {James}, D.~J. and {Kuehn}, K. and {Lahav}, O. and {Lee}, S. and {Marshall}, J.~L. and {Mena-Fern{\'a}ndez}, J. and {Miquel}, R. and {Muir}, J. and {Myles}, J. and {Palmese}, A. and {Plazas Malag{\'o}n}, A.~A. and {Porredon}, A. and {Samuroff}, S. and {Sanchez}, E. and {Sanchez Cid}, D. and {Sevilla-Noarbe}, I. and {Suchyta}, E. and {Tarle}, G. and {To}, C. and {Tucker}, D.~L. and {Vikram}, V. and {Walker}, A.~R. and {Weaverdyck}, N. and {Weller}, J.},
        title = "{Comparing the DES-SN5YR and Pantheon+ SN cosmology analyses: investigation based on 'evolving dark energy or supernovae systematics'?}",
      journal = {\mnras},
         year = 2025,
        month = aug,
       volume = {541},
       number = {3},
        pages = {2585-2593},
          doi = {10.1093/mnras/staf943},
archivePrefix = {arXiv},
       eprint = {2501.06664},
 primaryClass = {astro-ph.CO},
       adsurl = {https://ui.adsabs.harvard.edu/abs/2025MNRAS.541.2585V}
}

@ARTICLE{rix2021,
       author = {{Rix}, Hans-Walter and {Hogg}, David W. and {Boubert}, Douglas and {Brown}, Anthony G.~A. and {Casey}, Andrew and {Drimmel}, Ronald and {Everall}, Andrew and {Fouesneau}, Morgan and {Price-Whelan}, Adrian M.},
        title = "{Selection Functions in Astronomical Data Modeling, with the Space Density of White Dwarfs as a Worked Example}",
      journal = {\aj},
         year = 2021,
        month = oct,
       volume = {162},
       number = {4},
          eid = {142},
        pages = {142},
          doi = {10.3847/1538-3881/ac0c13},
archivePrefix = {arXiv},
       eprint = {2106.07653},
 primaryClass = {astro-ph.IM},
       adsurl = {https://ui.adsabs.harvard.edu/abs/2021AJ....162..142R}
}

@article{autenrieth2024improved,
  title={Improved weak lensing photometric redshift calibration via StratLearn and hierarchical modelling},
  author={Autenrieth, Maximilian and Wright, Angus H and Trotta, Roberto and Van Dyk, David A and Stenning, David C and Joachimi, Benjamin},
  journal={Monthly Notices of the Royal Astronomical Society},
  volume={534},
  number={4},
  pages={3808--3831},
  year={2024},
  publisher={Oxford University Press}
}

@ARTICLE{mitra2025,
       author = {{Mitra}, Ayan and {Kessler}, Richard and {Chen}, Rebecca C. and {Gagliano}, Alex and {Grayling}, Matthew and {More}, Surhud and {Narayan}, Gautham and {Qu}, Helen and {Raghunathan}, Srinivasan and {Malz}, Alex I. and {Lochner}, Michelle and {The LSST Dark Energy Science Collaboration}},
        title = "{A Fully Photometric Approach to Type Ia Supernova Cosmology in the LSST Era: Host Galaxy Redshifts and Supernova Classification}",
      journal = {ArXiv e-prints},
         year = 2025,
        month = dec,
          eid = {arXiv:2512.06319},
archivePrefix = {arXiv},
       eprint = {2512.06319},
 primaryClass = {astro-ph.CO},
       adsurl = {https://ui.adsabs.harvard.edu/abs/2025arXiv251206319M}
}

@ARTICLE{larison2024,
       author = {{Larison}, Conor and {Jha}, Saurabh W. and {Kwok}, Lindsey A. and {Camacho-Neves}, Yssavo},
        title = "{Environmental Dependence of Type Ia Supernovae in Low-redshift Galaxy Clusters}",
      journal = {\apj},
         year = 2024,
        month = feb,
       volume = {961},
       number = {2},
          eid = {185},
        pages = {185},
          doi = {10.3847/1538-4357/ad0e0f},
archivePrefix = {arXiv},
       eprint = {2306.01088},
 primaryClass = {astro-ph.HE},
       adsurl = {https://ui.adsabs.harvard.edu/abs/2024ApJ...961..185L}
}

@software{flowjax,
  title = {FlowJAX: Distributions and Normalizing Flows in Jax},
  author = {Daniel Ward},
  url = {https://github.com/danielward27/flowjax},
  version = {17.0.2},
  year = {2024},
  doi = {10.5281/zenodo.10402073},
}

@article{GWPopulation,
  author = {Colm Talbot and Amanda Farah and Shanika Galaudage and Jacob Golomb and Hui Tong},
  title = {GWPopulation: Hardware agnostic population inference for compact binaries and beyond},
  journal = {Journal of Open Source Software},
  doi = {10.21105/joss.07753},
  url = {https://doi.org/10.21105/joss.07753},
  year = {2025},
  publisher = {The Open Journal},
  volume = {10},
  number = {109},
  pages = {7753},
  archivePrefix = {arXiv},
  eprint = {2409.14143},
  primaryClass = {astro-ph.IM},
}

@ARTICLE{jax-cosmo,
       author = {{Campagne}, Jean-Eric and {Lanusse}, Fran{\c{c}}ois and {Zuntz}, Joe and {Boucaud}, Alexandre and {Casas}, Santiago and {Karamanis}, Minas and {Kirkby}, David and {Lanzieri}, Denise and {Peel}, Austin and {Li}, Yin},
        title = "{JAX-COSMO: An End-to-End Differentiable and GPU Accelerated Cosmology Library}",
      journal = {The Open Journal of Astrophysics},
         year = 2023,
        month = apr,
       volume = {6},
          eid = {15},
        pages = {15},
          doi = {10.21105/astro.2302.05163},
archivePrefix = {arXiv},
       eprint = {2302.05163},
 primaryClass = {astro-ph.CO},
       adsurl = {https://ui.adsabs.harvard.edu/abs/2023OJAp....6E..15C}
}

@ARTICLE{grayling2023,
       author = {{Grayling}, M. and {Guti{\'e}rrez}, C.~P. and {Sullivan}, M. and {Wiseman}, P. and {Vincenzi}, M. and {Galbany}, L. and {M{\"o}ller}, A. and {Brout}, D. and {Davis}, T.~M. and {Frohmaier}, C. and {Graur}, O. and {Kelsey}, L. and {Lidman}, C. and {Popovic}, B. and {Smith}, M. and {Toy}, M. and {Tucker}, B.~E. and {Zontos}, Z. and {Abbott}, T.~M.~C. and {Aguena}, M. and {Allam}, S. and {Andrade-Oliveira}, F. and {Annis}, J. and {Asorey}, J. and {Bacon}, D. and {Bertin}, E. and {Bocquet}, S. and {Brooks}, D. and {Carnero Rosell}, A. and {Carollo}, D. and {Carrasco Kind}, M. and {Carretero}, J. and {Costanzi}, M. and {da Costa}, L.~N. and {Pereira}, M.~E.~S. and {De Vicente}, J. and {Desai}, S. and {Diehl}, H.~T. and {Doel}, P. and {Everett}, S. and {Ferrero}, I. and {Friedel}, D. and {Frieman}, J. and {Garc{\'\i}a-Bellido}, J. and {Gatti}, M. and {Gruen}, D. and {Gschwend}, J. and {Gutierrez}, G. and {Hinton}, S.~R. and {Hollowood}, D.~L. and {Honscheid}, K. and {James}, D.~J. and {Kuehn}, K. and {Kuropatkin}, N. and {Lewis}, G.~F. and {Malik}, U. and {March}, M. and {Menanteau}, F. and {Miquel}, R. and {Morgan}, R. and {Ogando}, R.~L.~C. and {Palmese}, A. and {Paz-Chinch{\'o}n}, F. and {Pieres}, A. and {Plazas Malag{\'o}n}, A.~A. and {Rodriguez-Monroy}, M. and {Romer}, A.~K. and {Roodman}, A. and {Sanchez}, E. and {Scarpine}, V. and {Sevilla-Noarbe}, I. and {Suchyta}, E. and {Tarle}, G. and {To}, C. and {Tucker}, D.~L. and {Varga}, T.~N. and {DES Collaboration}},
        title = "{Core-collapse supernovae in the Dark Energy Survey: luminosity functions and host galaxy demographics}",
      journal = {\mnras},
         year = 2023,
        month = mar,
       volume = {520},
       number = {1},
        pages = {684-701},
          doi = {10.1093/mnras/stad056},
archivePrefix = {arXiv},
       eprint = {2207.08520},
 primaryClass = {astro-ph.HE},
       adsurl = {https://ui.adsabs.harvard.edu/abs/2023MNRAS.520..684G}
}

@article{beams,
  title = {Bayesian estimation applied to multiple species},
  author = {Kunz, Martin and Bassett, Bruce A. and Hlozek, Ren\'ee A.},
  journal = {Phys. Rev. D},
  volume = {75},
  issue = {10},
  pages = {103508},
  numpages = {12},
  year = {2007},
  month = {May},
  publisher = {American Physical Society},
  doi = {10.1103/PhysRevD.75.103508},
  url = {https://link.aps.org/doi/10.1103/PhysRevD.75.103508}
}

@ARTICLE{darvishi2026,
       author = {Darvishi, Roxana  and Stenning, David C. and Hippel, Ted von  and Ward, Owen G.},
        title = "{Estimating Complex Densities using Two-Stage Normalizing Flows}",
        eid = {arXiv:2603.06944},
      journal = {ArXiv e-prints},
         year = 2026,
        month = jan,
          eid = {arXiv:2603.06944},
          doi = {10.48550/arXiv.2603.06944},
          archivePrefix = {arXiv},
        eprint = {2603.06944},
archivePrefix = {arXiv},
 primaryClass = {astro-ph.CO},
       adsurl = {https://ui.adsabs.harvard.edu/abs/2026arXiv260306944D},
}

@ARTICLE{camilleri2024,
       author = {{Camilleri}, R. and {Davis}, T.~M. and {Vincenzi}, M. and {Shah}, P. and {Frieman}, J. and {Kessler}, R. and {Armstrong}, P. and {Brout}, D. and {Carr}, A. and {Chen}, R. and {Galbany}, L. and {Glazebrook}, K. and {Hinton}, S.~R. and {Lee}, J. and {Lidman}, C. and {M{\"o}ller}, A. and {Popovic}, B. and {Qu}, H. and {Sako}, M. and {Scolnic}, D. and {Smith}, M. and {Sullivan}, M. and {S{\'a}nchez}, B.~O. and {Taylor}, G. and {Toy}, M. and {Wiseman}, P. and {Abbott}, T.~M.~C. and {Aguena}, M. and {Allam}, S. and {Alves}, O. and {Annis}, J. and {Avila}, S. and {Bacon}, D. and {Bertin}, E. and {Bocquet}, S. and {Brooks}, D. and {Burke}, D.~L. and {Carnero Rosell}, A. and {Carretero}, J. and {Castander}, F.~J. and {da Costa}, L.~N. and {Pereira}, M.~E.~S. and {Desai}, S. and {Diehl}, H.~T. and {Doel}, P. and {Doux}, C. and {Everett}, S. and {Ferrero}, I. and {Flaugher}, B. and {Fosalba}, P. and {Garc{\'\i}a-Bellido}, J. and {Gatti}, M. and {Gaztanaga}, E. and {Giannini}, G. and {Gruen}, D. and {Hollowood}, D.~L. and {Honscheid}, K. and {James}, D.~J. and {Kuehn}, K. and {Lahav}, O. and {Lee}, S. and {Lewis}, G.~F. and {Marshall}, J.~L. and {Mena-Fern{\'a}ndez}, J. and {Miquel}, R. and {Muir}, J. and {Myles}, J. and {Ogando}, R.~L.~C. and {Pieres}, A. and {Malag{\'o}n}, A.~A. Plazas and {Porredon}, A. and {Rodriguez-Monroy}, M. and {Sanchez}, E. and {Sanchez Cid}, D. and {Schubnell}, M. and {Sevilla-Noarbe}, I. and {Suchyta}, E. and {Swanson}, M.~E.~C. and {Tarle}, G. and {Walker}, A.~R. and {Weaverdyck}, N. and {DES Collaboration}},
        title = "{The dark energy survey supernova program: investigating beyond-{\ensuremath{\Lambda}}CDM}",
      journal = {\mnras},
         year = 2024,
        month = sep,
       volume = {533},
       number = {3},
        pages = {2615-2639},
          doi = {10.1093/mnras/stae1988},
archivePrefix = {arXiv},
       eprint = {2406.05048},
 primaryClass = {astro-ph.CO},
       adsurl = {https://ui.adsabs.harvard.edu/abs/2024MNRAS.533.2615C}
}

@ARTICLE{montel2025,
       author = {{Anau Montel}, Noemi and {Alvey}, James and {Weniger}, Christoph},
        title = "{Tests for model misspecification in simulation-based inference: From local distortions to global model checks}",
      journal = {\prd},
         year = 2025,
        month = apr,
       volume = {111},
       number = {8},
          eid = {083013},
        pages = {083013},
          doi = {10.1103/PhysRevD.111.083013},
archivePrefix = {arXiv},
       eprint = {2412.15100},
 primaryClass = {astro-ph.IM},
       adsurl = {https://ui.adsabs.harvard.edu/abs/2025PhRvD.111h3013A}
}

@ARTICLE{ginolin2026,
       author = {{Ginolin}, Madeleine and {Grayling}, Matthew and {Mandel}, Kaisey S. and {Autenrieth}, Maximilian and {Boyd}, Benjamin M. and {Do}, Aaron and {Kelsey}, Lisa and {O'Callaghan}, Matthew},
        title = "{On the origin of the environmental step: A BayeSN view of the ZTF SN Ia DR2}",
      journal = {arXiv e-prints},
         year = 2026,
        month = may,
          eid = {arXiv:2605.06799},
        pages = {arXiv:2605.06799},
archivePrefix = {arXiv},
       eprint = {2605.06799},
 primaryClass = {astro-ph.CO},
       adsurl = {https://ui.adsabs.harvard.edu/abs/2026arXiv260506799G}
}



\newpage
\onecolumn
\appendix
\section{Propagation of Redshift and Peculiar Velocity Uncertainties}
\label{app:mu_err}

In this work we treat the cosmological parameter inference problem as a regression of distance modulus as a function of redshift. This means that any uncertainty associated with redshift $z_s$ or peculiar velocity $v^s_{\text{pec}}$ must be propagated as uncertainties on the distance modulus $\mu_s$.

The relationship between the observed CMB-redshift $\hat{z}_s$ and the true CMB-redshift $z_s$ is
\begin{equation}
   \hspace{5cm} \hat{z}_s = \frac{(1+\hat{z}^{\text{hel}}_s)}{\Big(1+\frac{v^s_{\odot}}{c}\Big)}-1 = \big(1 + z_s \big) \bigg(1 + \frac{v^s_{\text{pec}}}{c} \bigg) -1 + \epsilon^s_{z} ,
\end{equation}
where we assume the small spectroscopic measurement error noise term is drawn from a Gaussian distribution $\epsilon^s_{z} \sim N(0,\sigma^2_{z,s})$ and peculiar velocities are sampled from $v^s_{\text{pec}}~\sim~N(0,\sigma_{\text{pec}}^2)$. To first order, this can be approximated as
\begin{equation}
\hspace{6.5cm}
    \hat{z}_s \approx z_s + \frac{v^s_{\text{pec}}}{c} + \epsilon^s_{z} .
\end{equation}
Eq.~\eqref{eq:mu_func} implies we can write the true CMB-frame redshift $z_s$ as a function of the true distance modulus $\mu_s$, peculiar velocity $v^s_{\text{pec}}$ and motion relative to the CMB $v^s_{\odot}$ such that $z_s=z(\mu = \mu_s,v_{\text{pec}}=v^s_{\text{pec}},v_{\odot}=v^s_{\odot};\,\bm{C})$. The $v^s_{\odot}$ term is assumed to be known, however, there is further uncertainty on the peculiar velocity $P(v^s_{\text{pec}})=N(v^s_{\text{pec}}|\, 0, \sigma_{\text{pec}}^2)$. We approximate $\hat{z}_s$ further using a first order Taylor expansion of $z_s$ about $v^s_{\text{pec}} = 0 \text{ km s}^{-1}$
\begin{equation}
    \hat{z}_s \approx z(\mu=\mu_s,v_{\text{pec}}=0,v_{\odot}=v^s_{\odot};\,\bm{C})  +  \frac{\partial z}{\partial v_{\text{pec}}}\bigg|_{\substack{\mu=\mu_s\\ v_{\text{pec}}=0}} v^s_{\text{pec}} +  \frac{v^s_{\text{pec}}}{c} + \epsilon^s_{z} = z(\mu=\mu_s,v_{\text{pec}}=0,v_{\odot}=v^s_{\odot};\,\bm{C})  +  \Bigg(\frac{\partial z}{\partial v_{\text{pec}}}\bigg|_{\substack{\mu=\mu_s\\ v_{\text{pec}}=0}} +  \frac{1}{c}\Bigg)v^s_{\text{pec}} + \epsilon^s_{z}.
\end{equation}
Taking expectations and variances of the $\hat{z}_s$ approximation allows us to solve for the observed CMB-frame redshift marginal likelihood

\begin{equation}
\hspace{2cm}
    P(\hat{z}_s|\, \mu_s, \bm{C}) = \int^{\infty}_{-\infty}   P(\hat{z}_s|\, \mu_s,v^s_{\text{pec}}, \bm{C}) P(v^s_{\text{pec}}) \, \, \text{d} v^s_{\text{pec}}
    =N\bigg(\hat{z}_s \bigg|\, \mathbb{E}[\hat{z}_s|\, \mu_s,\bm{C}], \text{Var}[\hat{z}_s|\, \mu_s,\bm{C}]\bigg),
\end{equation}
where 
\begin{equation}
\hspace{5.5cm}
    \mathbb{E}[\hat{z}_s|\, \mu_s,\bm{C}] = z(\mu=\mu_s,v_{\text{pec}}=0,v_{\odot}=v^s_{\odot};\,\bm{C}),
\end{equation}
\vspace{-0.5cm}
\begin{equation}
\hspace{1cm}
    \text{Var}[\hat{z}_s|\, \mu_s,\bm{C}] = \Bigg(\frac{\partial z}{\partial v_{\text{pec}}}\bigg|_{\substack{\mu=\mu_s\\ v_{\text{pec}}=0}} +  \frac{1}{c}\Bigg)^2 \sigma^2_{\text{pec}} + \sigma^2_{z,s} = \sigma^2_{\text{pec}}\bigg|\frac{\partial z}{\partial v_{\text{pec}}}\bigg|^2_{\substack{\mu=\mu_s\\ v_{\text{pec}}=0}} + \frac{2\sigma^2_{\text{pec}}}{c} \frac{\partial z}{\partial v_{\text{pec}}}\bigg|_{\substack{\mu=\mu_s\\ v_{\text{pec}}=0}} + \frac{\sigma^2_{\text{pec}}}{c^2} +\sigma^2_{z,s}= \sigma^2_{z|\,\mu,s}.
\end{equation}

Using a similar approach to \cite{mandel2009} (section 2.2.2), we use Bayes rule to construct the posterior on $\mu_s$ given $\hat{z}_s$
\begin{equation}
\label{eq:mu_prop_int}
\hspace{5cm} P(\mu_s| \, \hat{z}_s, \bm{C}) \propto P(\hat{z}_s|\, \mu_s, \bm{C}) P(\mu_s |\, \bm{C}),
\end{equation}
where in practice we assume a flat prior on $P(\mu_s |\, \bm{C}) \propto 1$. The effect of assuming the flat prior on the distance modulus has been investigated in \cite{desmond2025}. In our context, since the assumption is only used in the propagation of relatively small uncertainties, the influence of the flat prior is negligible. 

We can derive the posterior using a further first order Taylor expansion of $z$ about the point $\mu^*_s$ that satisfies $\mu^*_s  =\mu(z=\hat{z}_s ,  v_{\text{pec}}=0,v_{\odot}=v^s_{\odot};\,\bm{C})$ and~$\hat{z}_s = z(\mu=\mu^*_s,v_{\text{pec}}=0,v_{\odot}=v^s_{\odot};\,\bm{C})$
\begin{multline}
    P(\mu_s |\, \hat{z}_s,\bm{C}) \propto P(\hat{z}_s |\, \mu_s,\bm{C})\\ \hspace{1cm}
    \propto \text{exp}\Bigg[-\frac{(\hat{z}_s-z(\mu=\mu_s,v_{\text{pec}}=0,v_{\odot}=v^s_{\odot};\,\bm{C}))^2}{2\sigma_{z|\,\mu,s}^2} \Bigg]\\
 \hspace{3cm} \approx \text{exp}\Bigg[-\frac{\big(\hat{z}_s- z(\mu=\mu^*_s,v_{\text{pec}}=0,v_{\odot}=v^s_{\odot};\,\bm{C})-\big|\frac{\partial z}{\partial \mu}\big|_{\mu =\mu^*_s,v_{\text{pec}}=0}(\mu_s-\mu^*_s)\big)^2}{2\sigma_{z|\,\mu*,s}^2} \Bigg]\\\hspace{6cm}
  = \text{exp}\Bigg[-\bigg|\frac{\partial z}{\partial \mu}\bigg|^2_{\substack{\mu =\mu^*_s \\ v_{\text{pec}}=0}}\frac{(\mu_s-\mu^*_s)^2}{2\sigma_{z|\,\mu*,s}^2} \Bigg] \\ 
  =\text{exp}\Bigg[-\frac{(\mu_s- \mu(z=\hat{z}_s ,  v_{\text{pec}}=0,v_{\odot}=v^s_{\odot};\,\bm{C}))^2}{2\sigma_{z|\,\mu*,s}^2\big|\frac{\partial \mu}{\partial z}\big|^2_{z =\hat{z}_s , v_{\text{pec}}=0}} \Bigg] .
\end{multline}
Leaving us with the posterior 
\begin{equation}
\hspace{4cm}
    P(\mu_s | \, \hat{z}_s, \bm{C})  \approx N\Bigg(\mu_s \Bigg| \,\mu(z=\hat{z}_s ,  v_{\text{pec}}=0,v_{\odot}=v^s_{\odot};\,\bm{C}) ,\sigma_{\mu|\,z,s}^2 = \bigg|\frac{\partial \mu}{\partial z}\bigg|^2_{\substack{z = \hat{z}_s\\ v_{\text{pec}}=0}} \sigma_{z|\,\mu*,s}^2 \Bigg),
\end{equation}
where we can use the cyclical rule of partial derivatives (triple product rule\footnote{\url{https://en.wikipedia.org/wiki/Triple_product_rule}}) $\frac{\partial A}{\partial B}\big|_C = - \frac{\partial C / \partial B|_A}{\partial C /\partial A|_B}$ to calculate  $\sigma_{\mu|\,z,s}^2$
\begin{multline}
    \sigma_{\mu|\,z,s}^2 = \bigg|\frac{\partial \mu}{\partial z}\bigg|^2_{\substack{z = \hat{z}_s\\ v_{\text{pec}}=0}} \sigma_{z|\,\mu^*,s}^2 \\ \hspace{2.5cm}= \bigg|\frac{\partial \mu}{\partial z}\bigg|^2_{\substack{z = \hat{z}_s\\ v_{\text{pec}}=0}} \Bigg[\sigma^2_{\text{pec}} \bigg|\frac{\partial z}{\partial v_{\text{pec}}}\bigg|^2_{\substack{\mu=\mu^*_s\\ v_{\text{pec}}=0}} + \frac{2\sigma^2_{\text{pec}}}{c} \frac{\partial z}{\partial v_{\text{pec}}}\bigg|_{\substack{\mu=\mu^*_s\\ v_{\text{pec}}=0}} + \frac{\sigma^2_{\text{pec}}}{c^2} +\sigma^2_{z,s}\Bigg] \\ \hspace{4.5cm} = \bigg|\frac{\partial \mu}{\partial z}\bigg|^2_{\substack{z = \hat{z}_s\\ v_{\text{pec}}=0}} \Bigg[  \frac{|\partial \mu / \partial v_{\text{pec}}|^2_{z=\hat{z}_s,v_{\text{pec}=0 
    }}}{|\partial \mu / \partial z|^2_{z=\hat{z}_s,v_{\text{pec}=0 
    }}} \sigma^2_{\text{pec}} - 2 \frac{\partial \mu / \partial v_{\text{pec}}|_{z=\hat{z}_s,v_{\text{pec}=0 
    }}}{\partial \mu / \partial z|_{z=\hat{z}_s,v_{\text{pec}=0 
    }}} \frac{\sigma^2_{\text{pec}}}{c} + \frac{\sigma^2_{\text{pec}}}{c^2}+ \sigma^2_{z,s} \Bigg] \\ = 
    \bigg| \frac{\partial\mu}{\partial z}\bigg|_{\substack{z=\hat{z}_s\\ v_{\text{pec}}=0}}^2 \bigg(\sigma_{z,s}^2 +\frac{\sigma_{\text{pec}}^2}{c^2}\bigg) + \bigg|\frac{\partial\mu}{\partial v_{\text{pec}}}\bigg|_{\substack{z=\hat{z}_s\\ v_{\text{pec}}=0}}^2\sigma_{\text{pec}}^2 
 - 2 \frac{\partial\mu}{\partial z}\bigg|_{\substack{z=\hat{z}_s\\ v_{\text{pec}}=0}} \frac{\partial\mu}{\partial v_{\text{pec}}}\bigg|_{\substack{z=\hat{z}_s\\ v_{\text{pec}}=0}} \frac{\sigma_{\text{pec}}^2}{c}  .  
\end{multline}

\section{Analytical Likelihood Solution to the Simplified Forward Model}
\label{app:toy_proof}

Here we derive the analytical likelihood solution for the simplified forward model defined in Section~\ref{sec:toy}. The selection probability in the model is defined using a Gaussian cumulative distribution function
\begin{equation}
\label{eq:cdf_app}
\hspace{5cm}P(I_s = 1 |\, \bm{\hat{d}}_s) = \Phi \Bigg(\frac{m_{\text{cut}} -(\hat{m}_s + a_{\text{cut}} \, \hat{x}_s + b_{\text{cut}} \, \hat{c}_s)}{\sigma_{\text{cut}}}\Bigg).
\end{equation} 
To begin, we reproduce the Eq.~\eqref{eq:full_aux} observed summary statistic likelihood  conditioned on the selection indicator $I_s=1$, distance modulus $\mu_s$, observed heliocentric redshift $\hat{z}_s^{\text{hel}}$ and SN population parameters $\bm{\Theta}_{\text{SN}}$ below
\begin{equation}
\label{eq:full_aux_app}
P\big(\bm{\hat{d}}_s\big|\,I_s=1,\mu_s,\hat{z}_s^{\text{hel}},\bm{\Theta}_{\text{SN}}\big)
=\frac{\int^\infty_{-\infty}P(I_s=1\big|\,\bm{\hat{d}}_s,\hat{z}_s^{\text{hel}})P(\bm{\hat{d}}_s|\, \bm{d}_s,\hat{z}_s^{\text{hel}})  P(\bm{d}_s|\ \mu_s, \bm{\Theta}_{\text{SN}} ) \, \text{d}  \bm{d}_s }{\int^\infty_{-\infty}\int^\infty_{-\infty}P(I_s=1\big|\,\bm{\hat{d}}_s,\hat{z}_s^{\text{hel}})P(\bm{\hat{d}}_s|\, \bm{d}_s,\hat{z}_s^{\text{hel}}) P(\bm{d}_s|\ \mu_s, \bm{\Theta}_{\text{SN}} )\, \text{d} \bm{d}_s \,\text{d}\bm{\hat{d}}_s 
}=g\big(\bm{\hat{d}}_s\big|\,m^s_0=M_0+\mu_s,\hat{z}_s^{\text{hel}},\bm{\Sigma}_s,\bm{\Theta}_{\text{SN}}\big) .
\end{equation}
For convenience, we write this as an auxiliary density $g$ and define the latent apparent magnitude $m^s_0=M_0+\mu_s$. In the simplified forward model, the population distribution $P(\bm{d}_s|\ \mu_s, \bm{\Theta}_{\text{SN}} )$ can be derived as a multivariate Gaussian distribution using the relationships described in Eq.~\eqref{eq:tripp}, Eq.~\eqref{eq:toy_xpop} and Eq.~\eqref{eq:toy_cpop}.  The influence of observation error on the summary statistics $P(\bm{\hat{d}}_s|, \bm{d}_s,\hat{z}_s^{\text{hel}})$ is also parametrised by a multivariate Gaussian distribution in Eq.~\eqref{eq:salt_like}. The marginalisation over latent variables $\bm{d}_s$ leads to the naive likelihood solution that does not take selection effects into account
\begin{equation}
\label{eq:marj_lik_app}
 \hspace{3.cm} P(\bm{\hat{d}}_s| \mu_s,\hat{z}_s^{\text{hel}}, \bm{\Theta_{\text{SN}}}) =\int^{\infty}_{-\infty} P(\bm{\hat{d}}_s| \bm{d}_s,\hat{z}_s^{\text{hel}})  P(\bm{d}_s|\ \mu_s, \bm{\Theta}_{\text{SN}} ) \ \text{d} \bm{d}_s = N(\bm{\hat{d}}_{s}|\,\bm{\rho_{d}},\bm{W_d}),
\end{equation}
where 
\[
 \hspace{5.cm} \bm{\rho_{d}} = 
\begin{pmatrix}
\mathbb{E}[\hat{m}_s] \\
\mathbb{E}[\hat{c}_s] \\
\mathbb{E}[\hat{x}_s]
\end{pmatrix}
=
\begin{pmatrix}
m_0^s + \alpha x_0 + \beta (\alpha_c x + c_0) \\
\alpha_c x + c_0 \\
x_0
\end{pmatrix}
\]
\[
 \hspace{5.cm} \bm{W_d} = \begin{pmatrix}
\text{Var}[\hat{m}_s] & \text{Cov}[\hat{m}_s,\hat{c}_s] &\text{Cov}[\hat{m}_s,\hat{x}_s]  \\
 \text{Cov}[\hat{m}_s,\hat{c}_s] & \text{Var}[\hat{c}_s]  &  \text{Cov}[\hat{c}_s,\hat{x}_s] \\
\text{Cov}[\hat{m}_s,\hat{x}_s]  & \text{Cov}[\hat{c}_s,\hat{x}_s]  & \text{Var}[\hat{x}_s]
\end{pmatrix}
\]

\begin{align*}
 \hspace{5.cm}   \text{Var}[\hat{m}_s] &= \sigma_{\text{res}}^2+\sigma_{\mu|\,z,s}^2+\beta^2\sigma_c^2+(\alpha+\beta\alpha_c)^2 \sigma_x^2+\Sigma^{mm}_s\\
    \text{Var}[\hat{c}_s] &= \sigma_c^2 + \alpha_c^2\sigma_x^2+\Sigma^{cc}_s\\
    \text{Var}[\hat{x}_s] &=\sigma_x^2+\Sigma^{xx}_s\\
    \text{Cov}[\hat{m}_s,\hat{c}_s] &= \alpha\alpha_c\sigma_x^2 +\beta(\sigma_c^2+\alpha_c^2\sigma_x^2)+\Sigma_s^{mc}\\
    \text{Cov}[\hat{m}_s,\hat{x}_s] &= \alpha\sigma_x^2 + \beta \alpha_c \sigma_x^2+\Sigma_s^{mx}\\
    \text{Cov}[\hat{c}_s,\hat{x}_s] &= \alpha_c \sigma_x^2 +\Sigma_s^{cx}.
\end{align*}
This simplifies Eq.~\eqref{eq:full_aux_app} to 
\begin{equation}\label{eq:genliknew}
\hspace{4cm}g \big(\bm{\hat{d}}_s\big|\, m^s_0,\bm{\Sigma}_s,\bm{\Theta}_{\text{SN}}\big) = \frac{ P(I_s=1\big|\,\bm{\hat{d}}_s,\hat{z}_s^{\text{hel}}) N\big(\bm{\hat{d}_{s}}\big|\,\bm{\rho_{d}},\bm{W_d}\big)}{\int^\infty_{-\infty} P(I_s=1\big|\,\bm{\hat{d}}_s,\hat{z}_s^{\text{hel}}) N\big(\bm{\hat{d}_{s}}\big|\,\bm{\rho_{d}},\bm{W_d}\big)\,\text{d}\bm{\hat{d}}_s}.
\end{equation}
We can easily evaluate the numerator of this, but still need to solve the integral in the denominator. The conditional properties of the multivariate Gaussian can be used to write
\begin{equation}
\hspace{2cm}N\big(\bm{\hat{d}_{s}}\big|\,\bm{\rho_{d}},\bm{W_d}\big) = N\big(\hat{m}_s \big|\, \mathbb{E}[\hat{m}_s|\,\hat{c}_s,\hat{x}_s],\text{Var} [\hat{m}_s|\,\hat{c}_s,\hat{x}_s]\big)N\big(\hat{c}_s \big|\, \mathbb{E}[\hat{c}_s|\,\hat{x}_s],\text{Var} [\hat{c}_s|\,\hat{x}_s]\big)N\big(\hat{x}_s \big|\, \mathbb{E}[\hat{x}_s],\text{Var} [\hat{x}_s]\big),
\end{equation}
where 
\begin{align*}
\hspace{5.5cm}
    \mathbb{E}[\hat{c}_s|\,\hat{x}_s]&= \mathbb{E}[\hat{c}_s] + \frac{\text{Cov}[\hat{c}_s,\hat{x}_s]}{\text{Var}[\hat{x}_s]}\big(\hat{x}_s -  \mathbb{E}[\hat{x}_s]\big) \\
 \text{Var}[\hat{c}_s|\,\hat{x}_s]&= \text{Var}[\hat{c}_s] - \frac{\text{Cov}[\hat{c}_s,\hat{x}_s]^2}{\text{Var}[\hat{x}_s]} \\   
\mathbb{E}[\hat{m}_s|\,\hat{x}_s] &= \mathbb{E}[\hat{m}_s] + \frac{\text{Cov}[\hat{m}_s,\hat{x}_s]}{\text{Var}[\hat{x}_s]}\big(\hat{x}_s -  \mathbb{E}[\hat{x}_s]\big)
 \\
 \text{Var}[\hat{m}_s|\,\hat{x}_s] &= \text{Var}[\hat{m}_s] - \frac{\text{Cov}[\hat{m}_s,\hat{x}_s]^2}{\text{Var}[\hat{x}_s]}\\
\text{Cov}[\hat{m}_s,\hat{c}_s|\,\hat{x}_s]&= \text{Cov}[\hat{m}_s,\hat{c}_s]-\frac{\text{Cov}[\hat{m}_s,\hat{x}_s]\text{Cov}[\hat{c}_s,\hat{x}_s]}{\text{Var}[\hat{x}_s]}
 \\
     \mathbb{E}[\hat{m}_s|\,\hat{c}_s,\hat{x}_s] &= \mathbb{E}[\hat{m}_s|\,\hat{x}_s] + \frac{\text{Cov}[\hat{m}_s,\hat{c}_s|\,\hat{x}_s]}{ \text{Var}[\hat{c}_s|\,\hat{x}_s]}\big(\hat{c}_s -  \mathbb{E}[\hat{c}_s|\,\hat{x}_s]\big)\\
\text{Var}[\hat{m}_s|\,\hat{c}_s,\hat{x}_s] &= \text{Var}[\hat{m}_s|\,\hat{x}_s] - \frac{\text{Cov}[\hat{m}_s,\hat{c}_s|\,\hat{x}_s]^2}{ \text{Var}[\hat{c}_s|\,\hat{x}_s]}.
\end{align*}
The denominator in Eq.~\eqref{eq:genliknew} can then be expanded into the integral over each dimension
\begin{multline}
   \int^{\infty}_{-\infty}\Phi \Bigg(\frac{m_{\text{cut}} -(\hat{m}_s + a_{\text{cut}} \hat{x}_s + b_{\text{cut}} \hat{c}_s)}{\sigma_{\text{cut}}}\Bigg)  N\big(\bm{\hat{d}_{s}}\big|\,\bm{\rho_{d}},\bm{W_d}\big) \,\text{d}\hat{d}_s \\ \hspace{1cm}
    = \int^{\infty}_{-\infty} \int^{\infty}_{-\infty} \int^{\infty}_{-\infty} \Bigg[\Phi \Bigg(\frac{m_{\text{cut}} -(\hat{m}_s + a_{\text{cut}} \hat{x}_s + b_{\text{cut}} \hat{c}_s)}{\sigma_{\text{cut}}}\Bigg) \times \\ \hspace{4cm}
   N\big(\hat{m}_s \big|\, \mathbb{E}[\hat{m}_s|\,\hat{c}_s,\hat{x}_s],\text{Var} [\hat{m}_s|\,\hat{c}_s,\hat{x}_s]\big)N\big(\hat{c}_s \big|\, \mathbb{E}[\hat{c}_s|\,\hat{x}_s],\text{Var} [\hat{c}_s|\,\hat{x}_s]\big)N\big(\hat{x}_s \big|\, \mathbb{E}[\hat{x}_s],\text{Var} [\hat{x}_s]\big)\Bigg]\,\text{d}\hat{m}_s\,\text{d}\hat{c}_s\,\text{d}\hat{x}_s .\\
\end{multline}
The three dimensions can be marginalised over using the identity
\begin{equation}
\label{eq:indentity}
\hspace{4cm}
\int^{\infty}_{-\infty} \Phi((\mu_1-x)/\sigma_1)N(x|\,\mu_2,\sigma_2^2)\,\text{d}x
= \Phi\Bigg(\frac{\mu_1 - \mu_2}{\sigma_1} \times \Big(1+\frac{\sigma_2^2}{\sigma_{1}^2}\Big)^{-\frac{1}{2}}\Bigg).
\end{equation}
The first dimension $\hat{m}_s$ is integrated to get 
\begin{multline}
    \int^{\infty}_{-\infty} \Phi \Bigg(\frac{m_{\text{cut}} -(\hat{m}_s + a_{\text{cut}} \hat{x}_s + b_{\text{cut}} \hat{c}_s)}{\sigma_{\text{cut}}}\Bigg) 
   N\big(\hat{m}_s \big|\, \mathbb{E}[\hat{m}_s|\,\hat{c}_s,\hat{x}_s],\text{Var} [\hat{m}_s|\,\hat{c}_s,\hat{x}_s]\big)\,\text{d}\hat{m}_s 
= \Phi \Bigg(\frac{m_{\text{cut}} -( \mathbb{E}[\hat{m}_s|\,\hat{c}_s,\hat{x}_s] + a_{\text{cut}} \hat{x}_s + b_{\text{cut}} \hat{c}_s)}{\sigma'_{\text{cut}}}\Bigg), 
\end{multline}
where 
\begin{equation*}
\hspace{6cm}
    \sigma'_{\text{cut}}= \sigma_{\text{cut}}\sqrt{1+ \frac{\text{Var}[\hat{m}_s|\,\hat{c}_s,\hat{x}_s]}{\sigma_{\text{cut}}^2}}.
\end{equation*}
The second dimension $\hat{c}_s$
is then integrated out to get
\begin{multline}
    \int^{\infty}_{-\infty} \Phi \Bigg(\frac{m_{\text{cut}} -(\mathbb{E}[\hat{m}_s|\,\hat{c}_s,\hat{x}_s] +a_{\text{cut}} \hat{x}_s + b_{\text{cut}} \hat{c}_s)}{\sigma'_{\text{cut}}}\Bigg) 
   N\big(\hat{c}_s \big|\, \mathbb{E}[\hat{c}_s|\,\hat{x}_s],\text{Var} [\hat{c}_s|\,\hat{x}_s]\big)\,\text{d}\hat{c}_s \\   = \int^{\infty}_{-\infty} \Phi \Bigg(\frac{m_{\text{cut}} -(\mathbb{E}[\hat{m}_s|\,\hat{x}_s]+\gamma_c (\hat{c}_s - \mathbb{E}[\hat{c}_s|\,\hat{x}_s]) +a_{\text{cut}} \hat{x}_s + b_{\text{cut}} \hat{c}_s)}{\sigma'_{\text{cut}}}\Bigg) 
   N\big(\hat{c}_s \big|\, \mathbb{E}[\hat{c}_s|\,\hat{x}_s],\text{Var} [\hat{c}_s|\,\hat{x}_s]\big)\,\text{d}\hat{c}_s \\
= \Phi \Bigg(\frac{m_{\text{cut}} -( \mathbb{E}[\hat{m}_s|\,\hat{x}_s] -\gamma_c \mathbb{E}[\hat{c}_s|\,\hat{x}_s] +a_{\text{cut}} \hat{x}_s + (b_{\text{cut}} + \gamma_c )\mathbb{E}[\hat{c}_s|\,\hat{x}_s])}{\sigma''_{\text{cut}}}\Bigg) \\
= \Phi \Bigg(\frac{m_{\text{cut}} -( \mathbb{E}[\hat{m}_s|\,\hat{x}_s] +a_{\text{cut}} \hat{x}_s + b_{\text{cut}}\mathbb{E}[\hat{c}_s|\,\hat{x}_s])}{\sigma''_{\text{cut}}}\Bigg) ,
\end{multline}
where 
\begin{align*}
\hspace{6cm}
    \sigma''_{\text{cut}}&= \sigma'_{\text{cut}}\sqrt{1+ \frac{(b_{\text{cut}} + \gamma_c)^2\text{Var}[\hat{c}_s|\,\hat{x}_s]}{{\sigma'_{\text{cut}}}^2}}
\\
\gamma_c &= \frac{\text{Cov}[\hat{m}_s,\hat{c}_s|\,\hat{x}_s]}{\text{Var}[\hat{c}_s|\,\hat{x}_s]}.
\end{align*}
Then the third dimension $\hat{x}_s$ is integrated to get the final result
\begin{multline}
\int^\infty_{-\infty} P(I_s=1\big|\,\bm{\hat{d}}_s,\hat{z}_s^{\text{hel}}) N\big(\bm{\hat{d}_{s}}\big|\,\bm{\rho_{d}},\bm{W_d}\big)\,\text{d}\bm{\hat{d}}_s \\
    =\int^{\infty}_{-\infty} \Phi \Bigg(\frac{m_{\text{cut}} -( \mathbb{E}[\hat{m}_s|\,\hat{x}_s] +a_{\text{cut}} \hat{x}_s + b_{\text{cut}}\mathbb{E}[\hat{c}_s|\,\hat{x}_s])}{\sigma''_{\text{cut}}}\Bigg)  
   N\big(\hat{x}_s \big|\, \mathbb{E}[\hat{x}_s],\text{Var} [\hat{x}_s]\big)\,\text{d}\hat{x}_s\\=
       \int^{\infty}_{-\infty} \Phi \Bigg(\frac{m_{\text{cut}} -( \mathbb{E}[\hat{m}_s]+\gamma_x(\hat{x}_s -\mathbb{E}[\hat{x}_s]) +a_{\text{cut}} \hat{x}_s + b_{\text{cut}}(\mathbb{E}[\hat{c}_s]+\delta_x(\hat{x}_s -\mathbb{E}[\hat{x}_s])))}{\sigma''_{\text{cut}}}\Bigg)  
   N\big(\hat{x}_s \big|\, \mathbb{E}[\hat{x}_s],\text{Var} [\hat{x}_s]\big)\,\text{d}\hat{x}_s\\
= \Phi \Bigg(\frac{m_{\text{cut}} -( \mathbb{E}[\hat{m}_s] - \gamma_x \mathbb{E}[\hat{x}_s] +(a_{\text{cut}}+\gamma_x + \delta_x b_{\text{cut}} )\mathbb{E}[\hat{x}_s]+ b_{\text{cut}} (\mathbb{E}[\hat{c}_s]-\delta_x \mathbb{E}[\hat{x}_s]))}{\sigma'''_{\text{cut}}}\Bigg) \\
    \hspace{4cm}=\Phi \Bigg(\frac{m_{\text{cut}} -( \mathbb{E}[\hat{m}_s] +a_{\text{cut}} \mathbb{E}[\hat{x}_s]+ b_{\text{cut}} \mathbb{E}[\hat{c}_s])}{\sigma'''_{\text{cut}}}\Bigg) ,
\end{multline}
where 
\begin{align*}
\hspace{5cm}
    \sigma'''_{\text{cut}}&= \sigma''_{\text{cut}}\sqrt{1+ \frac{(a_{\text{cut}}+\gamma_x + \delta_x b_{\text{cut}})^2\text{Var}[\hat{x}_s]}{{\sigma''_{\text{cut}}}^2}}\\
    \gamma_x &= \frac{\text{Cov}[\hat{m}_s,\hat{x}_s]}{\text{Var}[\hat{x}_s]}\\
     \delta_x &= \frac{\text{Cov}[\hat{c}_s,\hat{x}_s]}{\text{Var}[\hat{x}_s]}  .
\end{align*}

\newpage

\section{Simplified Forward Model Summary Fit Covariances}
\label{app:cov_samp}
In the simplified forward model described in Section~\ref{sec:toy_sims}, we must simulate realistic $3\times 3$ summary statistic fit covariances. Below we define simple empirical relations, motivated by the distributions in the Pantheon+ sample \citep{scolnic2022}, that we use to sample $\bm{\Sigma}_s \sim P(\bm{\Sigma})$ for each simulated SN. We use this same distribution in both the simplified model training and test sets.
\begin{align*}
\hspace{5cm}
\log \Sigma_s^{mm} &\sim N\big(0.2(m_s - 56),\,1.2^2\big) \\
\log \Sigma_s^{cc} &\sim N(-7,\,0.6^2) \\
\log \Sigma_s^{xx} &\sim N(-3,\,1^2) \\
\Sigma_s^{mc} &\sim N\Big(
5.2\times10^{-4} + 0.289\, \Sigma_s^{mm},\,
(0.126\, \Sigma_s^{mm})^2
\Big) \\
\Sigma_s^{mx} &\sim N\Big(
8.1\times10^{-4} + 0.0384\, \Sigma_s^{xx},\,
(0.0203\, \Sigma_s^{xx})^2
\Big) \\
\Sigma_s^{cx} &\sim N\Big(
1.69\times10^{-4} + 0.0136\, \Sigma_s^{xx},\,
(0.0118\, \Sigma_s^{xx})^2
\Big)
\end{align*}

\section{SNANA Colour and Stretch Population Distributions}
\label{app:cxpop}
Population colour and stretch distributions used in the SNANA simulations are from \cite{scolnic2016}, defined using asymmetric normal distributions such that
\begin{equation}
\hspace{4cm}
c_s \sim P(c|\ x_s, z_s,\bm{\theta}_c)   = \begin{cases}
      N(-0.054,0.043^2) & \text{if }-0.3\leq c <-0.054 \\
       N(-0.054,0.101^2) & \text{if }-0.054\leq c \leq 0.5\\
       0 & \text{if }c <  -0.3 \text{ or } c >  0.5
      \end{cases},
\end{equation}
\begin{equation}
\hspace{4cm}
x_s \sim P(x|\,  z_s,\bm{\theta_x})      = \begin{cases}
      N(0.973,1.472^2) & \text{if }-3.0\leq x < 0.973 \\
       N(0.973,0.222^2) & \text{if } 0.973\leq x \leq 2.0\\
       0 & \text{if }x <  -3.0 \text{ or } x >  2.0 
      \end{cases},
\end{equation}
where there is no assumed correlation between the latent colour and stretch.

\section{SNANA Selection Cuts}
\label{app:SNANA_cuts}
In the SNANA forward model, defined in Section~\ref{sec:snana_model}, we make a series of standard cuts depending on the quality and fit of the SALT3 \citep{kenworthy2021} light curve summary statistics. The cuts, taken from \cite{betoule2014}, are as follows
\begin{align*}
\hspace{7.5cm}
    &-3<\hat{x}_s<3 \\
    &-0.3<\hat{c}_s<0.3\\
    &\Sigma^{xx}_{s} <1.0\\
    &\Sigma^{cc}_{s} <1.0\\
&\sigma^s_{T^\text{max}} < 2.0 \text{ days}\\
&P^s_{\text{fit}} > 0.05,\\ 
\end{align*}
where $\sigma^s_{T^\text{max}}$ is the error on the estimated observer-frame time of peak $T^{\text{max}}_s$ and $P^s_{\text{fit}}$ is the fit probability in the maximum likelihood estimation.
\newpage
\section{Training Ranges and Priors}
In this section we describe the training ranges and priors used in the experiments. 

\subsection{Simplified Forward Model}
\label{app:toy_sims}

In Table~\ref{tab:toy_priors}, we define the priors on global parameters $P(\bm{\Theta})$ used in inferring the simplified model parameters. In the same table we also include the population parameter distributions sampled $\bm{\Theta}^t_{\text{SN}}\sim P_{\text{train}}(\bm{\Theta}_{\text{SN}})$ to train the normalising flow. In the training of the normalising flow, we sample latent apparent magnitude $m^t_0=M_0 +\mu_t$ directly to ensure conditional independence from $\bm{C}$. In practice we find oversampling dim SNe, that are more influenced by Malmquist bias, results in improved training. We therefore sample 90\% of training SNe from 
\begin{equation}
\hspace{7cm}
    m^t_0 \sim N(25,0.5^2)_{[-\infty,25]},
\end{equation}
where $N(0,1^2)_{[B_L,B_U]}$ represents a truncated normal distribution centred at 0 with a variance of 1, truncated between $B_L$ and $B_U$. The remaining 10\% of the training samples are sampled according to a uniform distribution $m^t_0 \sim U(17,24)$ to ensure the full apparent magnitude range is covered.
\begin{table*}
	\centering
	\caption{Global parameter priors $P(\bm{\Theta})$ used in our hierarchical  Bayesian model for the simplified forward model experiments. Detailed also are the distributions sampled $\bm{\Theta}^t_{\text{SN}} \sim P_{\text{train}}(\bm{\Theta}_{\text{SN}})$ in the normalising flow training. $U(B_L,B_U)$ is a uniform distribution between $B_L$ and $B_U$, while $N(0,1^2)_{[B_L,B_U]}$ represents a truncated normal distribution centred at 0 with a variance of 1, truncated between $B_L$ and $B_U$.}

	\label{tab:toy_priors}

\begin{tabular}{lcccccc}
\hline
Global Parameter & Prior &  Training Range\\
\hline
$w_0$ (Flat $w$CDM)    &  $U(-2,0)$ & - \\
$\Omega_{m0}$ (Flat $w$CDM) & $U(0,1)$& -\\
$\Omega_{m0}$ ($\Lambda$CDM)   &  $U(0,2)$ & -  \\
$\Omega_{\Lambda 0}$  ($\Lambda$CDM)   &  $U(0,2)$  & -  \\  
$M_0$ & $U(-\infty,\infty)$ & -\\
$\alpha$ & $U(-0.2,-0.1)$ & $U(-0.2,-0.1)$\\
$\beta$ & $U(2.5,3.5)$ & $U(2.5,3.5)$\\
$\sigma_{\text{res}}$ & $N(0,1^2)_{[0,0.2]}$& $U(0,0.2)$\\
$c_0$ & $U(-0.1,0)$& $U(-0.1,0)$\\
$\alpha_c$ & $U(-0.02,0)$& $U(-0.02,0)$\\
$\sigma_{c}$ & $N(0,1^2)_{[0,0.1]}$& $U(0,0.1)$ \\
$x_0$ & $U(-0.6,-0.2)$ &$U(-0.6,-0.2)$\\
$\sigma_{x}$ & $N(0,2^2)_{[0,1.5]}$ & $U(0,1.5)$\\

\hline
\end{tabular}
\end{table*}

\subsection{SNANA Simulations}
\label{app:SNANA_sims}

In Table~\ref{tab:snana_priors}  we define the priors on global parameters $P(\bm{\Theta})$ used in the SNANA experiments. In the same table we also include the population parameter distributions sampled $\bm{\Theta}^t_{\text{SN}}\sim P_{\text{train}}(\bm{\Theta}_{\text{SN}})$ to train the normalising flow. To train the normalising flow without being influenced by the assumed fiducial cosmology $\bm{C}_1$, we must define a latent apparent magnitude $m^t_0$ that includes  a smearing on the distance, such that
\begin{equation}
   \hspace{6cm} m_0^t = M_0 + \mu_t + \Delta \mu_t \hspace{0.3cm} \text{   where} \hspace{0.3cm} \Delta\mu_t \sim P_\text{train}(\Delta \mu).
\end{equation}
The $\Delta\mu_t \sim P_\text{train}(\Delta \mu)$ was sampled from a uniform distribution $U(-0.5,0.5)$ for 10\% of our training data using the \texttt{MUSHIFT: -0.1:0.1} setting in SNANA. The remaining 90\% of training data was sampled using a re-weighted polynomial distribution with \texttt{MUSHIFT: -0.6:2.05} and \texttt{DNDZ\_Z1POLY\_REWGT: 0,0,0,1} SNANA settings. The purpose of the polynomial distribution was to effectively sample selected SN distances associated with cosmological parameters within the prior volume roughly equally.

\begin{table*}
	\centering
	\caption{Global parameter priors $P(\bm{\Theta})$ used in our hierarchical  Bayesian model for the SNANA experiments. Detailed also are the distributions sampled $\bm{\Theta}^t_{\text{SN}} \sim P_{\text{train}}(\bm{\Theta}_{\text{SN}})$ in the normalising flow training. $U(B_L,B_U)$ is a uniform distribution between $B_L$ and $B_U$.}

	\label{tab:snana_priors}

\begin{tabular}{lcccccc}
\hline
Global Parameter & Prior &  Training Range\\
\hline
$w_0$     &  $U(-2,0)$ & - \\
$\Omega_{m0}$ & $U(0,1)$& -\\
$M_0$ & $U(-\infty,\infty)$ & -\\
$\alpha$ & $U(-0.2,-0.1)$ & $U(-0.2,-0.1)$\\
$\beta$ & $U(2.5,3.5)$ & $U(2.5,3.5)$\\
\hline
\end{tabular}
\end{table*}

\subsection{CMB Prior}
\label{app:cmb_prior}
Since we do not have a low redshift anchor set, we combine the SNe in the SNANA experiments with a CMB prior \citep{komatsu2009} to add additional constraining power, comparable to constraints from present day surveys.  

For each random realisation of SNANA survey simulations we sample the $\hat{R}$ ratio such that
\begin{equation}
\hspace{4cm}
    \hat{R} \sim  N( f_{\text{CMB}}(\bm{C}), \sigma_{R}^2) \text{\hspace{0.5cm} where\hspace{0.5cm}}     f_{\text{CMB}}(\bm{C}) =\frac{ d_c(z_{*}; \, \bm{C})H_0\sqrt{\Omega_{m0}}}{c},
\end{equation}
where $d_c$ is the transverse co-moving distance function in Mpc and the constant $z_* = 1089$ is the redshift at recombination. The CMB constraining power is determined by $\sigma_R = 0.007$. The variable $H_0$  is the fixed value  of the Hubble constant $H_0=70\, 
\text{kms}^{-1}\text{Mpc}^{-1}$ and $c$ is the speed of light in a vacuum in $\text{kms}^{-1}$. 
The Eq.~\eqref{eq:hmc_post} posterior sampled is then adapted to be
\begin{multline}
\label{eq:hmc_cmb_post}
P\big(\bm{\Theta}, \bm{\mu} | \, \bm{\hat{D}}_\text{obs}, \bm{\hat{Z}}_\text{obs}, \bm{I}_{\text{obs}},\hat{R}\big) \propto
\Bigg[ \prod^{N_{\text{SN}}^{\text{obs}}}_{s=1}  g\big(\bm{\hat{d}}_s\big|m^s_0=M_0+\mu_s,\hat{z}_s^{\text{hel}},\bm{\Sigma}_s,\bm{\Theta}_{\text{SN}}\big)P(\mu_s |\ \hat{z}_s, \bm{C})\Bigg] P( \hat{R} |\,\bm{\Theta})P(\bm{\Theta})\\
\text{ where }
    P( \hat{R} |\,\bm{\Theta}) = N(\hat{R}|\, f_{\text{CMB}}(\bm{C}),  \sigma_{R}^2).
\end{multline}

\section{Further Simplified Forward Model Results}
\begin{figure*}
\centering
	\includegraphics[width=0.95\textwidth]{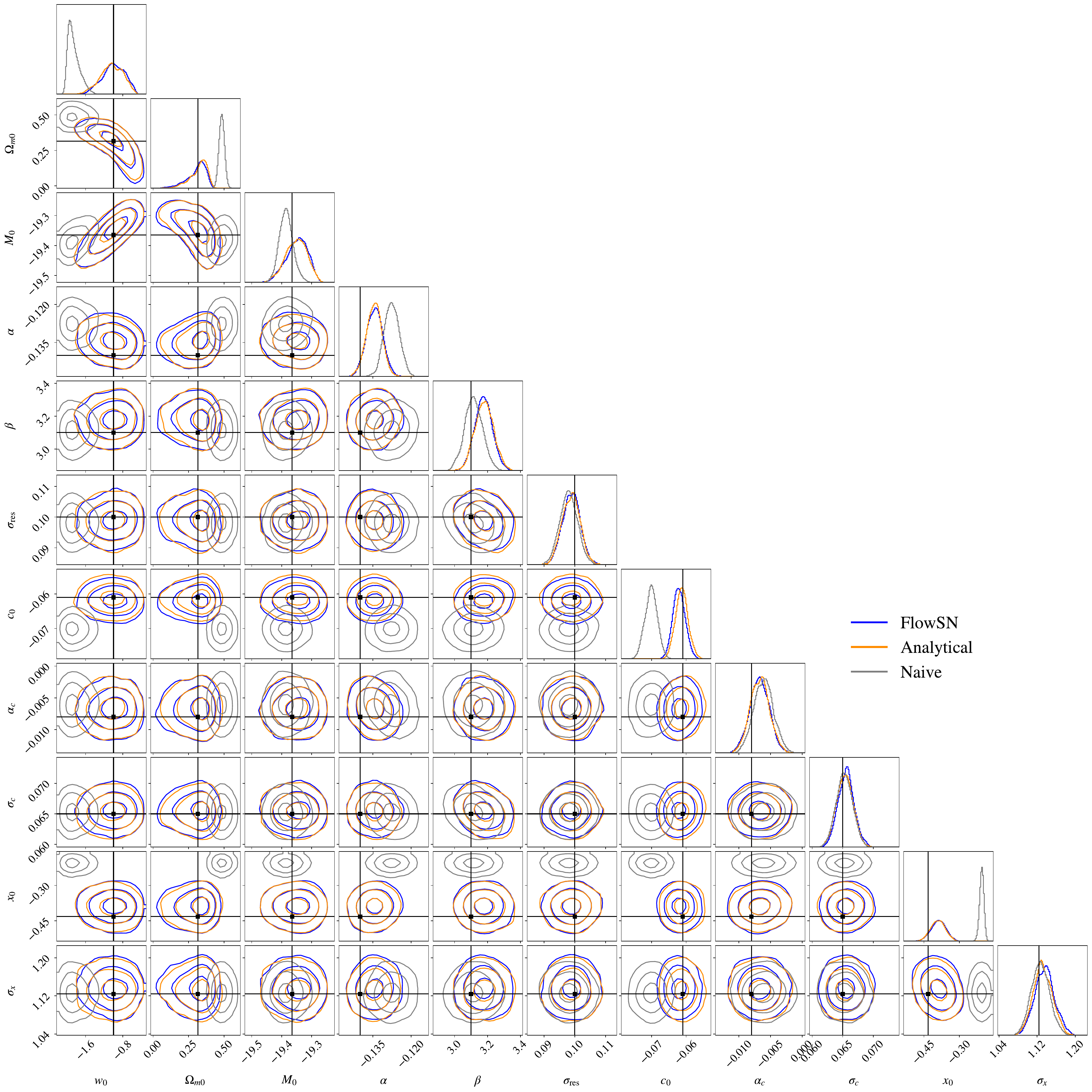}
    \vspace{0cm}
    \caption{Full posteriors on all eleven global parameters from the simplified forward modelling of 2000 spectroscopic SNe. We plot the \textit{FlowSN} posterior, the analytical likelihood solution posterior and  the naive solution posterior that does not account for selection effects  solution. Both the analytical and naive solutions are derived in Appendix~\ref{app:toy_proof}.\vspace{15cm}}
    \label{fig:toy_post_full}
\end{figure*}

\begin{figure*}
\centering
	\includegraphics[width=0.95\textwidth]{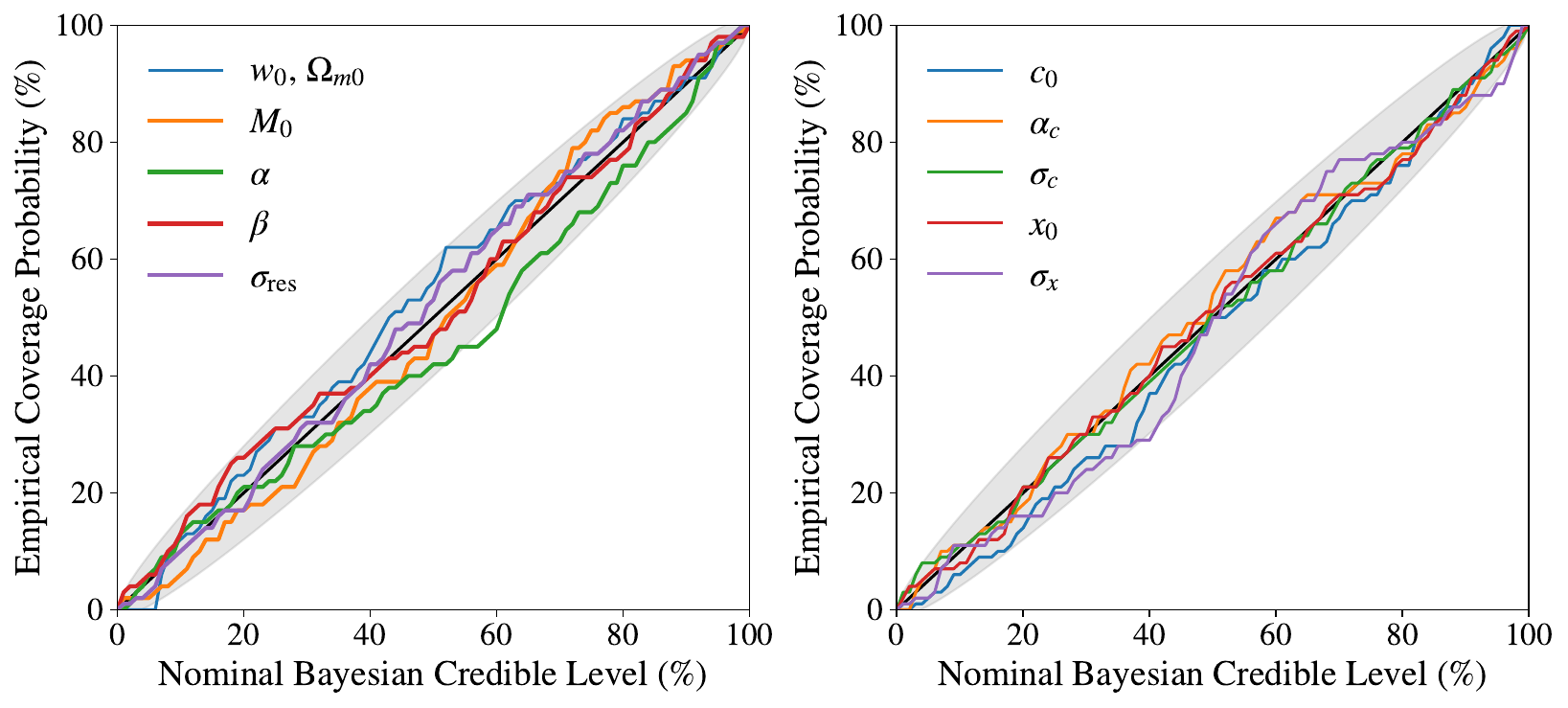}
    \vspace{0cm}
    \caption{\textit{FlowSN} frequentist calibration of posteriors for the simplified forward model experiment after 100 repeats on 2000 spectroscopic SNe. The grey shaded region shows the 95\% Monte Carlo confidence band for the empirical coverage under perfect calibration. \vspace{30cm}}
    \label{fig:toy_calib}
\end{figure*}

Here we include additional \textit{FlowSN} results on the simplified forward model described in Section~\ref{sec:toy}. Figure~\ref{fig:toy_post_full} compares the posteriors on all eleven global parameters for \textit{FlowSN}, the analytical likelihood solution defined in Eq.~\eqref{eq:genliknew} and the naive solution defined in Eq.~\eqref{eq:marj_lik_app} that does not account for selection effects. We see that the analytical solution and \textit{FlowSN} posteriors agree closely for all eleven parameters, while the naive solution leads to biased inference. In Figure~\ref{fig:toy_calib}, we demonstrate the \textit{FlowSN} posteriors have adequate frequentist calibration across all eleven  global parameters.

\newpage

\section{Evolving Dark Energy Inference on SNANA Simulations}
\label{app:w0wa}

\begin{figure}
    \centering
    \includegraphics[width=0.95\textwidth, keepaspectratio]{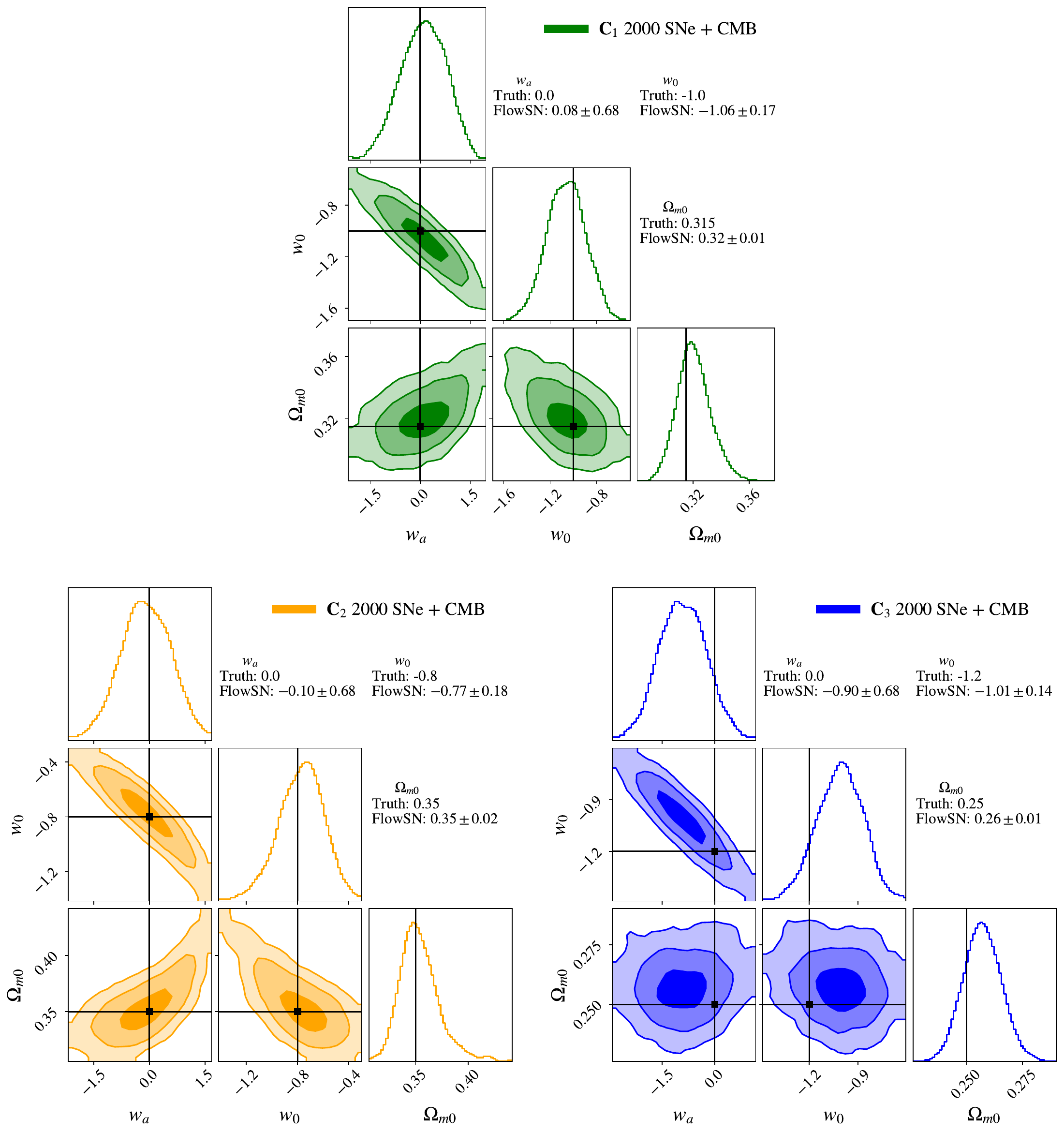}
    \caption{\textit{FlowSN} $w_0-w_a$ posteriors for three different SNANA simulations with differing cosmological parameter values. The true values in each corner plot are illustrated by solid black lines. These constraints were inferred without retraining the normalising flow.}
    \label{fig:w0wa_plot}
\end{figure}

Here we demonstrate that the current formulation of \textit{FlowSN} may be used for a $w_0 - w_a$ cosmology analysis without retraining the normalising flow. Figure~\ref{fig:w0wa_plot} shows the \textit{FlowSN} posteriors when inferring a $w_0w_a$CDM model for three SNANA simulations with different true cosmological parameter values. This is possible since the normalising flow is conditioned on latent $m_0^s = \mu(z_s; \, \bm{C}) + M_0$ rather than $\bm{C}$ directly. We find that \textit{FlowSN} is able to infer the truth within $2\sigma$ in all three cases. For these experiments, we use a CMB prior (as described in Appendix~\ref{app:cmb_prior}) and a uniform prior $P(w_a) = U(-3,3)$.

\bsp	
\label{lastpage}
\end{document}